\documentclass[prd,preprintnumbers,amsmath,amssymb,superscriptaddress]{revtex4}  
\usepackage{graphicx}
\usepackage{epsfig}
\usepackage{rotating}
\usepackage{dcolumn}
\usepackage{bm}
\newcommand{\Emiss}{\mbox{$\not \hspace{-0.10cm} E$ }}

\newcommand{\beq}{\begin{eqnarray}} 
\newcommand{\eeq}{\end{eqnarray}} 

\begin{document}

\preprint{LPT-ORSAY-12-104}

\title{Constraining extra-fermion(s) from the Higgs boson data}

\author{G.~Moreau \\
{\it Laboratoire de Physique Th\'eorique, B\^at. 210, CNRS,
Universit\'e Paris-sud 11 \\  F-91405 Orsay Cedex, France}}

\begin{abstract} 
First, we study the fit of the Higgs boson rates, based on all the latest collider data, in the effective framework for any Extra-Fermion(s) [EF].
The best-fit results are presented in a generic formalism allowing to apply those for the test of any EF scenario, under the assumption that
the corrections to the Higgs couplings are coming exclusively from EF effects. 
The variations of the fit with each one of the five fundamental parameters are described, and, the obtained fits can be better than in the Standard Model (SM). 
We show how the determination of the EF loop-contributions to the Higgs couplings with photons and gluons is relying
on the knowledge of the top and bottom Yukawa couplings (affected by EF mixings); for determining the latter coupling, the relevance of the
investigation of the Higgs production in association with bottom quarks is emphasized.
In the instructive approximation of a single EF, we find that the constraints from the fit already turn out to be quite predictive, 
in both cases of an EF mixed or not with SM fermions,
and especially when combined with the extra-quark (-lepton) mass bounds from direct EF searches at the LHC (LEP) collider. 
In the case of an unmixed extra-quark [in the same color representation as SM quarks], 
non-trivial fit constraints are pointed out on the Yukawa couplings for masses up to 
$\sim 200$~TeV. In particular, we define the {\it extra-dysfermiophilia}, which is predicted at $68.27\%{\rm C.L.}$ for any single 
extra-quark (independently of its electric charge). Another result is that, among any components of SM multiplet extensions, 
the extra-quark with a $-7/3$ electric charge is the one preferred by the present Higgs fit. 
\end{abstract}

\maketitle

\large

\section{Introduction}

Recently, based on the combined LHC data collected at the center-of-mass energies of $\sqrt{s}=7$~TeV and $8$~TeV,  
the ${\rm ATLAS}$~\cite{CERN4thATLAS} and ${\rm CMS}$~\cite{CERN4thCMS} Collaborations have independently announced   
the discovery at the $\sim 5 \sigma$ level of a new resonance -- with a mass close to $125$~GeV -- which can be identified 
as the missing Standard Model (SM) cornerstone~: the Higgs boson~\cite{Englert:1964et,Higgs:1964ia,Higgs:1964pj,Guralnik:1964eu}. 
The long list of measurements of the various Higgs boson rates provided during these last months by the two LHC Collaborations~\cite{ATLASweb,CMSweb} 
constitutes a new precious source of experimental results which can be exploited to test and constrain indirectly theories beyond the SM.

Most of the theories, underlying the SM and addressing the gauge hierarchy problem, predict the existence of new fermions, like
charginos/neutralinos in supersymmetry, fermionic Kaluza-Klein (KK) excitations in higher-dimensional scenarios ({\it e.g.} Gauge-Higgs unification 
frameworks as in Ref.~\cite{Carena:2006bn} or the warped extra-dimension setup~\cite{Gogberashvili:1998vx,Randall:1999ee} 
with matter in the bulk~\cite{Gherghetta:2000qt,Huber:2000ie,Huber:2001ug,Huber:2002gp,Huber:2003sf,Chang:2005ya,Moreau:2006np,Moreau:2005kz,Agashe:2004cp,Agashe:2004ay,Agashe:2006wa,Agashe:2006iy,delAguila:2008iz,Raidal:2008jk,Grossman:1999ra,Appelquist:2002ft,Gherghetta:2003he,Moreau:2004qe,Bouchart:2011va,Goertz:2011hj}), 
excited resonances of bounded states in the dual composite 
Higgs~\cite{Kaplan:1983fs,Kaplan:1983sm,Contino:2003ve,Agashe:2004rs,Contino:2006qr,Burdman:2007sx,DaRold:2010as,Azatov:2011qy} 
or composite top~\cite{Hill:1991at,Pomarol:2008bh} 
models and top quark multiplet components in the little Higgs context~\cite{ArkaniHamed:2001nc,ArkaniHamed:2002qx,ArkaniHamed:2002qy}.
Additional fermions could also arise as fourth generations~\cite{Ishiwata:2011hr}  
or as components embedded {\it e.g.} in simple ${\rm SU(5)}$ representations of gauge unification theories~\cite{Kilic:2010fs}.   

\hspace{0.2cm}

In the first part of this paper, 
we will combine all the Higgs rate measurements to constrain any model with extra-fermions [i.e. of any baryon/lepton number, Yukawa/gauge coupling]  
that are able to induce corrections to the Higgs couplings~\footnotemark[1]\footnotetext[1]{\label{FootMh}The 
extra-fermions are assumed to be heavier than the Higgs field to avoid new Higgs decay
openings (in particular invisible decays into stable particles) that would require special treatments.}.
We will assume that the presence of Extra-Fermion(s) [EF] constitutes the only origin of significant deviations to the Higgs interactions. 
Note that 
our results also apply to any model with extra scalar field(s) or vector boson(s) leading to significant Higgs interaction deviations, 
but not through their mixing(s) respectively with the Higgs boson or SM gauge bosons ({\it c.f.} end of Section~\ref{sec:EffLag}).
By using a generic parametrization, we will determine the corrections to the Higgs couplings -- coming from fermion mixing or new loop-level exchanges -- 
which are favored by the fits of the Higgs boson rates. 
We will show that the best Higgs rate fits obtained could be seen as first indirect indications of the presence of EF since those fits can be better than the SM fit;  
another way of seeing this indication will be to observe that the best-fit regions for the EF-induced corrections to the Higgs couplings 
do not contain the vanishing-correction point (SM point). 
\\
In the second part of the paper, the Higgs fit constraints will be applied to characteristic and well-motivated classes of single EF scenarios (extra-quark/lepton) 
and will reveal themselves to be already quite predictive. We will focus on single EF in same color representations as the SM quarks or leptons; various [including extreme]  
electric charges will be considered for the extra-quark whereas the extra-lepton will be assumed to have the same charge as the SM charged leptons. 

\hspace{0.2cm}

Let us close the introduction by comparing our analysis to the related literature.
The constraints from Higgs rate fits on corrections to the Higgs couplings, induced exclusively by EF, have been partly studied 
in analyses aimed at studying all the possible types of 
corrections~\cite{Barger:2012hv,Low:2012rj,Corbett:2012dm,Giardino:2012ww,Giardino:2012dp,Ellis:2012rx,Ellis:2012hz,Azatov:2012bz,Azatov:2012rd,Montull:2012ik,Espinosa:2012ir,Espinosa:2012im,Carmi:2012in,Banerjee:2012xc,Plehn:2012iz} (see Ref.~\cite{CONF-2012-127} for a statistical analysis by the ATLAS Collaboration). 
A first extension of the present work is to 
describe qualitatively and quantitatively the effect of varying the correction to the bottom-quark Yukawa coupling [parametrized here by $c_b$,
the ratio of the bottom Yukawa coupling over its SM prediction] on constraints for other Higgs couplings; similarly, we study the
dependence of the rate fit on $c_\tau$, namely the ratio of the tau-lepton Yukawa coupling over its SM value (without the simplifying assumption $c_\tau=c_b$).
Another extension is the inclusion of the data on the Higgs production in association with a top-quark pair (relying on the top ratio $c_t$) and on the Higgs decay channel
$h\to\bar\tau\tau$ (involving $c_\tau$) which can play a role in constraining fermion-mixings.
Because of the inclusion of the former data, we do not integrate out the top quark which allows us to explicitly study the $c_t$ parameter (and we do not take {\it e.g.}
$c_t=c_b$)~: we point out in particular that the $c_t$ variation leads to simple translations of the best-fit domains obtained.  
\\ 
Let us note that our fits are performed over the three free parameters $c_b$, $c_{gg}$ and $c_{\gamma\gamma}$ (related to the $hgg$ and $h\gamma\gamma$ coupling
corrections defined later) for characteristic fixed values of $c_\tau$ and $c_t$~\footnotemark[2]\footnotetext[2]{In 
order to explain clearly the influences of these five relevant parameters on the 
Higgs rate fit, we do not marginalize any of those parameters.}. In a second step, we fix $c_b$ for studying examples of EF scenarios.

\hspace{0.2cm}

In Section~\ref{modelbuild}, we discuss the theoretical context and the formalism used. 
Then the measurements of the Higgs boson rates are summarized in Section~\ref{Hdata} and confronted to the 
parameter space of EF scenarios in Section~\ref{fits}. In the part~\ref{procedure} we describe the fit procedure
and in the part~\ref{TRexpl} we present the numerical results while in the part~\ref{sec:ExEF} we
study the simplified case of a unique EF.
We conclude in Section~\ref{conclu}.

\section{Theoretical framework}
\label{modelbuild} 

\subsection{The physical context}

We consider the general framework with any EF able to modify the Higgs couplings.
In our context, no other source of physics beyond the SM is responsible for deviations of the Higgs couplings;
this choice allows to concentrate one's efforts on the class of models with EF and in turn to have a deeper analysis of the parameter space. 
In particular, we assume the Higgs scalar field to receive no coupling modifications due to
significant mixings with other scalars as it can occur {\it e.g.} in extended Higgs sectors.

For example, such a framework could be realized concretely in warped extra-dimension scenarios where some so-called custodians (fermionic KK 
modes)~\cite{Agashe:2003zs,Agashe:2006at,Djouadi:2006rk,Carena:2007ua,Djouadi:2007eg,Djouadi:2007fm,Ledroit:2007ik,Bouchart:2008vp,Bouchart:2009vq,Djouadi:2009nb,Casagrande:2010si,Djouadi:2011aj} would be  
below the TeV scale inducing {\it e.g.} large top mixings, while the decoupling KK gauge boson excitations would be much above $\sim 3$~TeV 
(the order of the lower bound from Electro-Weak (EW) precision tests~\cite{Agashe:2003zs,delAguila:2003bh,Cabrer:2011qb}) forbidding in particular 
significant corrections to the Higgs couplings with gauge bosons.  

From a more basic point of view, in a bottom-up approach without prejudice, this hypothesis that mainly EF affect the Higgs observables is one 
simple possibility, among others, to be considered. This possibility has been considered for instance in 
Ref.~\cite{Dawson:2012di,Carena:2012xa,Azatov:2012rj,Bonne:2012im,Joglekar:2012vc,Kearney:2012zi,Voloshin:2012tv,Batell:2012ca} where the sole effects 
from some EF species -- namely the vector-like fermions (which can arise in many SM extensions) -- on the Higgs production cross sections and
branching ratios were considered.  

In a different context from here, other sources of large Higgs coupling deviations could exist as well -- 
like extra-bosons below $\sim 10$~TeV as could be needed {\it e.g.} in a UV completion theory allowing a vacuum stability in the presence
of new fermions at the EW energy scale with large Yukawa couplings~\cite{ArkaniHamed:2012kq}; 
then the present results might be used to understand specifically the impact of EF on the Higgs rate fits.

\hspace{0.2cm}

Since we adopt a generic approach, we will not make assumptions in particular regarding the EF representations under the ${\rm SU(2)_L}$ gauge group.  
Hence it will not be possible to study EW precision tests on EF as those tests depend on the ${\rm SU(2)_L}$ isospins of EF.    
Such tests can be performed once a given EF model is chosen, like for instance in Ref.~\cite{Azatov:2012rj,Bonne:2012im,Joglekar:2012vc,Almeida:2012bq}
where it was shown that some EF models can pass the EW constraints.

\subsection{The effective Lagrangian}
\label{sec:EffLag}

In our framework, all the Higgs couplings receiving corrections can be written in the following effective Lagrangian, 
which allows to work out the current Higgs phenomenology at the LHC and Tevatron colliders~:   
\begin{eqnarray} 
{\cal L}_h  & = &  - \ c_t Y_t \ h \ \bar t_L \ t_R \ - c_b Y_b \ h \ \bar b_L \ b_R \ - c_\tau Y_\tau \ h \ \bar \tau_L \ \tau_R 
\nonumber \\  
& & + \ C_{h\gamma\gamma} \frac{\alpha}{\pi v} \ h \ F^{\mu\nu} F_{\mu\nu}   
\ + \ C_{hgg} \frac{\alpha_s}{12 \pi v} \ h \ G^{a\mu\nu} G^a_{\mu\nu} \  + \ {\rm h.c.}
\label{Eq:LagEff}
\end{eqnarray} 
where $Y_{t,b,\tau}$ are the SM Yukawa coupling constants of the associated fermions in the mass eigenbasis, $v$ is the Higgs vacuum expectation value,
the subscript $L/R$ indicates the fermion chirality   
and the tensor fields in the $h\gamma\gamma$ and $hgg$ coupling terms (following {\it e.g.} the normalization adopted in Ref.~\cite{Carmi:2012in})
are respectively the electromagnetic and gluon field strengths. The $c_{t,b,\tau}$ parameters -- taken real for simplicity -- 
are defined such that the limiting case $c_{t,b,\tau}\to 1$ corresponds to the SM;  
deviations from unity of those parameters can be caused by mixings of EF (like $t'$ states,\dots) with the SM fermions.  
Only the Yukawa couplings of the third generation are supposed to receive potentially important corrections from EF mixing effects 
since EF are closer in mass to the third generation and this heavy generation is in general more intimately connected to the ultraviolet physics, like the
top quark in warped/composite frameworks.

A few remarks are in order regarding terms absent from the Lagrangian~(\ref{Eq:LagEff}).  
First, we only consider tree-level (loop-level) corrections to couplings induced at the tree-level (loop-level) in the SM, i.e. we calculate exclusively the dominant corrections; 
in the absence of tree-level correction from EF origins for a certain SM tree-level induced coupling, we do not go to the next order so that the global analysis coherence is preserved. 
Secondly, we have not included in the Lagrangian the $hZ\gamma$ coupling~\cite{Djouadi:2005gi} 
as it is not constrained by a dedicated experimental analysis {\it e.g.} in the $Z\gamma$ channel, and, 
the EF-induced corrections to the relatively small $\Gamma(h\to Z\gamma)$ width 
are expected to be too weak to change significantly the total Higgs width (involved in all branching fractions). For similar reasons, we have not
considered flavor-changing Yukawa couplings (those are not excluded in some EF scenarios and could induce new partial Higgs decay widths). 

Ê
Let us make another comment about the Lagrangian~(\ref{Eq:LagEff}).  
Neglecting the mixings with the first two SM flavors, one gets, $- Y_{t,b,\tau} = m_{t,b,\tau} / v$
[the minus sign is due to the sign taken in front of the Yukawa couplings in Eq.~(\ref{Eq:LagEff})], 
where $m_{t,b,\tau}$ are the final masses generated after EW symmetry breaking. 
The EF mixing effect on the Yukawa couplings enters via the $c_{t,b,\tau}$ parameters. These parameter values also contain the $3\times 3$ 
SM flavor mixing effect in case it is not neglected. This $3\times 3$ mixing is considerable in the lepton sector (while CKM mixing angles~\cite{Beringer:1900zz} 
are typically small) but there a possibility is that the strongest mixing angles originate from the neutrino mass matrix. Now even if a Higgs decay channel into neutrinos is open,
like in the simple case of added right-handed neutrino singlets leading to neutrino Yukawa couplings, 
the partial width into neutrinos would typically be so tiny compared to others  
-- even for huge neutrino Yukawa coupling enhancements by say two orders of magnitude -- that it would not affect the Higgs fit analysis.

Summing over the dominant loop contributions, the coefficients of the dimension-five ope\-rators in Eq.~(\ref{Eq:LagEff}) can be written as, 
{\normalsize
\begin{eqnarray} 
C_{hgg} & = & 2 C(t) \ A[\tau(m_t)] \ (c_t + c_{gg}) \ + \ 2 C(b) \ A[\tau(m_b)] \ c_b \ + \ 2 C(c) \ A[\tau(m_c)], 
\label{Eq:Cglgldef}
\\ 
C_{h\gamma\gamma} & = & \frac{N^t_c}{6} Q_t^2 A[\tau(m_t)] \ (c_t + c_{\gamma\gamma}) + \frac{N^b_c}{6} Q_b^2 A[\tau(m_b)] \ c_b  
+  \frac{N^c_c}{6} Q_c^2 A[\tau(m_c)] 
+  \frac{N^\tau_c}{6} Q_\tau^2 A[\tau(m_\tau)] \ c_\tau +  \frac{1}{8} A_1[\tau(m_W)],  \nonumber \\ 
\label{Eq:Cgagadef}
\end{eqnarray} 
}
where $m_c$ ($m_W$) is the charm quark ($W^\pm$-boson) mass, 
$C({\rm r})$ is defined for the color representation, ${\rm r}$, by ${\rm Tr}(T^a_{\rm r}T^b_{\rm r})=C({\rm r})\delta^{ab}$ [$T^a$ denoting the eight generators of ${\rm SU(3)_c}$],  
$N^f_c$ is the number of colors for the fermion $f$, $Q_f$ is the electromagnetic charge for $f$, 
$A[\tau(m)]$ and $A_1[\tau(m)]$ are respectively the form factors for spin~1/2 and spin~1 particles~\cite{Djouadi:2005gi,Cacciapaglia:2009ky} 
normalized such that $A[\tau(m)\ll 1]\to 1$ and $A_1[\tau(m)\ll 1]\to -7$ with $\tau(m)=m_h^2/4m^2$ (for $m_h \simeq 125$~GeV one has
$A_1[\tau(m_W)] \simeq -8.3$ whereas $A[\tau(m>600\mbox{GeV})] \simeq 1.0$). The terms proportional to $c_t$, $c_b$ and $c_\tau$ account for 
the contributions from the fermionic triangular loops involving respectively the top, bottom quark and tau lepton Yukawa coupling.
The $A[\tau(m_c)]$ and $A_1[\tau(m_W)]$ terms are for the SM loop-exchanges of the charm quark and $W^\pm$-boson. The dimensionless 
$c_{gg}$ and $c_{\gamma\gamma}$ quantities -- vanishing in the SM -- parametrize the EF loop-exchange  
contributions to the $hgg$ and $h\gamma\gamma$ couplings. This choice of parametrization in Eq.~(\ref{Eq:Cglgldef}) with a common factor in 
front of $c_t$ and $c_{gg}$ [as well as for $c_t$ and $c_{\gamma\gamma}$ in Eq.~(\ref{Eq:Cgagadef})] makes easier the understanding of the $c_t$ 
influence on the best-fit $c_{gg}$ [or $c_{\gamma\gamma}$] ranges, that will be discussed in Section~\ref{TRexpl}. 
\\
Note also that extra scalar field(s), unmixed with the Higgs boson $h$ (like a squark in supersymmetry), 
or extra vector boson(s), unmixed with the SM gauge bosons,
could affect the Higgs couplings only through new loop-contributions to the $c_{gg}$ and $c_{\gamma\gamma}$ quantities studied here.

\subsection{Higgs rate modifications}
\label{HRModif}

Within the present context, let us write explicitly certain Higgs rates, normalized to their SM prediction, which will prove to be useful in the following.
The expression for the cross section of the gluon-gluon fusion mechanism of single Higgs production, over its SM prediction, reads as (for the LHC
or Tevatron),   
{\normalsize \begin{eqnarray} 
\frac{\sigma_{\rm gg \to h}}{\sigma_{\rm gg \to h}^{\rm SM}} \ \simeq \ 
\frac{\big \vert (c_t + c_{gg}) A[\tau(m_t)] + c_b A[\tau(m_b)] + A[\tau(m_c)] \big \vert^2}
{\big \vert A[\tau(m_t)] + A[\tau(m_b)] + A[\tau(m_c)] \big \vert^2} \ .
\label{Eq:fusionR}
\end{eqnarray} }  
The expression for the ratio of the diphoton partial decay width over the SM expectation is,   
{\normalsize \begin{eqnarray} 
\frac{\Gamma_{\rm h\to \gamma\gamma}}{\Gamma_{\rm h\to \gamma\gamma}^{\rm SM}} \ \simeq  \ 
\frac{\big \vert \frac{1}{4} A_1[\tau(m_W)] + (\frac{2}{3})^2 (c_t + c_{\gamma\gamma}) 
A[\tau(m_t)] + (-\frac{1}{3})^2 c_b A[\tau(m_b)] + (\frac{2}{3})^2 A[\tau(m_c)] + \frac{1}{3} c_\tau A[\tau(m_\tau)] \big \vert^2}
{\big \vert \frac{1}{4} A_1[\tau(m_W)] + (\frac{2}{3})^2 
A[\tau(m_t)] + (-\frac{1}{3})^2 A[\tau(m_b)] + (\frac{2}{3})^2 A[\tau(m_c)] + \frac{1}{3} A[\tau(m_\tau)] \big \vert^2} \ .
\label{Eq:diphoR}
\end{eqnarray} }
The ratios for the partial decay widths into the bottom quark and tau lepton pairs as well as for the cross section of Higgs production
in association with a top pair (LHC or Tevatron) are given by,
{\normalsize \begin{eqnarray} 
\frac{\Gamma_{\rm h\to \bar b b}}{\Gamma_{\rm h\to \bar b b}^{\rm SM}} \simeq  \vert c_b \vert^2 \ , \ \ \ 
\frac{\Gamma_{\rm h\to \bar \tau \tau}}{\Gamma_{\rm h\to \bar \tau \tau}^{\rm SM}} \simeq  \vert c_\tau \vert^2 \ , \ \ \ 
\frac{\sigma_{\rm h \bar t t}}{\sigma_{\rm h \bar t t}^{\rm SM}} \simeq  \vert c_t \vert^2 \ .
\label{Eq:otherR}
\end{eqnarray} }

Let us make a comment related to the mass insertion in the triangular loops of fermions inducing the $h\gamma\gamma$ and $hgg$ couplings. Strictly 
speaking, a factor $\epsilon_t$, equal to the ratio of the sign of $m_t$ in the SM over sign($m_t$) in the EF scenario, should
multiply $c_t$ in Eq.~(\ref{Eq:Cglgldef})-(\ref{Eq:Cgagadef}) or Eq.~(\ref{Eq:fusionR})-(\ref{Eq:diphoR}) [similarly for $\epsilon_b c_b$ and $\epsilon_\tau c_\tau$]; in other words,
if for instance $\epsilon_t=-1$ the values for $c_t$ obtained below would have to be interpreted instead as values for $-c_t$ (the observables
of Eq.~(\ref{Eq:otherR}) being insensitive to the $c_{t,b,\tau}$ signs).

\subsection{Ratio of $c_{\gamma\gamma}$ and $c_{gg}$}

For a better understanding of the above parametrization, we finally provide the examples of expressions for the $c_{gg}$ and $c_{\gamma\gamma}$ quantities,  
in the case of the existence of a $t'$ quark [same color number and electromagnetic charge as the top]
(possibly vector-like as {\it e.g.} in Ref.~\cite{Cacciapaglia:2010vn,Cacciapaglia:2011fx}), an exotic $q_{5/3}$ quark with 
electromagnetic charge $5/3$ and an additional $\ell'$ lepton (colorless), in terms of their physical Yukawa couplings and mass eigenvalues~: 
{\normalsize \begin{eqnarray} 
& c_{gg} & = \frac{1}{C(t) A[\tau(m_t)]/v} \bigg [ -C(t') \frac{Y_{t'}}{m_{t'}} A[\tau(m_{t'})] - C(q_{5/3}) \frac{Y_{q_{5/3}}}{m_{q_{5/3}}} A[\tau(m_{q_{5/3}})] + \dots \bigg ]  ,  
\label{Eq:cexamI}
\\ 
& c_{\gamma\gamma} & = \frac{1}{N^t_c Q_t^2 A[\tau(m_t)]/v} \bigg [ -3 \bigg ( \frac{2}{3} \bigg )^2 \frac{Y_{t'}}{m_{t'}} A[\tau(m_{t'})] 
- N^{q_{5/3}}_c \bigg ( \frac{5}{3} \bigg )^2 \frac{Y_{q_{5/3}}}{m_{q_{5/3}}} A[\tau(m_{q_{5/3}})] - Q_{\ell'}^2 \frac{Y_{\ell'}}{m_{\ell'}} A[\tau(m_{\ell'})] + \dots \bigg ]   . 
\nonumber \\ 
\label{Eq:cexamII}
\end{eqnarray} } 
The dots stand for any other EF loop-contributions. The mass assumption made in Footnote~[\ref{FootMh}] leads to real $A[\tau(m_{f'})]$ functions and thus real 
$c_{gg}$, $c_{\gamma\gamma}$ values, for real masses and Yukawa coupling constants, as appears clearly in the two above expressions.

It will turn out to be instructive to express the ratio of these parameters in the simplified scenario where a 
new single $q'$ quark is affecting the Higgs couplings; denoting its electromagnetic charge as $Q_{q'}$ and assuming the 
$q'$ to have the same color representation as the top quark, this ratio reads as~: 
{\normalsize \begin{eqnarray} 
\frac{c_{\gamma \gamma}}{c_{gg}} \bigg \vert_{q'} \  =  \  \frac{Q_{q'}^2}{(2/3)^2}  \ Ê.  
\label{Eq:Exqp}
\end{eqnarray} } 
This ratio takes indeed a simple form that will be exploited in Section~\ref{sec:ExEF}. In particular, notice that $c_{\gamma \gamma} \vert_{t'}  = c_{gg} \vert_{t'}$.
Clearly, $q'$ should have non-vanishing Yukawa couplings to satisfy Eq.~(\ref{Eq:Exqp}), otherwise $c_{\gamma \gamma} \vert_{q'}  = c_{gg} \vert_{q'} =0$.
In the specific case of a vector-like $q'_{L/R}$, this one could for example constitute a singlet under the ${\rm SU(2)_L}$ gauge group and have a 
Yukawa coupling with another $q''_{R/L}$ state of same $Q_{q'}$ charge but embedded in a ${\rm SU(2)_L}$ doublet; then 
the heaviest $q^{(2)}_{L/R}$ mass eigenstate, composed of $q'_{L/R}$ and $q''_{L/R}$, could decouple from the Higgs sector so that the orthogonal
$q^{(1)}_{L/R}$ composition would represent the considered unique new quark influencing significantly the Higgs couplings.

\section{The Higgs boson data}
\label{Hdata}

All the Higgs rates which have been measured at the Tevatron and LHC [for $\sqrt{s}=7$ and $8$~TeV] are defined in this section.
The references with their latest experimental values are also given below (these values have been summarized in Ref.~\cite{Espinosa:2012im}). 
\\
Generically, the measured observables are the signal strengths whose theoretical predictions read as 
(in the narrow width approximation as used in Ref.~\cite{CONF-2012-127}),
\begin{eqnarray}
\mu^p_{s,c,i} \ \simeq \ \frac{\sigma_{\rm gg\to h}\vert_{s} \ + \ \frac{\epsilon_{\rm hqq}}{\epsilon_{\rm gg\to h}}\vert^p_{s,c,i} \ \sigma^{\rm SM}_{\rm hqq}\vert_{s}
\ + \  \frac{\epsilon_{\rm hV}}{\epsilon_{\rm gg\to h}}\vert^p_{s,c,i} \ \sigma^{\rm SM}_{\rm hV}\vert_{s} 
\ + \  \frac{\epsilon_{\rm h\bar tt}}{\epsilon_{\rm gg\to h}}\vert^p_{s,c,i} \ \sigma_{\rm h\bar tt}\vert_{s}}
{\sigma_{\rm gg\to h}^{\rm SM}\vert_{s} \ + \  \frac{\epsilon_{\rm hqq}}{\epsilon_{\rm gg\to h}}\vert^p_{s,c,i} \ \sigma_{\rm hqq}^{\rm SM}\vert_{s}
\ + \  \frac{\epsilon_{\rm hV}}{\epsilon_{\rm gg\to h}}\vert^p_{s,c,i} \ \sigma_{\rm hV}^{\rm SM}\vert_{s} 
\ + \  \frac{\epsilon_{\rm h\bar tt}}{\epsilon_{\rm gg\to h}}\vert^p_{s,c,i} \ \sigma_{\rm h\bar tt}^{\rm SM}\vert_{s}} \ 
\frac{B_{\rm h\to XX}}{B_{\rm h\to XX}^{\rm SM}} \ ,    
\nonumber
\end{eqnarray}
{\normalsize  \begin{eqnarray}
\mbox{with,} \  \ Ê\  \  \  \  \  \  \   \  \ Ê\  \  \  \  \  \  \  
\sigma_{\rm gg\to h}\vert_{s} = \frac{\sigma_{\rm gg\to h}}{\sigma_{\rm gg\to h}^{\rm SM}} \sigma_{\rm gg\to h}^{\rm SM}\vert_{s} \ , \  \
\sigma_{\rm h\bar tt}\vert_{s} = \frac{\sigma_{\rm h\bar tt}}{\sigma_{\rm h\bar tt}^{\rm SM}} \sigma_{\rm h\bar tt}^{\rm SM}\vert_{s} \ , \  \
\nonumber \\  
\Gamma_{\rm h\to \gamma\gamma} = \frac{\Gamma_{\rm h\to \gamma\gamma}}{\Gamma_{\rm h\to \gamma\gamma}^{\rm SM}} \Gamma_{\rm h\to \gamma\gamma}^{\rm SM} \ , \ \
\Gamma_{\rm h\to \bar b b} = \frac{\Gamma_{\rm h\to \bar b b}}{\Gamma_{\rm h\to \bar b b}^{\rm SM}} \Gamma_{\rm h\to \bar b b}^{\rm SM} \ , \ \
\Gamma_{\rm h\to \bar \tau \tau} = \frac{\Gamma_{\rm h\to \bar \tau \tau}}{\Gamma_{\rm h\to \bar \tau \tau}^{\rm SM}} \Gamma_{\rm h\to \bar \tau \tau}^{\rm SM} \ , \ \ 
\label{Eq:mu}  
\end{eqnarray} } 
where the $p$-exponent labels the Higgs channel defined by its production and decay processes, 
the $s$-subscript represents the squared of the energy [we will note $\sqrt{s}=1.96,7,8$ in TeV] of the realized measurement, 
the $c$-subscript stands for the experimental collaboration (CDF and D0 at the Tevatron, ${\rm ATLAS}$ or ${\rm CMS}$ at LHC) having performed the measurement
and $i$ is an integer indicating the event cut category considered.   
$\sigma_{\rm hqq}$ is the predicted cross section for the Higgs production in association with a pair of light SM quarks and 
$\sigma_{\rm hV}$ is for the production in association with a gauge boson [$V\equiv Z^0,W^\pm$ bosons];
their $s$-subscript indicates the energy and in turn which collider is considered.   
The $B_{\rm h\to XX}$ (X stands for any possible final state particle) are the branching ratios defined from all the opened Higgs decay widths 
which are modified according to the second line of Eq.~(\ref{Eq:mu}) and taken as in the SM for the others.
The SM rates at LHC for a given energy, like $\sigma_{\rm gg\to h}^{\rm SM}\vert_{s}$, and the SM partial widths, $\Gamma_{\rm h\to XX}^{\rm SM}$, 
are taken from Ref.~\cite{XsectionWepPage} (including the cross section corrections at next-to-next-to leading order in QCD and
next-to leading order in the EW sector, except for $\sigma_{\rm h\bar tt}^{\rm SM}$ at next-to leading order in QCD), while
the SM rates at Tevatron are from Ref.~\cite{:2012zzl} (QCD corrections at next-to-next-to leading order). 
The cross section and partial width ratios in the second line of Eq.~(\ref{Eq:mu}) are those in the considered effective theory 
with EF expressed in Eq.~(\ref{Eq:fusionR})-(\ref{Eq:diphoR})-(\ref{Eq:otherR}). The EW/QCD corrections are expected typically   
to be compensated in these ratios (especially for heavy EF in the same gauge group representation as the SM fermions). Finally,  
$\epsilon_{\rm gg\to h}$, for the $\rm gg\to h$ reaction example, is the experimental efficiency [detector acceptance, particle identification, isolation,\dots]
including the (kinematical) selection cut effects;
the efficiency ratios entering Eq.~(\ref{Eq:mu}) are obtained by multiplying the SM cross section ratios by the ratios of expected Higgs reaction compositions (in $\%$) 
-- derived via simulations and provided in the relevant experimental papers [see just below]. 
These selection efficiencies, relying on the Higgs mass, are identical in the SM and in EF frameworks (i.e. in the denominator and numerator of $\mu^p_{s,c,i}$). 
\\ \\
Here is the list of Higgs channels that have been experimentally investigated (corresponding, once summed, to 55 measured signal strengths)~:
\\ \\
$\bullet$ \ For the process $I$, ${\rm pp\to h, \ h\to \gamma\gamma}$, the Higgs field is mainly produced by the gluon-gluon fusion; 
the signal strengths $\mu^I_{7/8,{\rm ATLAS/CMS},i}$ are proportional to $B_{\rm h\to \gamma\gamma}$ and depend on the efficiency ratios like {\it e.g.}  
$\epsilon_{\rm hqq}/\epsilon_{\rm gg\to h}\vert^I_{7/8,{\rm ATLAS/CMS},i}$ which can be derived from the reaction compositions provided in Ref.~\cite{CONF-2012-091}
(${\rm ATLAS}$) and Ref.~\cite{PAS-HIG-12-015} updated by Ref.~\cite{:2012gu} (${\rm CMS}$).
While for ${\rm ATLAS}$ nine cut categories ($i=1,\dots,9$) have been applied on the data collected at $\sqrt{s}=7$~TeV in 2011 ($4.8$~fb$^{-1}$) 
and $8$~TeV in 2012 ($5.8$~fb$^{-1}$) -- leading to a measured mass $m_h\simeq 126.0$~GeV after combination with other channels~\cite{:2012gk} --  
${\rm CMS}$ has chosen four cut classes ($j=0,\dots,3$) to treat the 2011 ($5.1$~fb$^{-1}$) and 2012 ($5.3$~fb$^{-1}$) data -- pointing out a mass 
$m_h\simeq 125.3$~GeV from combination with the $ZZ$ channel.
Note that in Eq.~(\ref{Eq:mu}), the terms,  
{\normalsize
$$\frac{\epsilon_{\rm hZ}}{\epsilon_{\rm gg\to h}}\vert^I_{7/8,{\rm ATLAS},i} \ \sigma^{\rm SM}_{\rm hZ}\vert_{7/8} 
+ \frac{\epsilon_{\rm hW}}{\epsilon_{\rm gg\to h}}\vert^I_{7/8,{\rm ATLAS},i} \ \sigma^{\rm SM}_{\rm hW}\vert_{7/8} \ ,$$ for the ${\rm ATLAS}$ data must be replaced by,  
$(\epsilon_{\rm hZ+hW}/\epsilon_{\rm gg\to h}\vert^I_{7/8,{\rm CMS},j})(\sigma^{\rm SM}_{\rm hZ}+\sigma^{\rm SM}_{\rm hW})\vert_{7/8}\ ,$ for ${\rm CMS}$} 
(a common efficiency is set).
\\ \\
$\bullet$ \ In the diphoton channel, other series of cuts have been employed to increase the vector boson fusion contribution, ${\rm pp\to hqq, \ h\to \gamma\gamma}$, 
defining the process noted $II$. The signal strengths $\mu^{II}_{7/8,{\rm ATLAS/CMS},i}$ rely on the efficiency ratios obtained 
from the reaction compositions in Ref.~\cite{CONF-2012-091} and Ref.~\cite{PAS-HIG-12-015,:2012gu}.
A unique cut category is selected by ${\rm ATLAS}$, to tag the dijet final state,  
whereas two of them ($i\equiv$tight,loose) are used with the ${\rm CMS}$ data at $\sqrt{s}=8$~TeV.
\\ \\
$\bullet$ \ The last diphoton channel analyzed, process $III$, is the inclusive Higgs production at the Tevatron, ${\rm p\bar p\to h, \ h\to \gamma\gamma}$. 
The $\mu^{III}_{1.96,{\rm CDF+D0}}$ strength is simply fixed by $\epsilon/\epsilon_{\rm gg\to h}\vert^{III}_{1.96,{\rm CDF+D0}}\simeq 1$~\cite{:2012zzl} 
for each Higgs production cross section in Eq.~(\ref{Eq:mu}). 
\\ \\
$\bullet$ \ For the process $IV$, ${\rm pp\to hV \ [V\to \mbox{leptons}], \ h\to \bar b b}$, all selection efficiencies vanish except,  
$\epsilon_{\rm hV} \vert^{IV}_{7/8,{\rm ATLAS/CMS}}$   $\simeq~1$~\cite{CONF-2012-093,:2012gk,PAS-HIG-12-020}  
(of course in such a case, one should not divide by $\epsilon_{\rm gg\to h}$ in Eq.~(\ref{Eq:mu})), 
so that, $\mu^{IV}_{7/8,{\rm ATLAS/CMS}} \simeq B_{\rm h\to \bar b b}/B_{\rm h\to \bar b b}^{\rm SM}$, since $\sigma^{\rm SM}_{\rm hV}$ 
does not receive corrections in the EF framework.
\\ \\
$\bullet$ \ Similarly, for the process $V$, ${\rm p \bar p\to hV \ [V\to \mbox{leptons}], \ h\to \bar b b}$, one has
$\mu^{V}_{1.96,{\rm CDF+D0}} = \mu^{IV}_{7/8,{\rm ATLAS/CMS}}$~\cite{:2012zzl}.
\\ \\
$\bullet$ \ The process $VI$, ${\rm pp\to h \bar t t, \ h \to \bar b b}$, is characterized by vanishing efficiencies except, 
$\epsilon_{\rm h \bar t t}\vert^{VI}_{7,{\rm CMS}} \simeq 1$, leading to,   
{\normalsize \begin{eqnarray}
\mu^{VI}_{7,{\rm CMS}} \ \simeq \ \frac{\sigma_{\rm h \bar t t}}{\sigma_{\rm h \bar t t}^{\rm SM}} \ \frac{B_{\rm h\to \bar b b}}{B_{\rm h\to \bar b b}^{\rm SM}} \ .
\label{eq:muttbb}
\end{eqnarray} }
The experimental value, which will be mentioned in next section, is $\mu^{VI}_{7,{\rm CMS}}\vert_{\rm exp} = -0.75^{+2}_{-1.8}$~\cite{PAS-HIG-12-020}.
\\ \\ $\bullet$ \ The reaction $VII$, ${\rm pp\to h, \ h\to ZZ}$, has a strength $\mu^{VII}_{7/8,{\rm ATLAS/CMS}}$ calculated according to selection efficiencies all equal to unity  
(for ${\rm CMS}$ see Ref.~\cite{PAS-HIG-12-020} and for ${\rm ATLAS}$ Ref.~\cite{CONF-2012-019} at $\sqrt{s}=7$~TeV or Ref.~\cite{:2012gk} at $8$~TeV).
\\ \\
$\bullet$ \ In the same way, for the reaction $VIII$, ${\rm pp\to h, \ h\to WW}$, the strength $\mu^{VIII}_{7/8,{\rm ATLAS}}$ is computed with efficiencies at unity
(see Ref.~\cite{CONF-2012-093} for $7$~TeV and Ref.~\cite{CONF-2012-098} for $8$~TeV, both updated by Ref.~\cite{:2012gk}),     
whereas $\mu^{VIII}_{7/8,{\rm CMS}}$ is based on vanishing efficiencies except $\epsilon_{\rm gg\to h}\vert^{VIII}_{7/8,{\rm CMS}}\simeq 1$~\cite{PAS-HIG-12-020}. 
\\ \\
$\bullet$ \ From analog considerations as in the channel $IV$, one predicts, 
$\mu^{IX}_{7/8,{\rm CMS}} \simeq \mu^{X}_{7,{\rm CMS}} \simeq B_{\rm h\to WW}/B_{\rm h\to WW}^{\rm SM}$, 
for the processes $IX$, ${\rm pp\to hqq, \ h\to WW}$, and $X$, ${\rm pp\to hV, \ h\to WW}$~\cite{PAS-HIG-12-020}. 
\\ \\
$\bullet$ \ The channel $XI$, ${\rm p\bar p\to h, \ h\to WW}$, has a strength $\mu^{XI}_{1.96,{\rm CDF+D0}}$ containing exclusively efficiencies at unity~\cite{:2012zzl}.
\\ \\ 
$\bullet$ \ As in channel $IV$, one has the theoretical predictions, 
$\mu^{XII}_{7/8,{\rm CMS}} \simeq \mu^{XIII}_{7,{\rm CMS}} \simeq B_{\rm h\to \bar \tau\tau}/B_{\rm h\to \bar \tau\tau}^{\rm SM}$, 
for the processes $XII$, ${\rm pp\to hqq, \ h\to \bar \tau\tau}$, and $XIII$, ${\rm pp\to hV, \ h\to \bar \tau\tau}$~\cite{PAS-HIG-12-020}. 
\\ \\
$\bullet$ \ Finally, for the process $XIV$, ${\rm pp\to h, \ h\to \bar \tau\tau}$, the strength $\mu^{XIV}_{7,{\rm ATLAS}}$ 
has the efficiencies equal to one~\cite{CONF-2012-093,:2012gk}
and $\mu^{XIV}_{7/8,{\rm CMS}}$ has all efficiencies equal to zero 
but $\epsilon_{\rm gg\to h}\vert^{XIV}_{7/8,{\rm CMS}}\simeq 1$~\cite{PAS-HIG-12-020}.

\section{The Higgs rate fits}
\label{fits}

\subsection{The fit procedure}
\label{procedure}

In order to analyze the fit of the Higgs boson data from colliders within the effective theory described above, 
we assume gaussian error statistics and we use the $\chi^2$ function,   
\begin{eqnarray}
\chi^2 \ = \ \sum_{p,s,c,i} \frac{(\mu^p_{s,c,i}-\mu^p_{s,c,i}\vert_{\rm exp})^2}{(\delta\mu^p_{s,c,i})^2} \ ,
\label{eq:Chi2def}
\end{eqnarray}
where the sum is taken over all the different channel observables defined in Section~\ref{Hdata} and  
$\mu^p_{s,c,i}\vert_{\rm exp}$ are the measured central values for the corresponding signal strengths. 
$\delta\mu^p_{s,c,i}$ are the uncertainties on these values and are obtained by symmetrizing the provided errors
below and above the central values~: $(\delta\mu^p_{s,c,i})^2 = [(\delta\mu^p_{s,c,i}\vert^+)^2+(\delta\mu^p_{s,c,i}\vert^-)^2]/2$.
$\mu^p_{s,c,i}\vert_{\rm exp}$ and $\delta\mu^p_{s,c,i}\vert^\pm$ are given in the experimental papers listed in
Section~\ref{Hdata} which contain the QCD error estimations.

The summation over all the signal strengths in Eq.~(\ref{eq:Chi2def}) allows to compare the maximum of available experimental information
with the theoretical predictions, in order to optimize the test of the effective EF theory. 
Note that the $i$-subscript in this summation corresponds to exclusive cut categories into which the event samples are split.

The global fit is performed without including the correlation coefficient effects which are currently not supplied in the experimental papers. 
Nevertheless, this does not affect the statistical and uncorrelated systematic errors.

\subsection{Fits in the $\{c_{gg},c_{\gamma\gamma},c_b\}$ space}
\label{TRexpl}

In Eq.~(\ref{eq:Chi2def}), $\chi^2=\chi^2(c_t,c_b,c_\tau,c_{gg},c_{\gamma\gamma})$ depends on the five effective parameters 
$c_t,c_b,c_\tau,c_{gg},c_{\gamma\gamma}$ through Eq.~(\ref{Eq:mu}) and Eq.~(\ref{Eq:fusionR})-(\ref{Eq:diphoR})-(\ref{Eq:otherR}).   
A priori, the fit analysis should be performed over these five free parameters but to still be able to draw plots of the whole parameter space 
(and in turn study it graphically) one has to restrict it to a three-dimensional space. In this section, 
we will indeed choose three freely varying parameters, $c_{gg},c_{\gamma\gamma},c_b$, and search for the
best-fit regions in this three-dimensional space. Then we will show slices of these regions at several chosen values of $c_b$ (i.e. in the plane 
$c_{\gamma\gamma}$ versus $c_{gg}$). This will be repeated for different fixed values of $c_t$ and $c_\tau$.     
\\ 
The motivation for fixing $c_t$ and $c_\tau$, among the five effective parameters, is the following one.  
First, the $\vert c_\tau \vert$ range compatible at $1\sigma$ with the Higgs data is known and turns out to be roughly $[0;\sim1.8]$
(for $c_t \approx 1$ and reasonable $c_b$ values described later on) because the measured values for $\mu^{XII}_{7/8,{\rm CMS}}$ 
are negative -- even with the errors -- so that $B_{\rm h\to \bar \tau \tau}$, 
and in turn $\Gamma_{\rm h\to \bar \tau \tau}$ and $\vert c_\tau \vert$, cannot be too large.
Hence, there is no need to apply the numerical global fit analysis on $c_\tau$, then treated as a free parameter, to find its relevant range.
Secondly, for the purpose of demonstrating the $c_t$ peculiarity (correlation with $c_{gg}$, $c_{\gamma\gamma}$) discussed in Section~\ref{ctDEP}, 
it is easier to choose ourselves its fixed values than to have those values dictated by the numerical best-fit search method. 

\hspace{0.2cm}

\begin{figure}[t]
\begin{center}
\begin{tabular}{cc}
\includegraphics[width=0.47\textwidth,height=8.cm]{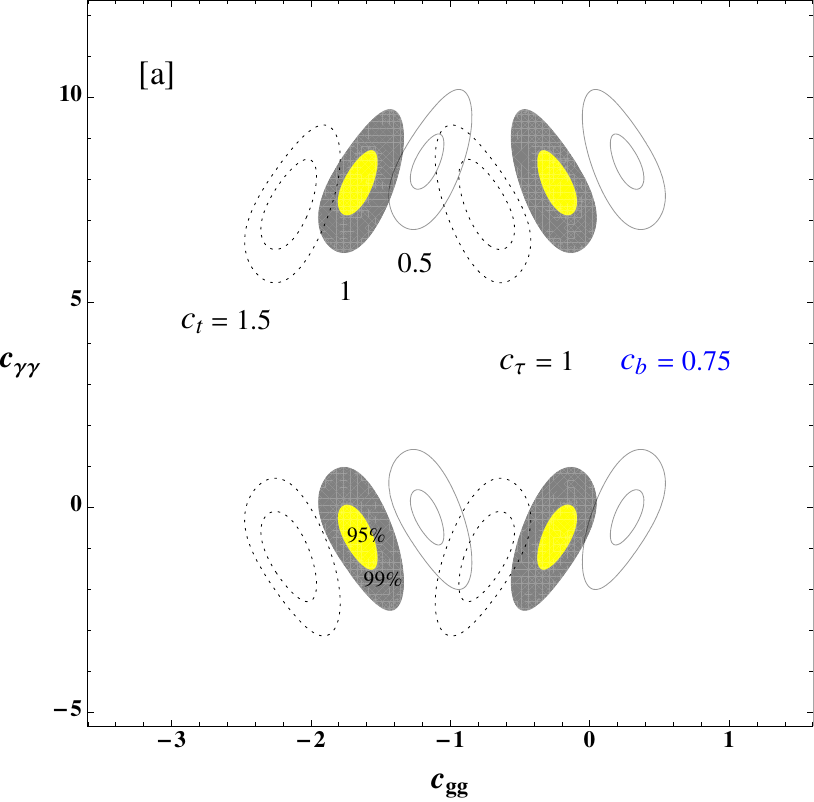}
\hspace{0.5cm}
\includegraphics[width=0.47\textwidth,height=8.cm]{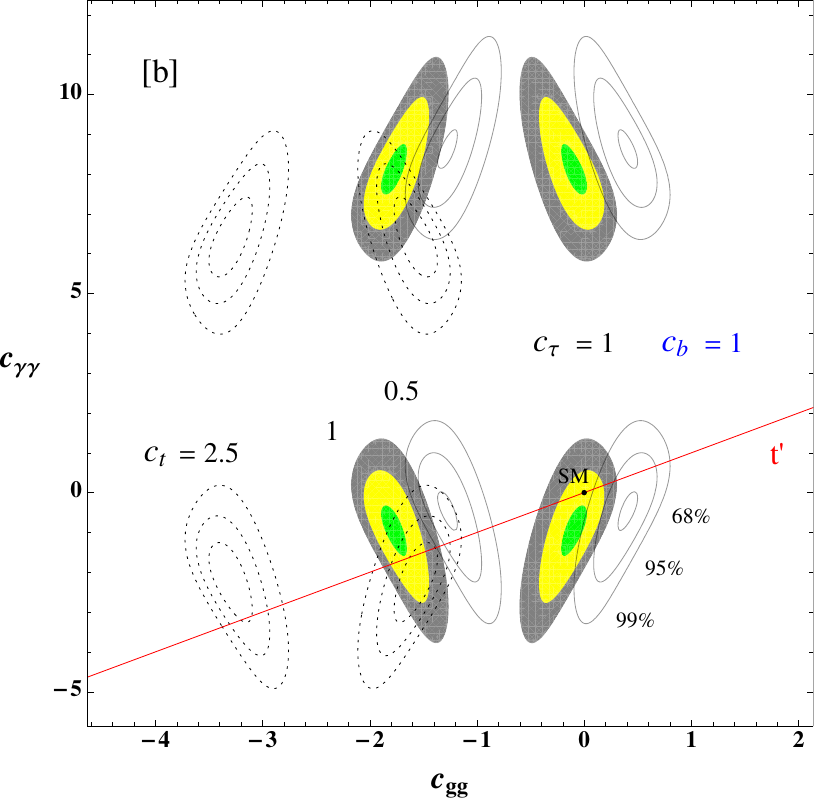}
\vspace{0.1cm}
\\
\includegraphics[width=0.47\textwidth,height=8cm]{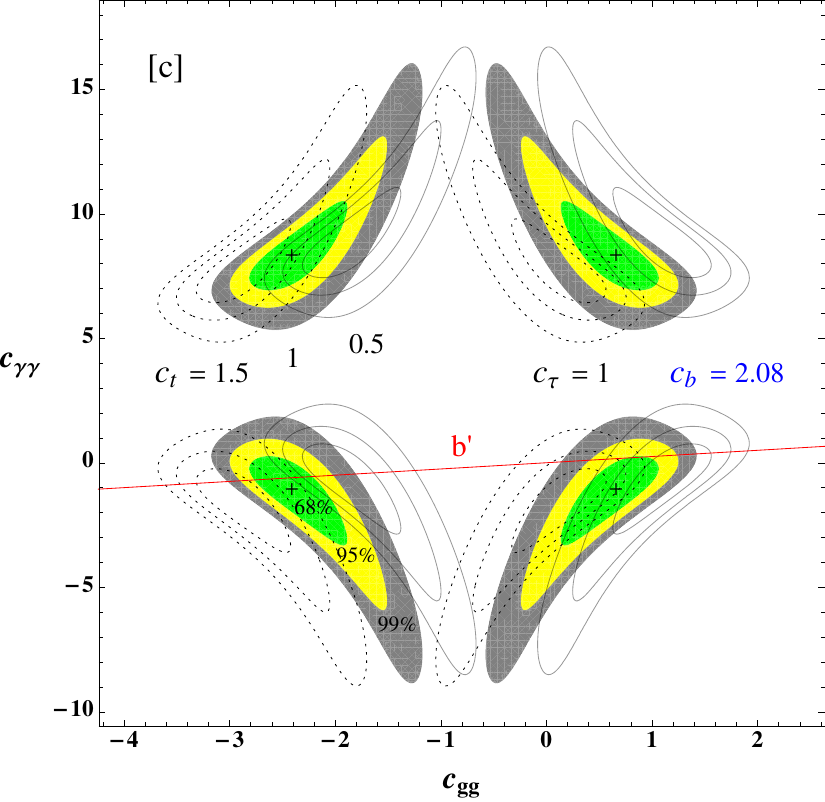}
\hspace{0.5cm}
\includegraphics[width=0.47\textwidth,height=8cm]{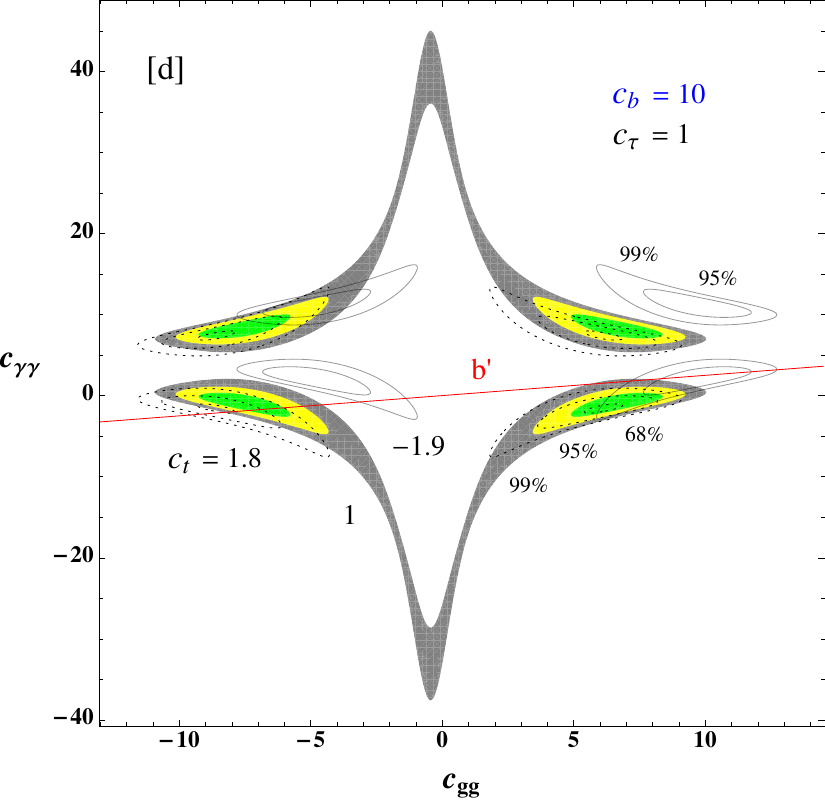}
\\
\end{tabular}
\caption{Best-fit regions at $68.27\%{\rm C.L.}$ (in green), $95.45\%{\rm C.L.}$ (yellow) and $99.73\%{\rm C.L.}$ (grey) in the plane
$c_{\gamma\gamma}$ versus $c_{gg}$, for $c_\tau=1$. Each one of the four figures [a,b,c,d] is associated to a certain $c_b$ value 
written (in blue) on the figure itself. 
In each figure, the regions are drawn for three $c_t$ values, the corresponding value being indicated nearby the relevant region;
the regions for the lowest, intermediate, highest $c_t$ values are respectively shown by the plain contours, colored filled domains, dotted contours.
The SM (black) point, at $c_t=c_b=c_\tau=1$, $c_{\gamma\gamma}=c_{gg}=0$, is shown on the plot [b]. 
Finally, the four best-fit point locations are indicated by crosses in the plot [c].
The theoretically predicted lines for extra-quarks of type $b'$ (plot [c,d]) and $t'$ (plot [b]) are also represented (in red).}
\label{Fig:cbVAR}
\end{center}
\end{figure}

\begin{figure}[t]
\begin{center}
\begin{tabular}{cc}
\includegraphics[width=0.33\textwidth,height=6.cm]{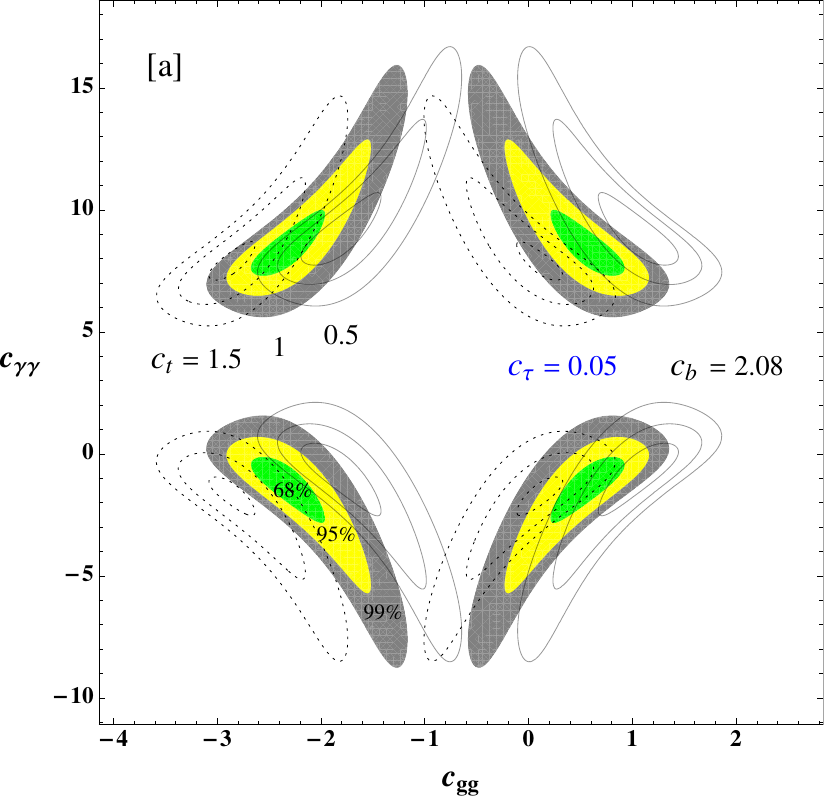}
\includegraphics[width=0.33\textwidth,height=6.cm]{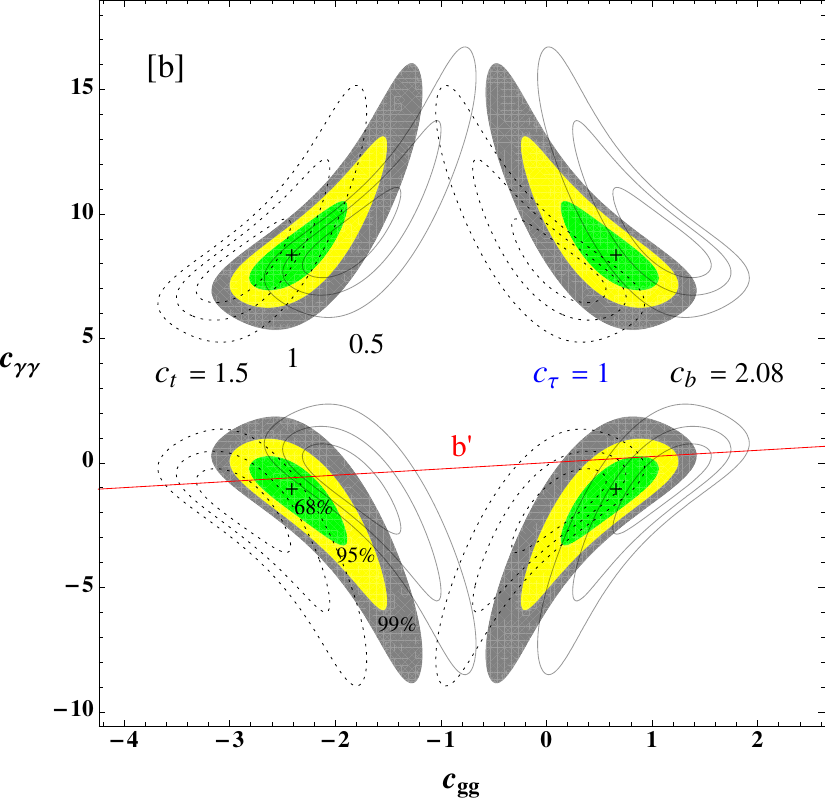}
\includegraphics[width=0.33\textwidth,height=6.cm]{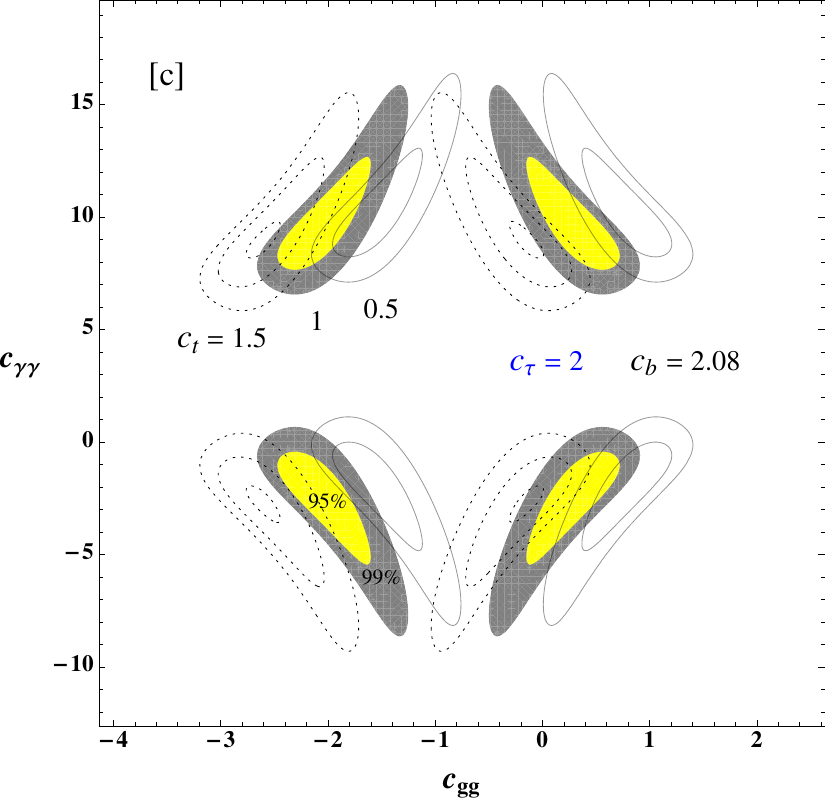}
\\
\end{tabular}
\caption{Best-fit regions at $68.27\%{\rm C.L.}$ (green), $95.45\%{\rm C.L.}$ (yellow) and $99.73\%{\rm C.L.}$ (grey) in the plane
$c_{\gamma\gamma}$ versus $c_{gg}$, for $c_b=2.08$. Each one of the three figures is obtained for a $c_\tau$ value which is
indicated (in blue). In each figure, the regions are drawn for three $c_t$ values [same conventions as in Fig.(\ref{Fig:cbVAR})].
The predicted (red) line for an extra-quark of type $b'$ is also represented in the plot [b].}
\label{Fig:ctauVAR}
\end{center}
\end{figure}

So now, having the three free parameters, $c_{gg},c_{\gamma\gamma},c_b$, 
we are going to show the best-fit domains in this three-dimensional space at $68.27\%{\rm C.L.}$ ($1\sigma$), $95.45\%{\rm C.L.}$ ($2\sigma$) and $99.73\%{\rm C.L.}$
($3\sigma$) which correspond to established values of $\Delta \chi^2 = \chi^2 - \chi^2_{\rm min}$ ($\chi^2_{\rm min}$ being the minimum $\chi^2$
value reached in the $\{c_{gg},c_{\gamma\gamma},c_b\}$ space) [see for instance Ref.~\cite{Beringer:1900zz}].  
\\ 
In Fig.(\ref{Fig:cbVAR}), we present four slices of these three-dimensional best-fit regions 
at four $c_b$ values ($c_b=0.75$ [a]; $1$ [b]; $2.08$ [c]; $10$ [d]) in the plane $c_{\gamma\gamma}$ versus $c_{gg}$. 
These regions are shown for three different fixed values of $c_t$ but for the unique choice $c_\tau=1$. The $c_\tau$ parameter is varied in the several  
plots of Fig.(\ref{Fig:ctauVAR}) (again for three $c_t$ values) where the behavior of the domain-slice at $c_b=2.08$ is still shown in the $\{c_{\gamma\gamma},c_{gg}\}$ plane; 
note that Fig.(\ref{Fig:cbVAR})[c] has also been included in Fig.(\ref{Fig:ctauVAR}) [see plot [b]] for an easier comparison with Fig.(\ref{Fig:ctauVAR})[a,c].
All these plots of Fig.(\ref{Fig:cbVAR})-(\ref{Fig:ctauVAR}) are discussed in the following subsections.

\subsubsection{The $c$-ranges}
\label{ChoiceRanges}

A few comments are in order with respect to the reasonable choice of parameter ranges in Fig.(\ref{Fig:cbVAR})-(\ref{Fig:ctauVAR}).
The naive perturbativity condition $\vert c_t Y_t \vert \lesssim 4\pi$ leads to $\vert c_t \vert \lesssim 18$ since $\vert Y_t \vert \simeq \vert m_t/v \vert$.
The similar theoretical constraints for $\vert c_b \vert$ and $\vert c_\tau \vert$ are even less stringent due to the smaller $m_{b,\tau}$ values.
The perturbativity considerations on $c_{\gamma\gamma}$ and $c_{gg}$ are model-dependent;
for example, in the case of a $t'$ state with $m_{t'}$ of the order of $m_t$, Eq.~(\ref{Eq:cexamI})-(\ref{Eq:cexamII}) show that
$c_{\gamma\gamma}$ and $c_{gg}$ would typically set the $t'$ Yukawa
coupling (relatively to $Y_t$) and would thus have to satisfy roughly the same condition as $c_t$~: $\vert c_{\gamma\gamma} \vert \lesssim 18$, $\vert c_{gg} \vert \lesssim 18$.
For the sake of generality, we consider the whole ranges of $c_{\gamma\gamma}$, $c_{gg}$ values pointed out by the Higgs rate fits.
\\
The $c_{t,b,\tau}$ choice is also related to the generation of fermion masses through Yukawa couplings. 
In the SM, the top quark mass determines $Y_t$ up to CKM mixing angles. For large deviations with respect to the SM Yukawa coupling, 
i.e. for $c_t$ values very different from unity, the physical top mass may be recovered by new strong mixing effects like in $t-t'$ mixings.
$\vert c_t\vert$ values different from unity by a factor $\sim 5$ would certainly already require strong $t-t'$ mixings, to be predicted by specific scenarios.
Similar comments hold for $c_{b}$ and $c_{\tau}$. From this point of view, the value of $c_b=10$ in Fig.(\ref{Fig:cbVAR})[d], and $c_\tau=0.05$ in Fig.(\ref{Fig:ctauVAR})[a], 
are respectively large and tiny; those have been chosen for the purpose of explaining the behavior of the best-fit domains in the large $c_b$ and low $c_\tau$ regimes.

\subsubsection{Best-fit points}
\label{BFpoints}

The best-fit points reachable, when varying the three free parameters, $c_b,c_{gg},c_{\gamma\gamma}$, 
for fixed values $c_t=1$ and $c_\tau=1$, are at $c_b=2.08$ and the $c_{gg}$, $c_{\gamma\gamma}$ values
corresponding to the four crosses drawn in Fig.(\ref{Fig:cbVAR})[c]. Since there are exact symmetries along the 
$c_{gg}$ and $c_{\gamma\gamma}$ axes (see discussion below), those four cross-points are all associated to the same $\chi^2_{\rm min}=52.36$.
\\Ê
For comparison, the best-fit point reachable, when varying the five effective parameters, $c_t,c_b,c_\tau,c_{gg},c_{\gamma\gamma}$, is 
$\{c_t=0.0;c_b=1.13;c_\tau=0.0;c_{gg}=-0.79;c_{\gamma\gamma}=-0.11\}$ leading to $\chi^2=50.26$.   
A vanishing $c_t$ (a top-phobic Higgs boson) imposes $\mu^{VI}_{7,{\rm CMS}}=0$ [via Eq.~(\ref{Eq:otherR})] which lies inside the $1\sigma$ experimental interval  
and is even the possible value the closest to the measured negative central value [given just after Eq.~(\ref{eq:muttbb})].  
Similarly, $c_\tau=0$ (tau-phobic) induces $\mu^{XII}_{7/8,{\rm CMS}}=0$ which is the closest value to the negative experimental central values.
In view of the generation of fermion masses through the Yukawa couplings, one could require say $\vert c_\tau\vert >0.3$ and $\vert c_t\vert >0.3$ 
which leads instead to the best-fit point $\{c_t=0.3;c_b=1.18;c_\tau=-0.3;c_{gg}=0.67;c_{\gamma\gamma}=-0.42\}$ having $\chi^2=50.44$.  
\\Ê
All these minimal $\chi^2$ values are smaller than the SM one, $\chi^2_{\rm SM}=57.10$ [from taking all the strength predictions at unity in Eq.~(\ref{eq:Chi2def})].
The regions at $68.27\%{\rm C.L.}$ in Fig.(\ref{Fig:cbVAR})[b] do not even contain the SM point ($\{c_t=1;c_b=1;c_\tau=1;c_{gg}=0;c_{\gamma\gamma}=0\}$).

\hspace{0.2cm}

Let us interpret the $c$-values of the best-fit points obtained in Fig.(\ref{Fig:cbVAR})[c] (or equivalently Fig.(\ref{Fig:ctauVAR})[b]). 
For example the best-fit point at, $c_b=2.08$, $c_{gg}=0.66$ and $c_{\gamma\gamma}=-1.09$, shown on Fig.(\ref{Fig:cbVAR})[c] (for fixed $c_t=c_\tau=1$)
indicates in particular that an increase of the diphoton partial width is favored by the data. Indeed, a negative $c_{\gamma\gamma}$ implies a constructive interference 
between EF loops and the main SM $W^\pm$-boson exchange, as shows Eq.~(\ref{Eq:diphoR}). Interestingly, the preferred $c_{\gamma\gamma}$ value approximatively 
cancel the top-loop 
contribution. The obtained indication for a $\Gamma_{\rm h\to \gamma\gamma}$ enhancement is not surprising as most of the measured strengths in the diphoton channel --
described in Section~\ref{Hdata} -- are above their SM expectations (even significantly for some of those).
\\
The best-fit value, $c_{gg}=0.66$, also outlines the preference for a $\sigma_{\rm gg \to h}$ increase [see Eq.~(\ref{Eq:fusionR})] 
related to the excesses with respect to the SM rates of the experimental values for some of the diphoton rates.  
\\ 
Finally, a $\Gamma_{\rm h\to \bar b b}$ increase is favored (see Eq.~(\ref{Eq:otherR}) with $c_b=2.08$) which tends to enhance the $\mu^{V}_{1.96,{\rm CDF+D0}}$ strength
and suppress $\mu^{XII}_{7/8,{\rm CMS}}$ relatively to the SM, as indicated by the experimental results (all at more than $1\sigma$ from the SM).
\\
The three other best-fit points of Fig.(\ref{Fig:cbVAR})[c] can be obtained through the symmetries described in the next subsection and are thus interpretable with the
same physical arguments about the Higgs rates.

\subsubsection{The symmetries}
\label{SYM}

Some exact reflection symmetries with respect to vertical and horizontal axes appear clearly on Fig.(\ref{Fig:cbVAR}) and Fig.(\ref{Fig:ctauVAR}).
Indeed, for a $c_{\gamma\gamma}$ value giving rise to a certain $\Delta \chi^2$, there always exists a $c_{\gamma\gamma}$ partner value leading 
to the opposite-sign ${\rm h\to \gamma\gamma}$ amplitude [squared in Eq.~(\ref{Eq:diphoR})] and in turn to the same $\Delta \chi^2$. 
The same kind of symmetry occurs for $c_{gg}$ entering the ${\rm h\to gg}$ (or ${\rm gg\to h}$) amplitude.
\\
Another type of symmetry is constituted by the transformation, $c_b \to - c_b$, leaving invariant the ${\rm \bar b b}$ partial width [{\it c.f.} Eq.~(\ref{Eq:otherR})].   
This symmetry is approximative due to the dependence of $\sigma_{\rm gg \to h}$ and $\Gamma_{\rm h\to \gamma\gamma}$ on $c_b$;
for $c_b$ values such that the bottom-exchange contributions to $\sigma_{\rm gg \to h}$ and $\Gamma_{\rm h\to \gamma\gamma}$ remain sub-leading (as in the SM), 
the transformation, $c_b \to - c_b$, keeps unchanged, at the percent level, the $c_{\gamma\gamma}$, $c_{gg}$ values associated to a given $\Delta \chi^2$. 
The similar symmetry arises for, $c_\tau \to - c_\tau$.

\subsubsection{Dependence of the best-fit regions on $c_{gg}$ and $c_{\gamma\gamma}$}
\label{SHAPE}

At this level, one is able to interpret the typical shapes of the obtained best-fit regions. The typical oblique direction (diagonal positioning) 
of the best-fit domains, for example for the fixed value, $c_t=1$, around the best-fit point, at $c_{gg}=0.66$ and $c_{\gamma\gamma}=-1.09$ 
in Fig.(\ref{Fig:cbVAR})[c], can be understood as follows -- the orientations of the three other best-fit region groups are then deduced through the 
reflection symmetries along $c_{gg}$ and $c_{\gamma\gamma}$.  
Starting from this best-fit point and decreasing $c_{gg}$ tends to decrease $\sigma_{\rm gg \to h}$ and hence to degrade the fits for diphoton rates,
a degradation which must be compensated by the $c_{\gamma\gamma}$ decrease ($\vert c_{\gamma\gamma}\vert$ increase 
enhancing $\Gamma_{\rm h\to \gamma\gamma}$) to remain below $68.27\%{\rm C.L.}$  

\subsubsection{Dependence of the best-fit regions on $c_t$}
\label{ctDEP}

We discuss now the modifications of the best-fit domains as the effective parameter, $c_t$, is varying.
We observe separately on Fig.(\ref{Fig:cbVAR})[a,b,c] and Fig.(\ref{Fig:ctauVAR}) that a $c_t$ variation of amount, $\delta c_t$, leads in a good approximation to a 
translation (no domain 
shape modification) of $-\delta c_t$ along both the $c_{\gamma\gamma}$ and $c_{gg}$ axes, for each one of the three best-fit regions. It is particularly clear
in Fig.(\ref{Fig:cbVAR})[b] where a large $\delta c_t$ is exhibited.
\\
Indeed, considering a given Confidence Level, the $\Delta \chi^2 = \chi^2 - \chi^2_{\rm min}$ value is fixed which determines [{\it c.f.} Eq.~(\ref{eq:Chi2def})] 
in particular the $c_t$ correction factor for the major top loop-exchanges and the parameters for EF loop-contributions, $c_{gg}$, $c_{\gamma\gamma}$, 
entering the predicted strengths [{\it c.f.} Eq.~(\ref{Eq:mu})] through the sums $(c_t + c_{gg})$ and $(c_t +c_{\gamma\gamma})$ [{\it c.f.} Eq.~(\ref{Eq:fusionR})-(\ref{Eq:diphoR})]. 
Hence for a $\delta c_t$ parameter variation, since the $\chi^2_{\rm min}$ value 
is unchanged (for similar compensation reasons to the following one), the induced $\chi^2$ modification should be exactly
compensated by variations, $\delta c_{gg} = \delta c_{\gamma\gamma} = -\delta c_t$.
\\
Note that for different $c_t$, $c_{gg}$ and $c_{\gamma\gamma}$ definitions from here (then distinguished by a prime), 
say generalizing to effective parameters entering Eq.~(\ref{Eq:fusionR})-(\ref{Eq:diphoR}) via 
$(\alpha_g c'_t + \beta_g c'_{gg})$ and $(\alpha_\gamma c'_t +\beta_\gamma c'_{\gamma\gamma})$ with new constants $\alpha_{g,\gamma},\beta_{g,\gamma}$, 
the translations would be instead of 
$$\delta c'_{gg} =  - \frac{\alpha_g}{\beta_g} \ \delta c'_t \ , \  \mbox{and,} \ \ \delta c'_{\gamma\gamma} =  -  \frac{\alpha_\gamma}{\beta_\gamma} \ \delta c'_t \ .$$
\\
The measured signal strength of Eq.~(\ref{eq:muttbb}) is also sensitive to $c_t$~\footnotemark[3]\footnotetext[3]{Other signal strengths, like in the diphoton channel, are also sensitive
to $c_t$ [{\it c.f.} Eq.~(\ref{Eq:mu})] but less, due to the experimental 
selection efficiencies and the smallness of $\sigma_{\rm h\bar tt}$ relatively to the dominant Higgs production reactions.} 
and there is no possible $\delta c_t$ compensation in it, as shows Eq.~(\ref{Eq:otherR}), which invalidates the above argumentation strictness.
Nevertheless, since the error bar on this measured rate is quite large, the above translation estimations remain a good approximation up to relatively large $\vert c_t \vert$ 
values where the three reference best-fit domain sizes start to decrease -- before disappearing. This is visible for instance in Fig.(\ref{Fig:cbVAR})[d]; in fact these more central, 
i.e. more fit-favored, domains in the $\{c_{\gamma\gamma},c_{gg}\}$ plane mainly allow to balance the degradation of the $\mu^{VI}_{7,{\rm CMS}}$ fit due to larger  
$\vert c_t \vert$ values (tending to increase too much the ${\rm h\bar tt}$ production cross section). 
This effect of decreasing domain widths appears in Fig.(\ref{Fig:cbVAR})[d] for smaller $\vert c_t \vert$ values than in all the other figures because, 
for this extremely large $c_b=10$ enhancing $B_{\rm h\to \bar b b}$, $\mu^{VI}_{7,{\rm CMS}}$ is getting above its $1\sigma$ range faster as
$\vert c_t \vert$ increases.

\hspace{0.2cm}

To conclude on this part,
this strong parameter interdependence implies that in order to determine experimentally the $c_{\gamma\gamma}$ and $c_{gg}$ quantities, it is crucial to determine as well  
the $c_t$ Yukawa correction whose measurement is essentially relying on the $\mu^{VI}$ analysis; now this analysis requires in particular
good efficiencies for the challenging simultaneous reconstruction of the top and bottom quark pairs in the final state.

\subsubsection{Dependence of the best-fit regions on $c_b$}
\label{cbDEP}

Concerning the $c_b$ variation (for fixed $c_t=c_\tau=1$), we first explain the impact of the $c_b$ increase on the typically allowed $c_{\gamma\gamma}$, $c_{gg}$ values
-- starting from the best-fit domains around the best-fit point, $\{c_b=2.08;c_{gg}=0.66;c_{\gamma\gamma}=-1.09\}$, in Fig.(\ref{Fig:cbVAR})[c] -- 
and the reasons why huge values up to $c_b \simeq 50$ could still agree with present Higgs rate fits. For such a $c_b$ increase, the strengths
$\mu^{VII, VIII}_{7/8,{\rm ATLAS/CMS}}$, $\mu^{XI}_{1.96,{\rm CDF+D0}}$ and $\mu^{XIV}_{7/8,{\rm CMS}}$  
are reduced via $\Gamma_{\rm h\to \bar b b}$, a reduction which has to be compensated by a $\sigma_{\rm gg \to h}$ increase through a $c_{gg}$ enhancement to
conserve a satisfactory $\chi^2$ (or equivalently here, $\Delta \chi^2$).
This explains the shift of the considered best-fit domains, around $\{c_b=2.08;c_{gg}=0.66;c_{\gamma\gamma}=-1.09\}$ in Fig.(\ref{Fig:cbVAR})[c], to higher $c_{gg}$  
values in the plot [d] where $c_b=10$ (still with $c_t=1$).
This necessary compensation between the $\Gamma_{\rm h\to \bar b b}$ and $\sigma_{\rm gg \to h}$ increases also guarantees the stability of diphoton rates (there is also 
a significant gluon-gluon fusion contribution in the three dijet-tagged final states) letting the $\chi^2$ at the same level, without $c_{\gamma\gamma}$ modifications 
-- explaining nearly identical $c_{\gamma\gamma}$ values for the studied regions in Fig.(\ref{Fig:cbVAR})[c] and [d].  
The $\Gamma_{\rm h\to \bar b b}$ increase leads to enhancements of the strengths 
$\mu^{IV}_{7/8,{\rm ATLAS/CMS}}$, $\mu^{V}_{1.96,{\rm CDF+D0}}$ and $\mu^{VI}_{7,{\rm CMS}}$ without major consequences on the fit;  
a $c_b$ increase up to $\sim 50$ [leading to $\Gamma_{\rm h\to \bar b b} \lesssim 5$~GeV] would 
still leave existing domains at $68.27\%{\rm C.L.}$ since in the theoretical limit, $c_b \to \infty$, $B_{\rm h\to \bar b b}$ 
tends obviously to a finite value compatible with data~: $B_{\rm h\to \bar b b}\to 1$. 
Similarly, the $\Gamma_{\rm h\to \bar b b}$ induced decrease of $\mu^{IX,X,XII,XIII}_{7/8,{\rm CMS}}$ does not affect significantly the global fit;
in the limit, $c_b \to \infty$, all these signal strengths tend to zero (via the involved branching ratios) which is clearly 
in agreement at $1\sigma$ with their experimental central value [and $\mu^{XII}_{7/8,{\rm CMS}}\vert_{\rm exp}$ is negative]. 
\\
There is another effect induced by the $c_b$ enhancement;
as $c_b$ is increasing, its contribution to $\sigma_{\rm gg \to h}$ renders softer the $\sigma_{\rm gg \to h}$ evolution with $c_{gg}$ so that the $c_{gg}$ interval,  
spanning the $\sigma_{\rm gg \to h}$ range allowed by the fit, gets larger; this can be seen by comparing the considered best-fit domain widths along the 
$c_{gg}$ axis in Fig.(\ref{Fig:cbVAR})[c] and [d]. 
\\
Now in the other direction, when $c_b$ decreases from its value in Fig.(\ref{Fig:cbVAR})[c] down to its values in the plots [b] and finally [a], 
the dominant effect of surface area diminution (and disappearance) for the best-fit regions is related to $\mu^{V}_{1.96,{\rm CDF+D0}}$ which is reduced 
and thus moved away from its best-fit value.   

\hspace{0.2cm}

{\it What is the experimental impact of the above $c_b$ variation analysis ?} The present experimental results do not prevent $c_b$ from taking    
extremely large values -- due in particular to Higgs rate compensations. 
In order to put a more stringent experimental upper limit on it, one could of course if possible improve the accuracies 
on the signal strengths involving $\sigma_{\rm gg \to h}$ and $\Gamma_{\rm h\to \bar b b}$. A new possibility to measure $c_b$ (or equivalently the bottom 
Yukawa coupling constant) would be to investigate the processes, 
${\rm \bar q q \to h\bar b b}$ and ${\rm gg \to h\bar b b}$ (or ${\rm \bar b b \to h}$ and ${\rm bg \to hb}$), followed by the decay, ${\rm h\to \bar b b}$. Indeed, here both
the production and decay rates should increase with $c_b$ ($\Gamma_{\rm h\to \bar b b}$ being the dominant partial width) so that compensations should not occur; 
then too large $c_b$ values would be experimentally ruled out. This Higgs production in association with bottom quarks could have significant cross sections
for high LHC luminosities and enhanced $c_b$ values compared to the SM~\cite{Dawson:2010yz} as the present fit points out. The sensitivity to such a reaction  
relies deeply on the b-tagging capability~\cite{Djouadi:2005gi}. This reaction suffers from large QCD backgrounds
but new search strategies have been developed for such a bottom final state topology, as in Ref.~\cite{Carena:2012rw}.

\subsubsection{Dependence of the best-fit regions on $c_\tau$}
\label{ctauDEP}

Finally, to complete our discussion on the parameter variations, we describe the $c_\tau$ influence on the best-fit domains.
\\
If the fixed $c_\tau$ parameter is chosen at a larger value, like in Fig.(\ref{Fig:ctauVAR})[c] compared to the plot [b], 
the induced best-fit $c_b$ value, obtained by $\chi^2$ minimization, is modified.   
The best-fit $\mu^{XII}_{7/8,{\rm CMS}}$ value, minimizing $\chi^2$, can involve (via $B_{\rm h\to \bar \tau \tau}$) a larger best-fit $c_b$ 
value in the case of [c] than in case [b], to compensate the higher $c_\tau$ (also entering $B_{\rm h\to \bar \tau \tau}$). 
In consequence, along the $c_b$ axis, the distance of the regions in the plot [c] (at $c_b=2.08$) to the best-fit point at $c_b>2.08$ is larger than 
the distance of the domain-slices in [b] (also at $c_b=2.08$) to the best-fit point at $c_b=2.08$ [indicated by the cross(es) on the figure].
Along the $c_{\gamma\gamma}$ and $c_{gg}$ axes, the typical distances of contours at a given Confidence Level to the respective central best-fit points   
are shorter in [c] than in [b]. In other terms, best-fit regions in [c] are smaller than in [b].
\\ 
The $c_\tau$ decrease from Fig.(\ref{Fig:ctauVAR})[b] to [a] leads to a softer region size reduction [in the limit $c_\tau\to 0$,
$\mu^{XII}_{7/8,{\rm CMS}}\to 0$ which is the preferred strength].

\subsection{The case of single EF scenarios}
\label{sec:ExEF}

In this Section~\ref{sec:ExEF}, we apply the constraints from the Higgs rate fit to examples of simple scenarios where a unique  
EF state significantly affects the Higgs interactions.

\subsubsection{An EF mixed with SM fermions}
\label{MIX}

\begin{figure}[t]
\begin{center}
\includegraphics[width=0.44\textwidth,height=7.7cm]{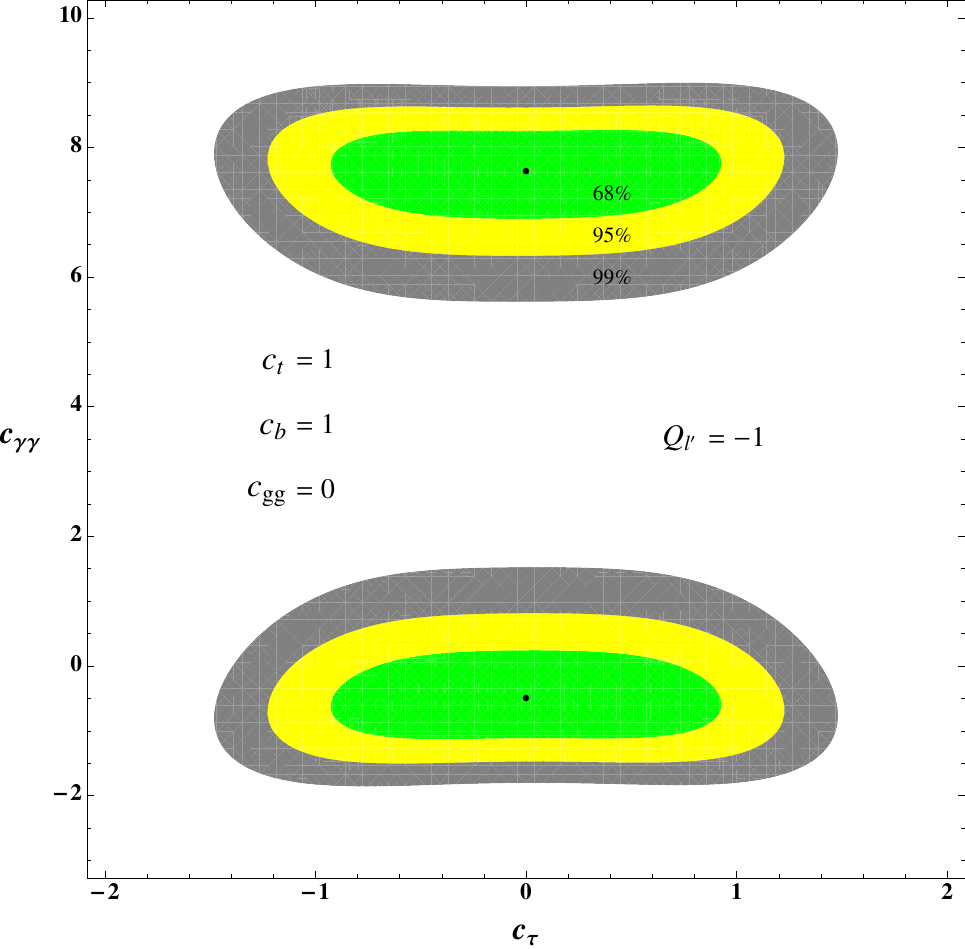}
\caption{Best-fit regions at $68.27\%{\rm C.L.}$, $95.45\%{\rm C.L.}$ and $99.73\%{\rm C.L.}$ in the plane
$c_{\gamma\gamma}$ versus $c_\tau$, for the case of an extra-lepton with electric charge, $Q_{\ell'}=-1$, corresponding to $c_t=c_b=1$, $c_{gg}=0$. 
The two best-fit points are indicated (in black).}
\label{Fig:EFcgg-lepton}
\end{center}
\end{figure}

For instance, a single $b'$ state [same color representation and electromagnetic charge as the bottom quark], 
that could be a light custodian top-partner in warped/composite frameworks,
would lead to a ratio in Eq.~(\ref{Eq:Exqp}), $(c_{\gamma \gamma}/c_{gg})\vert_{b'} = 1/4$, corresponding to the straight line drawn on Fig.(\ref{Fig:cbVAR})[c,d].  
Generically, a $b'$ would be mixed with the SM bottom quark so that possibly, $c_b \neq 1$, whereas one would have, $c_t=c_\tau=1$ -- like in Fig.(\ref{Fig:cbVAR})[c,d].
These figures show that there exist $c_{\gamma \gamma}$, $c_{gg}$ and $c_b$ values for which the predicted $b'$ line crosses the $68.27\%{\rm C.L.}$ region.
The simultaneous knowledge of the exact position on the $b'$ line and the $c_b$ value fixing the ${\rm C.L.}$ regions, necessary to determine the goodness of fit, 
requires the specification of the bottom mass matrix and hence of the considered model.

\hspace{0.2cm}

The other example of EF candidate able to be mixed with SM quarks is the $t'$ state, possibly constituted {\it e.g.} by a light top-partner in little Higgs models.
For a dominant $t'$ state, the ratio of Eq.~(\ref{Eq:Exqp}) tends to one which corresponds to the straight line on Fig.(\ref{Fig:cbVAR})[b].  
Since a $t'$ field can mix with the top quark, $c_t \neq 1$, but in the context of a single $t'$ one should have, $c_b=c_\tau=1$, as in Fig.(\ref{Fig:cbVAR})[b].
The predicted $t'$ line crosses two $95.45\%{\rm C.L.}$ regions {\it e.g.} for, $c_t=0.5$, as well as two $68.27\%{\rm C.L.}$ regions exclusively
in the range, $c_t\sim 1.1 \leftrightarrow  2.6$ (above $\sim 2.6$ the region sizes decrease as explained in Section~\ref{ctDEP}).

\hspace{0.2cm}

These discussions on the $b'$ and $t'$ states illustrate the fact that 
it is useful to study the best-fit domains in the $\{c_{\gamma\gamma},c_{gg}\}$ plane as, in simplified models, the theoretical prediction for the 
$c_{\gamma\gamma}/c_{gg}$ ratio takes a simple form independent of the extra-quark masses and Yukawa couplings.

\hspace{0.2cm}

For a single extra-lepton (colorless)
with charge, $Q_{\ell'}=-1$, potentially mixed with the SM $\tau$-lepton, the parameters, $c_b=c_t=1$, $c_{gg}=0$ [see Eq.~(\ref{Eq:cexamI})], are fixed
and there remain two free effective parameters, namely $c_{\gamma\gamma}$ and $c_\tau$. The best-fit regions for such a two-dimensional fit are presented in 
Fig.(\ref{Fig:EFcgg-lepton}). The two best-fit points shown in this figure correspond to, $\chi^2_{\rm min}=52.54$.

\subsubsection{An EF unmixed with SM fermions}
\label{UNMIX}

It is also possible theoretically that the new single $t'$ (or $b'$) particle does not mix with the SM top (bottom) quark.
This would be the case as well for additional $q'$ quarks with exotic electric charges.  
For illustration, let us first concentrate on the components of possible extensions of the SM quark multiplets under ${\rm SU(2)_L}$, 
as in warped/composite frameworks where SM multiplets are promoted to representations of the custodial 
symmetry~\cite{Agashe:2003zs,Agashe:2006at,Djouadi:2006rk,Carena:2007ua,Djouadi:2007eg,Djouadi:2007fm,Ledroit:2007ik,Bouchart:2008vp,Bouchart:2009vq,Djouadi:2009nb,Casagrande:2010si,Djouadi:2011aj}.
The charges for such $q'$ components obey the relation, ${\cal Y}_{q'}=Q_{q'}-I^{q'}_{3L}$ 
(${\cal Y}\equiv$~hypercharge, $I_{3L}\equiv {\rm SU(2)_L}$ isospin). We will consider the electric charges of smallest absolute values,  
 $Q_{q'}=-1/3$, $2/3$, $-4/3$, $5/3$, $-7/3$ and $8/3$, keeping in mind that the naive perturbative limit on the electric charge reads as, 
 $\vert Q_{q'} \vert \lesssim \sqrt{4\pi /\alpha}\simeq 40$ ($\alpha\equiv$~fine-structure constant~\cite{Beringer:1900zz}). 
The $q'$ states are in the same color representation as the SM quarks.

\begin{figure}[t]
\begin{center}
\includegraphics[width=0.47\textwidth,height=8cm]{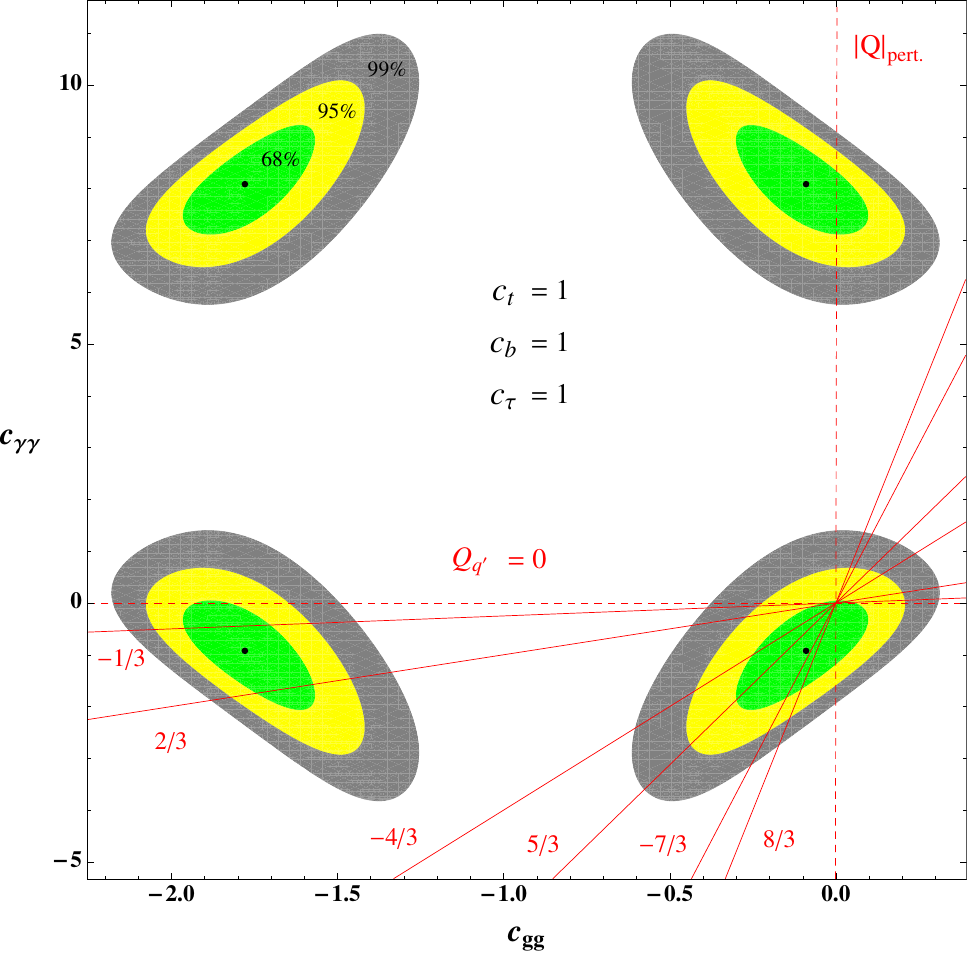}
\caption{Best-fit regions at $68.27\%{\rm C.L.}$, $95.45\%{\rm C.L.}$ and $99.73\%{\rm C.L.}$ in the plane
$c_{\gamma\gamma}$ versus $c_{gg}$, for $c_t=c_b=c_\tau=1$. Also represented are the predicted (red plain) lines for
extra-quarks with the several electric charges, $Q_{q'}= -1/3$, $2/3$, $-4/3$, $5/3$, $-7/3$ and $8/3$. 
The extreme (red dashed) lines for, $Q_{q'}=0$, and, $\vert Q_{q'} \vert =\vert Q_{q'} \vert_{\rm pert.} = \sqrt{4\pi /\alpha}$, are shown as well. 
The four best-fit points are indicated (in black).}
\label{Fig:EFcgg}
\end{center}
\end{figure}

In the case of the presence of such a $q'$ quark, unmixed with SM quarks, while $c_t=c_b=c_\tau=1$, one has $c_{\gamma \gamma} \neq 0$ 
and $c_{gg} \neq 0$ if the $q'$ state possesses non-zero Yukawa couplings; the best-fit domains for a two-dimensional fit keeping the fixed parameters,
$c_t=c_b=c_\tau=1$, are shown in Fig.(\ref{Fig:EFcgg}) together with the four best-fit points associated to, $\chi^2_{\rm min}=55.04$.
On this plot, we also represent the theoretically predicted regions in the cases of a single $q'$ quark with electric charge $Q_{q'}$~: these regions
are the straight lines defined by Eq.~(\ref{Eq:Exqp}). All the predicted lines -- whatever is the $Q_{q'}$ charge -- cross the SM point which is reached in the decoupling 
limit, $c_{\gamma \gamma} \to 0$, $c_{gg} \to 0$. The first result is that the upper-left best-fit regions, around $c_{\gamma \gamma} \sim 8$, $c_{gg} \sim -1.8$,  
cannot be explored in single $q'$ models [no line can reach it]. We also observe on Fig.(\ref{Fig:EFcgg}) that the predicted line being the closest to a best-fit point is for, 
$Q_{q'} = - 7/3$. This result means that, among any possible SM multiplet extension component, the fit prefers the $q_{-7/3}$ state compared for example to
a $t'$ or $q_{5/3}$ state. For instance, this latter $q_{5/3}$ state leads to a smaller $\vert c_{\gamma\gamma}/c_{gg} \vert_{q'}$ ratio ($\propto Q_{q'}^2$) which is less favored by 
the data due in particular to the observed diphoton rate enhancements. 
\\
A possibility in the future is that, as the measurements of the Higgs signal strengths will improve their accuracies -- leading typically to smaller best-fit regions in plots such as 
Fig.(\ref{Fig:EFcgg}) -- some absolute charges like for example, $\vert Q_{q'} \vert = 2/3$, could get excluded at $68.27\%{\rm C.L.}$ 
(the overlaps of the associated line with any $1\sigma$ region could disappear).
This kind of exclusion would be quite powerful in the sense that it would be independent of the $Y_{q'}$ Yukawa coupling constants,  
the $q'$ mass values ($m_{q'}$) and the $q'$ representations under ${\rm SU(2)_L}$. This is due to the simplifications occurring in the ratio of Eq.~(\ref{Eq:Exqp})
or in other terms to the correlations between $c_{\gamma \gamma}$ and $c_{gg}$ [see Eq.~(\ref{Eq:cexamI})-(\ref{Eq:cexamII})].

\hspace{0.2cm}

\begin{figure}[t]
\begin{center}
\begin{tabular}{cc}
\includegraphics[width=0.42\textwidth,height=6.7cm]{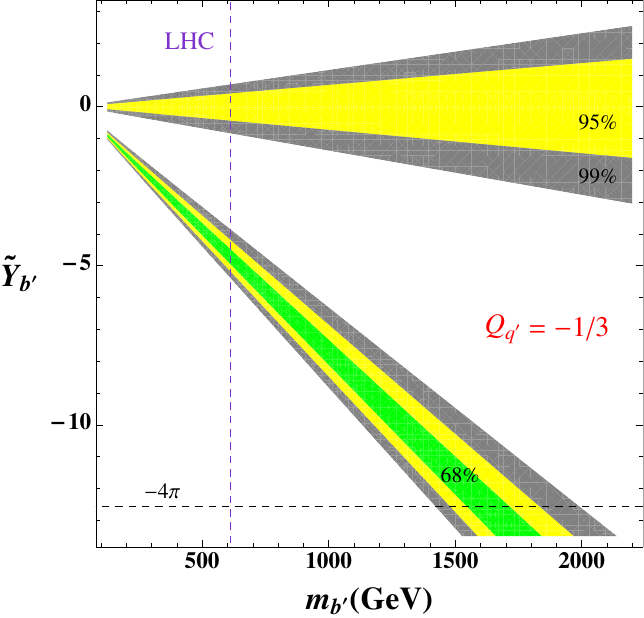}
\hspace{0.6cm}
\includegraphics[width=0.42\textwidth,height=6.7cm]{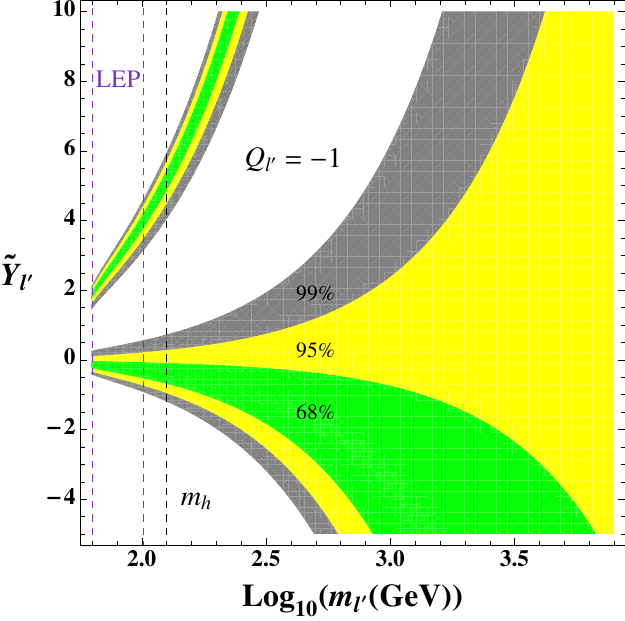}
\vspace{0.5cm}
\\
\includegraphics[width=0.42\textwidth,height=6.7cm]{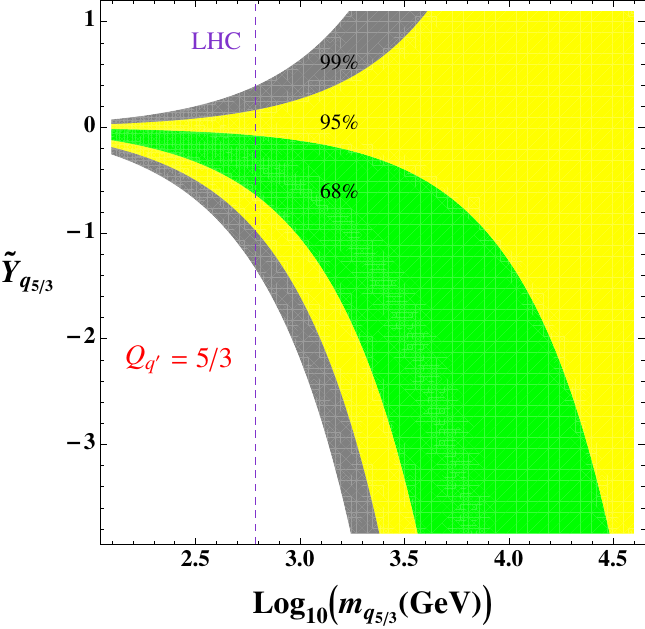}
\hspace{0.6cm}
\includegraphics[width=0.45\textwidth,height=6.7cm]{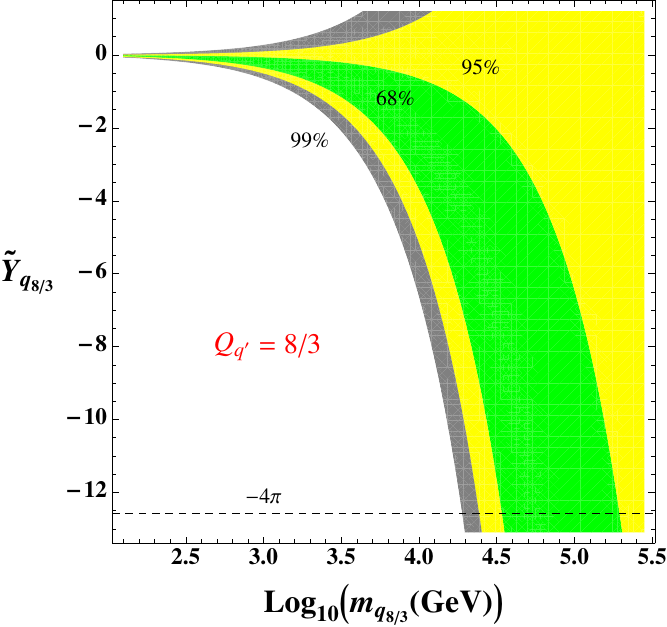}
\\
\end{tabular}
\caption{Best-fit regions at $68.27\%{\rm C.L.}$, $95.45\%{\rm C.L.}$ and $99.73\%{\rm C.L.}$ in the plane of the 
$\vert m_{f'} \vert$ absolute mass (in GeV) versus the $\tilde Y_{f'}$ coupling, with $c_t=c_b=c_\tau=1$, for the cases $Q_{q'}= - 1/3$, $5/3$, $8/3$ and $Q_{\ell'} =-1$. 
Also represented as dashed lines are the lower perturbative limit on Yukawa couplings, 
$-4\pi$ [black horizontal lines], and the direct LHC bounds, $m_{b'}>611$~GeV, $m_{q_{5/3}}>611$~GeV, 
or LEP constraint, $m_{\ell'}>63.5$~GeV ($>101.9$~GeV) for $m_{\ell'}-m_{\nu'}>7$~GeV ($>15$~GeV) [purple vertical lines].
The mass ranges start at the Higgs mass, $m_h$ (see Footnote~[\ref{FootMh}]), except for the $\ell'$-lepton case (see text) where the 
$m_h$ value is indicated by a [black dashed vertical] line.}
\label{Fig:Ym}
\end{center}
\end{figure}

\noindent {\bf $\bullet$ Best-fit domains in the plane of the Yukawa coupling versus the EF mass:}
Now we determine the physical parameters corresponding typically to an overlap between a given line in Fig.(\ref{Fig:EFcgg}) and the best-fit regions;  
we consider the characteristic examples of the charges, $Q_{q'}= - 1/3$, $5/3$ and $8/3$. More precisely, we plot in Fig.(\ref{Fig:Ym}) 
the regions in the plane $\vert m_{q'} \vert$ versus $\tilde Y_{q'} = - Y_{q'}/\mbox{sign}(m_{q'})$  
which correspond [see Eq.~(\ref{Eq:cexamI})-(\ref{Eq:cexamII})] to $c_{\gamma\gamma}$, $c_{gg}$ quantities giving rise to the best $\Delta \chi^2$ values
in the case of one free effective parameter, say $c_{gg}$ (related to $c_{\gamma\gamma}$ through the fixed ratio 
$c_{\gamma\gamma}/c_{gg} \vert_{q'} \propto Q_{q'}^2$).
\\
In Fig.(\ref{Fig:Ym}), we also illustrate the case of a single additional $\ell '$ lepton (colorless) without significant mixing to SM leptons [$c_\tau = 1$], as may be justified
by exotic $Q_{\ell'}$ charges or the large mass difference between the SM and extra-leptons. Here we choose, $Q_{\ell'} =-1$, being quite common for extra-lepton scenarios 
(as for instance recently in Ref.~\cite{Joglekar:2012vc}). There is, again, a unique free effective parameter, $c_{\gamma\gamma}$, since $c_{gg}=0$.

Let us discuss the results shown in Fig.(\ref{Fig:Ym}). 
For a given Confidence Level, the linear dependence of $\tilde Y_{b'}$ on $\vert m_{b'} \vert$ appearing clearly on the upper left plot   
is explained by the expressions (\ref{Eq:cexamI})-(\ref{Eq:cexamII}) and the constant limit, $A[\tau(m_{b'}\gg m_h)]\to 1$ [described after Eq.~(\ref{Eq:Cgagadef})]. 
This linear behavior also holds for the three other cases illustrated in this figure, even if for those it is hidden by the chosen logarithmic scale 
(allowing for a better view of the couplings in the small mass ranges).
Eq.~(\ref{Eq:cexamI})-(\ref{Eq:cexamII}) show that increasing $Q_{q'}$ leads to a slower evolution of $\vert \tilde Y_{q'} \vert$ with $\vert m_{q'} \vert$
(perturbative limit, $-4\pi$, reached for higher $\vert m_{q'} \vert$) and a smaller allowed $\tilde Y_{q'}$ range at fixed $\vert m_{q'} \vert$ as can 
be observed by comparing $Q_{q'}=5/3$ and $8/3$ in Fig.(\ref{Fig:Ym}). Comparing a $\ell'$ extra-lepton with the $b'$ extra-quark, it appears in Eq.~(\ref{Eq:cexamII}) that 
the smaller $N^{\ell'}_c=1$ color number tends to compensate the larger $Q_{\ell'}^2=1$ squared charge (the favored $c_{\gamma\gamma}\vert_{f'}$ interval
size also affects the $\tilde Y_{f'}$ range width). The two unconnected $95.45\%{\rm C.L.}$ regions in the $\{\vert m_{b'} \vert,\tilde Y_{b'}\}$ plane correspond basically to 
the two overlaps between the $95.45\%{\rm C.L.}$ domains and the $b'$ line in Fig.(\ref{Fig:EFcgg}).

We now describe the direct experimental constraints indicated on the various plots of Fig.(\ref{Fig:Ym}). 
The LHC bound, $m_{b'}>611$~GeV, illustrated on the upper left plot is the strongest direct experimental constraint on a $b'$ state;
this bound is based on the QCD $b'$ pair production and it is less stringent for a branching ratio, $B_{\rm b'\to tW^-}<1$~\cite{Chatrchyan:2012yea}. 
The bound for, $B_{\rm b'\to tW^-}=1$, combined with the $68.27\%{\rm C.L.}$ region push the Yukawa couplings towards large absolute values, 
as Fig.(\ref{Fig:Ym}) is demonstrating. 
The experimental bounds from investigations of other decay channels, like ${\rm b'\to bZ}$ or ${\rm b'\to bh}$, are not relevant in the context of a 
$b'$ field unmixed with SM quarks.  
\\
The bound, $m_{q_{5/3}}>611$~GeV, from the LHC shown in Fig.(\ref{Fig:Ym}) is imposed by the search for the same decay final state, ${\rm q_{5/3}\to tW^+}$, following
the $q_{5/3}$ pair production; this bound is obtained for, $B_{\rm  q_{5/3}\to tW^+}=1$~\cite{Chatrchyan:2012yea}, and it leaves a possible  
region at $68.27\%{\rm C.L.}$ in Fig.(\ref{Fig:Ym}). 
Concerning the $q_{8/3}$ particle which could decay as, ${\rm  q_{8/3} \to t W^+ W^+}$, there have been no experimental searches so far.
\\
There exist bounds on extra-leptons from the LEP collider; those read as, $m_{\ell'}>63.5$~GeV ($m_{\ell'}>101.9$~GeV) for $m_{\ell'}-m_{\nu'}>7$~GeV 
($>15$~GeV)~\cite{Beringer:1900zz,Joglekar:2012vc}, in the case of the existence of an additional $\nu'$ neutrino (which would have no effects on the Higgs 
couplings to charged fermions or gauge bosons). These constraints have been 
obtained from investigating the channel, ${\rm  \ell' \to W^{(\star)} \nu' \to \ell + \Emiss}$, where $\ell$ denotes
a SM charged lepton and $\Emiss$ stands for missing energy, assuming a stable $\nu'$ on collider time-scales.  
The results for the domain, $m_h>m_{\ell'}$, shown in Fig.(\ref{Fig:Ym}) are valid in the absence of new significant partial Higgs decay widths 
(see Footnote~[\ref{FootMh}]).

To conclude on all these aspects of Fig.(\ref{Fig:Ym}), 
one can say that the collider constraints from Higgs rate measurements on representative single EF models are already significant,  
especially in the low-mass regime where the allowed range for Yukawa coupling constants can be quite predictive. The constraints are sensitive to 
larger masses in cases of higher electric charges, as expected, and this indirect sensitivity on EF candidates can reach large mass scales up to $\sim 200$~TeV.

\hspace{0.2cm}

\noindent {\bf $\bullet$ Constraints on the signs of fundamental parameters:}
Concerning the constraints on the signs, as shown in Fig.(\ref{Fig:Ym}) based on the present Higgs data,  
the sign, $\tilde Y_{q'}<0$ [leading to $c_{\gamma\gamma}<0$], is preferred at $68.27\%{\rm C.L.}$ [except with absolute charges, $\vert Q_{q'} \vert \gtrsim 7$, i.e. in a range
close to the $\vert Q_{q'} \vert_{\rm pert.}$ limit as illustrated in Fig.(\ref{Fig:EFcgg})] 
for any single extra-quark as it creates a constructive interference with the $W^\pm$-boson exchange increasing the diphoton rates. 
The specific sign configuration, $\tilde Y_{q'}<0$, 
is selected by the two relevant best-fit points which pin down, $c_{\gamma\gamma}<0$, as obtained for extra-quarks in Fig.(\ref{Fig:EFcgg}).
This predicted condition means that the Yukawa coupling constant [$-Y_{q'}$ in our conventions] must have a sign opposite 
to $m_{q'}$ which could be written,   
{\normalsize
\begin{eqnarray} 
\mbox{sign}\bigg (\frac{- Y_{q'}}{m_{q'}}\bigg ) \ < \ 0 \ . 
\label{Eq:ExtraDYS}
\end{eqnarray} 
}

Related to this condition, there are comments on the configuration denoted as {\it dysfermiophilia} in the literature. 
As described at the end of Section~\ref{HRModif}, strictly speaking the $c_{t,b,\tau}$ parameters entering Eq.~(\ref{Eq:fusionR})-(\ref{Eq:diphoR})
-- whose values are generally given in best-fit plots such as the present ones in Fig.(\ref{Fig:cbVAR}) -- should in fact be understood as being,
{\normalsize $$
\epsilon_tc_t \ = \ \frac{\mbox{sign}(m_t)}{\mbox{sign}(m^{\rm EF}_t)} \ c_t \ 
= \ \frac{\mbox{sign}(m_t)}{\mbox{sign}(m^{\rm EF}_t)} \ \frac{\mbox{sign}(-Y^{\rm EF}_t)}{\mbox{sign}(-Y_t)} \ \vert c_t\vert \ 
= \ \frac{\mbox{sign}(-Y^{\rm EF}_t)}{\mbox{sign}(m^{\rm EF}_t)} \ \vert c_t \vert \
= \ \mbox{sign}\bigg (\frac{-Y^{\rm EF}_t}{m^{\rm EF}_t} \bigg ) \ \bigg \vert \frac{Y^{\rm EF}_t}{Y_t} \bigg \vert \ ,
$$ } 
in our conventions of Lagrangian~(\ref{Eq:LagEff}), and similarly for $\epsilon_{b,\tau}c_{b,\tau}$;
here the EF-exponent indicates that the parameter is considered within the context of an EF model
(and remind that $m_t$, $Y_t$ are in the SM). 
Therefore, the {\it dysfermiophilia} property of increasing, $\Gamma_{\rm h\to \gamma\gamma}/\Gamma_{\rm h\to \gamma\gamma}^{\rm SM}$,
via changing the top Yukawa sign is in fact relying on the possibility to have, $\epsilon_tc_t<0$, or equivalently, 
$\mbox{sign}(-Y^{\rm EF}_t/m^{\rm EF}_t)<0$. This makes sense as it is the sign of, $-Y^{\rm EF}_t/m^{\rm EF}_t$, which has a physical
meaning and appears in $\Gamma_{\rm h\to \gamma\gamma}$ [see Eq.~(\ref{Eq:cexamII}) for an analogy with the $t'$-loop].
\\
The other comment is that the {\it dysfermiophilia} possibility of having, $\epsilon_tc_t<0$, can indeed gives rise to an acceptable 
agreement with the Higgs data (see {\it e.g.} Fig.(\ref{Fig:cbVAR})[d]) but it is not necessary to achieve a good agreement
({\it c.f.} Fig.(\ref{Fig:EFcgg}) where $\epsilon_tc_t=1$)
since the constructive interference with the $W^\pm$-loop increasing the diphoton rates can be realized
with an EF-loop inducing, $c_{\gamma\gamma}<0$. 
\\ 
Hence the above condition (\ref{Eq:ExtraDYS}) can be called an {\it extra-dysfermiophilia} as it is exactly the same as for the top quark
transposed to an EF. Besides, this condition (\ref{Eq:ExtraDYS}) leads to a decrease of, $\sigma_{\rm gg\to h}/\sigma_{\rm gg\to h}^{\rm SM}$, 
for a single EF [see Eq.~(\ref{Eq:cexamI})] through negative $c_{gg}$ values [{\it c.f.} Fig.(\ref{Fig:EFcgg})]. 
Generally speaking, an {\it extra-dysfermiophilia} is probably easier to realize than a {\it dysfermiophilia} due to the potentially higher degree of freedom 
(allowing to de-correlate EF masses and Yukawa couplings) which can come {\it e.g.} from additional mass terms not induced by EW symmetry breaking, 
like KK masses.

\section{Conclusions}
\label{conclu}

We have learnt from varying the effective parameters of the Higgs rate fit that shifts of the correction factor affecting the top quark Yukawa coupling,
$c_t$, lead to translations of the best-fit domains in the $\{ c_{\gamma\gamma},c_{gg} \}$ plane 
($c_{\gamma\gamma}$ and $c_{gg}$ parametrize respectively new loop-contributions to the $h\gamma\gamma$ and $hgg$ vertex) 
proportional to $\delta c_t$. This means that to constrain precisely the new loop-contributions to the 
$hgg$ and $h\gamma\gamma$ couplings, one has to determine simultaneously the top Yukawa coupling which might be an experimental
challenge. 
\\
The $c_{gg}$ determination relies as well significantly on the correction factor affecting the bottom quark Yukawa coupling, namely $c_b$, 
for which extremely large values are not ruled out by the combination of present Higgs data; for that purpose, new Higgs reactions,
like ${\rm gg \to h\bar b b}$, ${\rm h\to \bar b b}$, would be interesting to investigate experimentally.

\hspace{0.1cm}

We have then considered the effective case of a single EF affecting the Higgs rates. It could for example be  
the lightest KK mode of some higher-dimensional theory and have dominant effects on collider physics; 
the lightest KK state effects are generically at least the strongest ones so assuming this state
to be the sole one is quantitatively a good (starting) approximation. In contrast, within theories containing several crucial EF, 
one could of course combine the (different) single EF effects described here and there could be compensations. 
\\
In this basic single EF framework, significant constraints have been placed on extra-leptons.
We have also found that the Higgs rate measurements put non-trivial constraints on $c_{\gamma\gamma}$ and $c_{gg}$ 
for $b'$, $t'$ states able to mix with the SM $b$, $t$ quarks. Regarding unmixed EF candidates [still with same color number as SM quarks], 
it is remarkable that, due to the $c_{\gamma\gamma}-c_{gg}$ correlations, the Higgs fit can potentially constrain  
intervals of absolute electric charges independently of the ${\rm SU(2)_L}$ representations, Yukawa couplings and masses for the EF. 
Another related result is that, among any possible components of SM quark multiplet extensions, the $q_{-7/3}$ field is the one preferred by the fit. 
The Higgs rate fit also allows to constrain significantly the EF Yukawa couplings for $m_{q'}$ values up to $\sim 200$~TeV, and,  
points out at $68.27\%{\rm C.L.}$ an {\it extra-dysfermiophilia} [condition (\ref{Eq:ExtraDYS})] for any single $q'$ quark 
(independently of $Q_{q'}$ as long as it does not approach non-perturbative couplings). 

\hspace{0.1cm}

Finally, let us note that any model with EF predicts certain values for the parameters, $c_t$, $c_b$, $c_\tau$ 
($c_\tau$~: correction factor for the $\tau$-lepton Yukawa coupling) and $c_{\gamma\gamma}$, $c_{gg}$ [easily calculable
through Eq.~(\ref{Eq:cexamI})-(\ref{Eq:cexamII})], which can then be located on the best-fit plots obtained in this paper in order to determine
the degree of compatibility with the Higgs data. Anyone could also use the synthesized
fit informations contained in Fig.(\ref{Fig:EFcgg}) to constrain one's extra-quark electric charge, 
and, in Fig.(\ref{Fig:Ym}) to study the $\{\vert m_{f'} \vert,\tilde Y_{f'}\}$ plane of one's single $f'$ model.
\\
\\
{\bf Acknowledgements:}
The author thanks A.~Azatov, H.~Bachacou, O.~J.~P.~\'Eboli, J.~R.~Espinosa, A.~Falkowski, S.~Gopalakrishna, 
C.~Grojean, V.~Sanz, A.~Strumia and T.~Tath for useful discussions as well as the CERN TH-Division where this paper has been initiated. 
This work is supported by the ``Institut Universitaire de France''. The author also acknowledges support from  
the ANR {\it CPV-LFV-LHC} under project \textbf{NT09-508531}.

 \bibliography{REFefit.sub_li}

\begin{thebibliography}{117}
\expandafter\ifx\csname natexlab\endcsname\relax\def\natexlab#1{#1}\fi
\expandafter\ifx\csname bibnamefont\endcsname\relax
  \def\bibnamefont#1{#1}\fi
\expandafter\ifx\csname bibfnamefont\endcsname\relax
  \def\bibfnamefont#1{#1}\fi
\expandafter\ifx\csname citenamefont\endcsname\relax
  \def\citenamefont#1{#1}\fi
\expandafter\ifx\csname url\endcsname\relax
  \def\url#1{\texttt{#1}}\fi
\expandafter\ifx\csname urlprefix\endcsname\relax\def\urlprefix{URL }\fi
\providecommand{\bibinfo}[2]{#2}
\providecommand{\eprint}[2][]{\url{#2}}

\bibitem[{\citenamefont{{F. Gianotti}}(2012)}]{CERN4thATLAS}
\bibinfo{author}{\bibnamefont{{F. Gianotti}}} (\bibinfo{year}{2012}),
  \eprint{CERN Seminar, {\it Update on the Standard Model Higgs searches in
  ATLAS}, July, 4, 2012, {\tt
  http://indico.cern.ch/conferenceDisplay.py?confId=197461}}.

\bibitem[{\citenamefont{{J. Incandela}}(2012)}]{CERN4thCMS}
\bibinfo{author}{\bibnamefont{{J. Incandela}}} (\bibinfo{year}{2012}),
  \eprint{CERN Seminar, {\it Update on the Standard Model Higgs searches in
  CMS}, July, 4 2012, {\tt
  http://indico.cern.ch/conferenceDisplay.py?confId=197461}}.

\bibitem[{\citenamefont{Englert and Brout}(1964)}]{Englert:1964et}
\bibinfo{author}{\bibfnamefont{F.}~\bibnamefont{Englert}} \bibnamefont{and}
  \bibinfo{author}{\bibfnamefont{R.}~\bibnamefont{Brout}},
  \bibinfo{journal}{Phys.Rev.Lett.} \textbf{\bibinfo{volume}{13}},
  \bibinfo{pages}{321} (\bibinfo{year}{1964}).

\bibitem[{\citenamefont{Higgs}(1964{\natexlab{a}})}]{Higgs:1964ia}
\bibinfo{author}{\bibfnamefont{P.~W.} \bibnamefont{Higgs}},
  \bibinfo{journal}{Phys.Lett.} \textbf{\bibinfo{volume}{12}},
  \bibinfo{pages}{132} (\bibinfo{year}{1964}{\natexlab{a}}).

\bibitem[{\citenamefont{Higgs}(1964{\natexlab{b}})}]{Higgs:1964pj}
\bibinfo{author}{\bibfnamefont{P.~W.} \bibnamefont{Higgs}},
  \bibinfo{journal}{Phys.Rev.Lett.} \textbf{\bibinfo{volume}{13}},
  \bibinfo{pages}{508} (\bibinfo{year}{1964}{\natexlab{b}}).

\bibitem[{\citenamefont{Guralnik et~al.}(1964)\citenamefont{Guralnik, Hagen,
  and Kibble}}]{Guralnik:1964eu}
\bibinfo{author}{\bibfnamefont{G.}~\bibnamefont{Guralnik}},
  \bibinfo{author}{\bibfnamefont{C.}~\bibnamefont{Hagen}}, \bibnamefont{and}
  \bibinfo{author}{\bibfnamefont{T.}~\bibnamefont{Kibble}},
  \bibinfo{journal}{Phys.Rev.Lett.} \textbf{\bibinfo{volume}{13}},
  \bibinfo{pages}{585} (\bibinfo{year}{1964}).

\bibitem[{\citenamefont{{ATLAS web page}}(2012)}]{ATLASweb}
\bibinfo{author}{\bibnamefont{{ATLAS web page}}}, \bibinfo{journal}{{\tt
  https://twiki.cern.ch/twiki/bin/view/AtlasPublic}}  (\bibinfo{year}{2012}).

\bibitem[{\citenamefont{{CMS web page}}(2012)}]{CMSweb}
\bibinfo{author}{\bibnamefont{{CMS web page}}}, \bibinfo{journal}{{\tt
  https://twiki.cern.ch/twiki/bin/view/CMSPublic/PhysicsResults}}
  (\bibinfo{year}{2012}).

\bibitem[{\citenamefont{Carena et~al.}(2006)\citenamefont{Carena, Ponton,
  Santiago, and Wagner}}]{Carena:2006bn}
\bibinfo{author}{\bibfnamefont{M.~S.} \bibnamefont{Carena}},
  \bibinfo{author}{\bibfnamefont{E.}~\bibnamefont{Ponton}},
  \bibinfo{author}{\bibfnamefont{J.}~\bibnamefont{Santiago}}, \bibnamefont{and}
  \bibinfo{author}{\bibfnamefont{C.~E.~M.} \bibnamefont{Wagner}},
  \bibinfo{journal}{Nucl.Phys.} \textbf{\bibinfo{volume}{B759}},
  \bibinfo{pages}{202} (\bibinfo{year}{2006}), \eprint{hep-ph/0607106}.

\bibitem[{\citenamefont{Gogberashvili}(2002)}]{Gogberashvili:1998vx}
\bibinfo{author}{\bibfnamefont{M.}~\bibnamefont{Gogberashvili}},
  \bibinfo{journal}{Int.J.Mod.Phys.} \textbf{\bibinfo{volume}{D11}},
  \bibinfo{pages}{1635} (\bibinfo{year}{2002}), \eprint{hep-ph/9812296}.

\bibitem[{\citenamefont{Randall and Sundrum}(1999)}]{Randall:1999ee}
\bibinfo{author}{\bibfnamefont{L.}~\bibnamefont{Randall}} \bibnamefont{and}
  \bibinfo{author}{\bibfnamefont{R.}~\bibnamefont{Sundrum}},
  \bibinfo{journal}{Phys.Rev.Lett.} \textbf{\bibinfo{volume}{83}},
  \bibinfo{pages}{3370} (\bibinfo{year}{1999}), \eprint{hep-ph/9905221}.

\bibitem[{\citenamefont{Gherghetta and Pomarol}(2000)}]{Gherghetta:2000qt}
\bibinfo{author}{\bibfnamefont{T.}~\bibnamefont{Gherghetta}} \bibnamefont{and}
  \bibinfo{author}{\bibfnamefont{A.}~\bibnamefont{Pomarol}},
  \bibinfo{journal}{Nucl.Phys.} \textbf{\bibinfo{volume}{B586}},
  \bibinfo{pages}{141} (\bibinfo{year}{2000}), \eprint{hep-ph/0003129}.

\bibitem[{\citenamefont{Huber and Shafi}(2001{\natexlab{a}})}]{Huber:2000ie}
\bibinfo{author}{\bibfnamefont{S.~J.} \bibnamefont{Huber}} \bibnamefont{and}
  \bibinfo{author}{\bibfnamefont{Q.}~\bibnamefont{Shafi}},
  \bibinfo{journal}{Phys.Lett.} \textbf{\bibinfo{volume}{B498}},
  \bibinfo{pages}{256} (\bibinfo{year}{2001}{\natexlab{a}}),
  \eprint{hep-ph/0010195}.

\bibitem[{\citenamefont{Huber and Shafi}(2001{\natexlab{b}})}]{Huber:2001ug}
\bibinfo{author}{\bibfnamefont{S.~J.} \bibnamefont{Huber}} \bibnamefont{and}
  \bibinfo{author}{\bibfnamefont{Q.}~\bibnamefont{Shafi}},
  \bibinfo{journal}{Phys.Lett.} \textbf{\bibinfo{volume}{B512}},
  \bibinfo{pages}{365} (\bibinfo{year}{2001}{\natexlab{b}}),
  \eprint{hep-ph/0104293}.

\bibitem[{\citenamefont{Huber and Shafi}(2002)}]{Huber:2002gp}
\bibinfo{author}{\bibfnamefont{S.~J.} \bibnamefont{Huber}} \bibnamefont{and}
  \bibinfo{author}{\bibfnamefont{Q.}~\bibnamefont{Shafi}},
  \bibinfo{journal}{Phys.Lett.} \textbf{\bibinfo{volume}{B544}},
  \bibinfo{pages}{295} (\bibinfo{year}{2002}), \eprint{hep-ph/0205327}.

\bibitem[{\citenamefont{Huber and Shafi}(2004)}]{Huber:2003sf}
\bibinfo{author}{\bibfnamefont{S.~J.} \bibnamefont{Huber}} \bibnamefont{and}
  \bibinfo{author}{\bibfnamefont{Q.}~\bibnamefont{Shafi}},
  \bibinfo{journal}{Phys.Lett.} \textbf{\bibinfo{volume}{B583}},
  \bibinfo{pages}{293} (\bibinfo{year}{2004}), \eprint{hep-ph/0309252}.

\bibitem[{\citenamefont{Chang et~al.}(2006)\citenamefont{Chang, Kim, and
  Yamaguchi}}]{Chang:2005ya}
\bibinfo{author}{\bibfnamefont{S.}~\bibnamefont{Chang}},
  \bibinfo{author}{\bibfnamefont{C.}~\bibnamefont{Kim}}, \bibnamefont{and}
  \bibinfo{author}{\bibfnamefont{M.}~\bibnamefont{Yamaguchi}},
  \bibinfo{journal}{Phys.Rev.} \textbf{\bibinfo{volume}{D73}},
  \bibinfo{pages}{033002} (\bibinfo{year}{2006}), \eprint{hep-ph/0511099}.

\bibitem[{\citenamefont{Moreau and
  Silva-Marcos}(2006{\natexlab{a}})}]{Moreau:2006np}
\bibinfo{author}{\bibfnamefont{G.}~\bibnamefont{Moreau}} \bibnamefont{and}
  \bibinfo{author}{\bibfnamefont{J.}~\bibnamefont{Silva-Marcos}},
  \bibinfo{journal}{JHEP} \textbf{\bibinfo{volume}{0603}}, \bibinfo{pages}{090}
  (\bibinfo{year}{2006}{\natexlab{a}}), \eprint{hep-ph/0602155}.

\bibitem[{\citenamefont{Moreau and
  Silva-Marcos}(2006{\natexlab{b}})}]{Moreau:2005kz}
\bibinfo{author}{\bibfnamefont{G.}~\bibnamefont{Moreau}} \bibnamefont{and}
  \bibinfo{author}{\bibfnamefont{J.}~\bibnamefont{Silva-Marcos}},
  \bibinfo{journal}{JHEP} \textbf{\bibinfo{volume}{0601}}, \bibinfo{pages}{048}
  (\bibinfo{year}{2006}{\natexlab{b}}), \eprint{hep-ph/0507145}.

\bibitem[{\citenamefont{Agashe et~al.}(2005{\natexlab{a}})\citenamefont{Agashe,
  Perez, and Soni}}]{Agashe:2004cp}
\bibinfo{author}{\bibfnamefont{K.}~\bibnamefont{Agashe}},
  \bibinfo{author}{\bibfnamefont{G.}~\bibnamefont{Perez}}, \bibnamefont{and}
  \bibinfo{author}{\bibfnamefont{A.}~\bibnamefont{Soni}},
  \bibinfo{journal}{Phys.Rev.} \textbf{\bibinfo{volume}{D71}},
  \bibinfo{pages}{016002} (\bibinfo{year}{2005}{\natexlab{a}}),
  \eprint{hep-ph/0408134}.

\bibitem[{\citenamefont{Agashe et~al.}(2004)\citenamefont{Agashe, Perez, and
  Soni}}]{Agashe:2004ay}
\bibinfo{author}{\bibfnamefont{K.}~\bibnamefont{Agashe}},
  \bibinfo{author}{\bibfnamefont{G.}~\bibnamefont{Perez}}, \bibnamefont{and}
  \bibinfo{author}{\bibfnamefont{A.}~\bibnamefont{Soni}},
  \bibinfo{journal}{Phys.Rev.Lett.} \textbf{\bibinfo{volume}{93}},
  \bibinfo{pages}{201804} (\bibinfo{year}{2004}), \eprint{hep-ph/0406101}.

\bibitem[{\citenamefont{Agashe et~al.}(2007)\citenamefont{Agashe, Perez, and
  Soni}}]{Agashe:2006wa}
\bibinfo{author}{\bibfnamefont{K.}~\bibnamefont{Agashe}},
  \bibinfo{author}{\bibfnamefont{G.}~\bibnamefont{Perez}}, \bibnamefont{and}
  \bibinfo{author}{\bibfnamefont{A.}~\bibnamefont{Soni}},
  \bibinfo{journal}{Phys.Rev.} \textbf{\bibinfo{volume}{D75}},
  \bibinfo{pages}{015002} (\bibinfo{year}{2007}), \eprint{hep-ph/0606293}.

\bibitem[{\citenamefont{Agashe et~al.}(2006{\natexlab{a}})\citenamefont{Agashe,
  Blechman, and Petriello}}]{Agashe:2006iy}
\bibinfo{author}{\bibfnamefont{K.}~\bibnamefont{Agashe}},
  \bibinfo{author}{\bibfnamefont{A.~E.} \bibnamefont{Blechman}},
  \bibnamefont{and}
  \bibinfo{author}{\bibfnamefont{F.}~\bibnamefont{Petriello}},
  \bibinfo{journal}{Phys.Rev.} \textbf{\bibinfo{volume}{D74}},
  \bibinfo{pages}{053011} (\bibinfo{year}{2006}{\natexlab{a}}),
  \eprint{hep-ph/0606021}.

\bibitem[{\citenamefont{del Aguila et~al.}(2008)\citenamefont{del Aguila,
  Aguilar-Saavedra, Allanach, Alwall, Andreev et~al.}}]{delAguila:2008iz}
\bibinfo{author}{\bibfnamefont{F.}~\bibnamefont{del Aguila}},
  \bibinfo{author}{\bibfnamefont{J.}~\bibnamefont{Aguilar-Saavedra}},
  \bibinfo{author}{\bibfnamefont{B.}~\bibnamefont{Allanach}},
  \bibinfo{author}{\bibfnamefont{J.}~\bibnamefont{Alwall}},
  \bibinfo{author}{\bibfnamefont{Y.}~\bibnamefont{Andreev}},
  \bibnamefont{et~al.}, \bibinfo{journal}{Eur.Phys.J.}
  \textbf{\bibinfo{volume}{C57}}, \bibinfo{pages}{183} (\bibinfo{year}{2008}),
  \eprint{0801.1800}.

\bibitem[{\citenamefont{Raidal et~al.}(2008)\citenamefont{Raidal, van~der
  Schaaf, Bigi, Mangano, Semertzidis et~al.}}]{Raidal:2008jk}
\bibinfo{author}{\bibfnamefont{M.}~\bibnamefont{Raidal}},
  \bibinfo{author}{\bibfnamefont{A.}~\bibnamefont{van~der Schaaf}},
  \bibinfo{author}{\bibfnamefont{I.}~\bibnamefont{Bigi}},
  \bibinfo{author}{\bibfnamefont{M.}~\bibnamefont{Mangano}},
  \bibinfo{author}{\bibfnamefont{Y.~K.} \bibnamefont{Semertzidis}},
  \bibnamefont{et~al.}, \bibinfo{journal}{Eur.Phys.J.}
  \textbf{\bibinfo{volume}{C57}}, \bibinfo{pages}{13} (\bibinfo{year}{2008}),
  \eprint{0801.1826}.

\bibitem[{\citenamefont{Grossman and Neubert}(2000)}]{Grossman:1999ra}
\bibinfo{author}{\bibfnamefont{Y.}~\bibnamefont{Grossman}} \bibnamefont{and}
  \bibinfo{author}{\bibfnamefont{M.}~\bibnamefont{Neubert}},
  \bibinfo{journal}{Phys.Lett.} \textbf{\bibinfo{volume}{B474}},
  \bibinfo{pages}{361} (\bibinfo{year}{2000}), \eprint{hep-ph/9912408}.

\bibitem[{\citenamefont{Appelquist et~al.}(2002)\citenamefont{Appelquist,
  Dobrescu, Ponton, and Yee}}]{Appelquist:2002ft}
\bibinfo{author}{\bibfnamefont{T.}~\bibnamefont{Appelquist}},
  \bibinfo{author}{\bibfnamefont{B.~A.} \bibnamefont{Dobrescu}},
  \bibinfo{author}{\bibfnamefont{E.}~\bibnamefont{Ponton}}, \bibnamefont{and}
  \bibinfo{author}{\bibfnamefont{H.-U.} \bibnamefont{Yee}},
  \bibinfo{journal}{Phys.Rev.} \textbf{\bibinfo{volume}{D65}},
  \bibinfo{pages}{105019} (\bibinfo{year}{2002}), \eprint{hep-ph/0201131}.

\bibitem[{\citenamefont{Gherghetta}(2004)}]{Gherghetta:2003he}
\bibinfo{author}{\bibfnamefont{T.}~\bibnamefont{Gherghetta}},
  \bibinfo{journal}{Phys.Rev.Lett.} \textbf{\bibinfo{volume}{92}},
  \bibinfo{pages}{161601} (\bibinfo{year}{2004}), \eprint{hep-ph/0312392}.

\bibitem[{\citenamefont{Moreau}(2005)}]{Moreau:2004qe}
\bibinfo{author}{\bibfnamefont{G.}~\bibnamefont{Moreau}},
  \bibinfo{journal}{Eur.Phys.J.} \textbf{\bibinfo{volume}{C40}},
  \bibinfo{pages}{539} (\bibinfo{year}{2005}), \eprint{hep-ph/0407177}.

\bibitem[{\citenamefont{Bouchart et~al.}(2011)\citenamefont{Bouchart, Knochel,
  and Moreau}}]{Bouchart:2011va}
\bibinfo{author}{\bibfnamefont{C.}~\bibnamefont{Bouchart}},
  \bibinfo{author}{\bibfnamefont{A.}~\bibnamefont{Knochel}}, \bibnamefont{and}
  \bibinfo{author}{\bibfnamefont{G.}~\bibnamefont{Moreau}},
  \bibinfo{journal}{Phys.Rev.} \textbf{\bibinfo{volume}{D84}},
  \bibinfo{pages}{015016} (\bibinfo{year}{2011}), \eprint{1101.0634}.

\bibitem[{\citenamefont{Goertz et~al.}(2012)\citenamefont{Goertz, Haisch, and
  Neubert}}]{Goertz:2011hj}
\bibinfo{author}{\bibfnamefont{F.}~\bibnamefont{Goertz}},
  \bibinfo{author}{\bibfnamefont{U.}~\bibnamefont{Haisch}}, \bibnamefont{and}
  \bibinfo{author}{\bibfnamefont{M.}~\bibnamefont{Neubert}},
  \bibinfo{journal}{Phys.Lett.} \textbf{\bibinfo{volume}{B713}},
  \bibinfo{pages}{23} (\bibinfo{year}{2012}), \eprint{1112.5099}.

\bibitem[{\citenamefont{Kaplan and Georgi}(1984)}]{Kaplan:1983fs}
\bibinfo{author}{\bibfnamefont{D.~B.} \bibnamefont{Kaplan}} \bibnamefont{and}
  \bibinfo{author}{\bibfnamefont{H.}~\bibnamefont{Georgi}},
  \bibinfo{journal}{Phys.Lett.} \textbf{\bibinfo{volume}{B136}},
  \bibinfo{pages}{183} (\bibinfo{year}{1984}).

\bibitem[{\citenamefont{Kaplan et~al.}(1984)\citenamefont{Kaplan, Georgi, and
  Dimopoulos}}]{Kaplan:1983sm}
\bibinfo{author}{\bibfnamefont{D.~B.} \bibnamefont{Kaplan}},
  \bibinfo{author}{\bibfnamefont{H.}~\bibnamefont{Georgi}}, \bibnamefont{and}
  \bibinfo{author}{\bibfnamefont{S.}~\bibnamefont{Dimopoulos}},
  \bibinfo{journal}{Phys.Lett.} \textbf{\bibinfo{volume}{B136}},
  \bibinfo{pages}{187} (\bibinfo{year}{1984}).

\bibitem[{\citenamefont{Contino et~al.}(2003)\citenamefont{Contino, Nomura, and
  Pomarol}}]{Contino:2003ve}
\bibinfo{author}{\bibfnamefont{R.}~\bibnamefont{Contino}},
  \bibinfo{author}{\bibfnamefont{Y.}~\bibnamefont{Nomura}}, \bibnamefont{and}
  \bibinfo{author}{\bibfnamefont{A.}~\bibnamefont{Pomarol}},
  \bibinfo{journal}{Nucl.Phys.} \textbf{\bibinfo{volume}{B671}},
  \bibinfo{pages}{148} (\bibinfo{year}{2003}), \eprint{hep-ph/0306259}.

\bibitem[{\citenamefont{Agashe et~al.}(2005{\natexlab{b}})\citenamefont{Agashe,
  Contino, and Pomarol}}]{Agashe:2004rs}
\bibinfo{author}{\bibfnamefont{K.}~\bibnamefont{Agashe}},
  \bibinfo{author}{\bibfnamefont{R.}~\bibnamefont{Contino}}, \bibnamefont{and}
  \bibinfo{author}{\bibfnamefont{A.}~\bibnamefont{Pomarol}},
  \bibinfo{journal}{Nucl.Phys.} \textbf{\bibinfo{volume}{B719}},
  \bibinfo{pages}{165} (\bibinfo{year}{2005}{\natexlab{b}}),
  \eprint{hep-ph/0412089}.

\bibitem[{\citenamefont{Contino et~al.}(2007)\citenamefont{Contino, Da~Rold,
  and Pomarol}}]{Contino:2006qr}
\bibinfo{author}{\bibfnamefont{R.}~\bibnamefont{Contino}},
  \bibinfo{author}{\bibfnamefont{L.}~\bibnamefont{Da~Rold}}, \bibnamefont{and}
  \bibinfo{author}{\bibfnamefont{A.}~\bibnamefont{Pomarol}},
  \bibinfo{journal}{Phys.Rev.} \textbf{\bibinfo{volume}{D75}},
  \bibinfo{pages}{055014} (\bibinfo{year}{2007}), \eprint{hep-ph/0612048}.

\bibitem[{\citenamefont{Burdman and Da~Rold}(2007)}]{Burdman:2007sx}
\bibinfo{author}{\bibfnamefont{G.}~\bibnamefont{Burdman}} \bibnamefont{and}
  \bibinfo{author}{\bibfnamefont{L.}~\bibnamefont{Da~Rold}},
  \bibinfo{journal}{JHEP} \textbf{\bibinfo{volume}{0712}}, \bibinfo{pages}{086}
  (\bibinfo{year}{2007}), \eprint{arXiv:0710.0623 [hep-ph]}.

\bibitem[{\citenamefont{Da~Rold}(2011)}]{DaRold:2010as}
\bibinfo{author}{\bibfnamefont{L.}~\bibnamefont{Da~Rold}},
  \bibinfo{journal}{JHEP} \textbf{\bibinfo{volume}{1102}}, \bibinfo{pages}{034}
  (\bibinfo{year}{2011}), \eprint{1009.2392}.

\bibitem[{\citenamefont{Azatov and Galloway}(2012)}]{Azatov:2011qy}
\bibinfo{author}{\bibfnamefont{A.}~\bibnamefont{Azatov}} \bibnamefont{and}
  \bibinfo{author}{\bibfnamefont{J.}~\bibnamefont{Galloway}},
  \bibinfo{journal}{Phys.Rev.} \textbf{\bibinfo{volume}{D85}},
  \bibinfo{pages}{055013} (\bibinfo{year}{2012}), \eprint{1110.5646}.

\bibitem[{\citenamefont{Hill}(1991)}]{Hill:1991at}
\bibinfo{author}{\bibfnamefont{C.~T.} \bibnamefont{Hill}},
  \bibinfo{journal}{Phys.Lett.} \textbf{\bibinfo{volume}{B266}},
  \bibinfo{pages}{419} (\bibinfo{year}{1991}).

\bibitem[{\citenamefont{Pomarol and Serra}(2008)}]{Pomarol:2008bh}
\bibinfo{author}{\bibfnamefont{A.}~\bibnamefont{Pomarol}} \bibnamefont{and}
  \bibinfo{author}{\bibfnamefont{J.}~\bibnamefont{Serra}},
  \bibinfo{journal}{Phys.Rev.} \textbf{\bibinfo{volume}{D78}},
  \bibinfo{pages}{074026} (\bibinfo{year}{2008}), \eprint{0806.3247}.

\bibitem[{\citenamefont{Arkani-Hamed et~al.}(2001)\citenamefont{Arkani-Hamed,
  Cohen, and Georgi}}]{ArkaniHamed:2001nc}
\bibinfo{author}{\bibfnamefont{N.}~\bibnamefont{Arkani-Hamed}},
  \bibinfo{author}{\bibfnamefont{A.~G.} \bibnamefont{Cohen}}, \bibnamefont{and}
  \bibinfo{author}{\bibfnamefont{H.}~\bibnamefont{Georgi}},
  \bibinfo{journal}{Phys. Lett.} \textbf{\bibinfo{volume}{B513}},
  \bibinfo{pages}{232} (\bibinfo{year}{2001}), \eprint{hep-ph/0105239}.

\bibitem[{\citenamefont{Arkani-Hamed
  et~al.}(2002{\natexlab{a}})}]{ArkaniHamed:2002qx}
\bibinfo{author}{\bibfnamefont{N.}~\bibnamefont{Arkani-Hamed}}
  \bibnamefont{et~al.}, \bibinfo{journal}{JHEP} \textbf{\bibinfo{volume}{08}},
  \bibinfo{pages}{021} (\bibinfo{year}{2002}{\natexlab{a}}),
  \eprint{hep-ph/0206020}.

\bibitem[{\citenamefont{Arkani-Hamed
  et~al.}(2002{\natexlab{b}})\citenamefont{Arkani-Hamed, Cohen, Katz, and
  Nelson}}]{ArkaniHamed:2002qy}
\bibinfo{author}{\bibfnamefont{N.}~\bibnamefont{Arkani-Hamed}},
  \bibinfo{author}{\bibfnamefont{A.}~\bibnamefont{Cohen}},
  \bibinfo{author}{\bibfnamefont{E.}~\bibnamefont{Katz}}, \bibnamefont{and}
  \bibinfo{author}{\bibfnamefont{A.}~\bibnamefont{Nelson}},
  \bibinfo{journal}{JHEP} \textbf{\bibinfo{volume}{0207}}, \bibinfo{pages}{034}
  (\bibinfo{year}{2002}{\natexlab{b}}), \eprint{hep-ph/0206021}.

\bibitem[{\citenamefont{Ishiwata and Wise}(2011)}]{Ishiwata:2011hr}
\bibinfo{author}{\bibfnamefont{K.}~\bibnamefont{Ishiwata}} \bibnamefont{and}
  \bibinfo{author}{\bibfnamefont{M.~B.} \bibnamefont{Wise}},
  \bibinfo{journal}{Phys.Rev.} \textbf{\bibinfo{volume}{D84}},
  \bibinfo{pages}{055025} (\bibinfo{year}{2011}), \eprint{1107.1490}.

\bibitem[{\citenamefont{Kilic et~al.}(2011)\citenamefont{Kilic, Kopp, and
  Okui}}]{Kilic:2010fs}
\bibinfo{author}{\bibfnamefont{C.}~\bibnamefont{Kilic}},
  \bibinfo{author}{\bibfnamefont{K.}~\bibnamefont{Kopp}}, \bibnamefont{and}
  \bibinfo{author}{\bibfnamefont{T.}~\bibnamefont{Okui}},
  \bibinfo{journal}{Phys.Rev.} \textbf{\bibinfo{volume}{D83}},
  \bibinfo{pages}{015006} (\bibinfo{year}{2011}), \eprint{arXiv:1008.2763
  [hep-ph]}.

\bibitem[{\citenamefont{Barger et~al.}(2012)\citenamefont{Barger, Ishida, and
  Keung}}]{Barger:2012hv}
\bibinfo{author}{\bibfnamefont{V.}~\bibnamefont{Barger}},
  \bibinfo{author}{\bibfnamefont{M.}~\bibnamefont{Ishida}}, \bibnamefont{and}
  \bibinfo{author}{\bibfnamefont{W.-Y.} \bibnamefont{Keung}}
  (\bibinfo{year}{2012}), \eprint{1203.3456}.

\bibitem[{\citenamefont{Low et~al.}(2012)\citenamefont{Low, Lykken, and
  Shaughnessy}}]{Low:2012rj}
\bibinfo{author}{\bibfnamefont{I.}~\bibnamefont{Low}},
  \bibinfo{author}{\bibfnamefont{J.}~\bibnamefont{Lykken}}, \bibnamefont{and}
  \bibinfo{author}{\bibfnamefont{G.}~\bibnamefont{Shaughnessy}}
  (\bibinfo{year}{2012}), \eprint{1207.1093}.

\bibitem[{\citenamefont{Corbett et~al.}(2012)\citenamefont{Corbett, Eboli,
  Gonzalez-Fraile, and Gonzalez-Garcia}}]{Corbett:2012dm}
\bibinfo{author}{\bibfnamefont{T.}~\bibnamefont{Corbett}},
  \bibinfo{author}{\bibfnamefont{O.}~\bibnamefont{Eboli}},
  \bibinfo{author}{\bibfnamefont{J.}~\bibnamefont{Gonzalez-Fraile}},
  \bibnamefont{and}
  \bibinfo{author}{\bibfnamefont{M.}~\bibnamefont{Gonzalez-Garcia}}
  (\bibinfo{year}{2012}), \eprint{1207.1344}.

\bibitem[{\citenamefont{Giardino
  et~al.}(2012{\natexlab{a}})\citenamefont{Giardino, Kannike, Raidal, and
  Strumia}}]{Giardino:2012ww}
\bibinfo{author}{\bibfnamefont{P.~P.} \bibnamefont{Giardino}},
  \bibinfo{author}{\bibfnamefont{K.}~\bibnamefont{Kannike}},
  \bibinfo{author}{\bibfnamefont{M.}~\bibnamefont{Raidal}}, \bibnamefont{and}
  \bibinfo{author}{\bibfnamefont{A.}~\bibnamefont{Strumia}}
  (\bibinfo{year}{2012}{\natexlab{a}}), \eprint{1203.4254}.

\bibitem[{\citenamefont{Giardino
  et~al.}(2012{\natexlab{b}})\citenamefont{Giardino, Kannike, Raidal, and
  Strumia}}]{Giardino:2012dp}
\bibinfo{author}{\bibfnamefont{P.~P.} \bibnamefont{Giardino}},
  \bibinfo{author}{\bibfnamefont{K.}~\bibnamefont{Kannike}},
  \bibinfo{author}{\bibfnamefont{M.}~\bibnamefont{Raidal}}, \bibnamefont{and}
  \bibinfo{author}{\bibfnamefont{A.}~\bibnamefont{Strumia}}
  (\bibinfo{year}{2012}{\natexlab{b}}), \eprint{1207.1347}.

\bibitem[{\citenamefont{Ellis and You}(2012{\natexlab{a}})}]{Ellis:2012rx}
\bibinfo{author}{\bibfnamefont{J.}~\bibnamefont{Ellis}} \bibnamefont{and}
  \bibinfo{author}{\bibfnamefont{T.}~\bibnamefont{You}}
  (\bibinfo{year}{2012}{\natexlab{a}}), \eprint{1204.0464}.

\bibitem[{\citenamefont{Ellis and You}(2012{\natexlab{b}})}]{Ellis:2012hz}
\bibinfo{author}{\bibfnamefont{J.}~\bibnamefont{Ellis}} \bibnamefont{and}
  \bibinfo{author}{\bibfnamefont{T.}~\bibnamefont{You}}
  (\bibinfo{year}{2012}{\natexlab{b}}), \eprint{1207.1693}.

\bibitem[{\citenamefont{Azatov et~al.}(2012{\natexlab{a}})\citenamefont{Azatov,
  Contino, and Galloway}}]{Azatov:2012bz}
\bibinfo{author}{\bibfnamefont{A.}~\bibnamefont{Azatov}},
  \bibinfo{author}{\bibfnamefont{R.}~\bibnamefont{Contino}}, \bibnamefont{and}
  \bibinfo{author}{\bibfnamefont{J.}~\bibnamefont{Galloway}},
  \bibinfo{journal}{JHEP} \textbf{\bibinfo{volume}{1204}}, \bibinfo{pages}{127}
  (\bibinfo{year}{2012}{\natexlab{a}}), \eprint{1202.3415}.

\bibitem[{\citenamefont{Azatov et~al.}(2012{\natexlab{b}})\citenamefont{Azatov,
  Contino, Del~Re, Galloway, Grassi et~al.}}]{Azatov:2012rd}
\bibinfo{author}{\bibfnamefont{A.}~\bibnamefont{Azatov}},
  \bibinfo{author}{\bibfnamefont{R.}~\bibnamefont{Contino}},
  \bibinfo{author}{\bibfnamefont{D.}~\bibnamefont{Del~Re}},
  \bibinfo{author}{\bibfnamefont{J.}~\bibnamefont{Galloway}},
  \bibinfo{author}{\bibfnamefont{M.}~\bibnamefont{Grassi}},
  \bibnamefont{et~al.} (\bibinfo{year}{2012}{\natexlab{b}}),
  \eprint{1204.4817}.

\bibitem[{\citenamefont{Montull and Riva}(2012)}]{Montull:2012ik}
\bibinfo{author}{\bibfnamefont{M.}~\bibnamefont{Montull}} \bibnamefont{and}
  \bibinfo{author}{\bibfnamefont{F.}~\bibnamefont{Riva}}
  (\bibinfo{year}{2012}), \eprint{1207.1716}.

\bibitem[{\citenamefont{Espinosa
  et~al.}(2012{\natexlab{a}})\citenamefont{Espinosa, Grojean, Muhlleitner, and
  Trott}}]{Espinosa:2012ir}
\bibinfo{author}{\bibfnamefont{J.}~\bibnamefont{Espinosa}},
  \bibinfo{author}{\bibfnamefont{C.}~\bibnamefont{Grojean}},
  \bibinfo{author}{\bibfnamefont{M.}~\bibnamefont{Muhlleitner}},
  \bibnamefont{and} \bibinfo{author}{\bibfnamefont{M.}~\bibnamefont{Trott}},
  \bibinfo{journal}{JHEP} \textbf{\bibinfo{volume}{1205}}, \bibinfo{pages}{097}
  (\bibinfo{year}{2012}{\natexlab{a}}), \eprint{1202.3697}.

\bibitem[{\citenamefont{Espinosa
  et~al.}(2012{\natexlab{b}})\citenamefont{Espinosa, Grojean, Muhlleitner, and
  Trott}}]{Espinosa:2012im}
\bibinfo{author}{\bibfnamefont{J.}~\bibnamefont{Espinosa}},
  \bibinfo{author}{\bibfnamefont{C.}~\bibnamefont{Grojean}},
  \bibinfo{author}{\bibfnamefont{M.}~\bibnamefont{Muhlleitner}},
  \bibnamefont{and} \bibinfo{author}{\bibfnamefont{M.}~\bibnamefont{Trott}}
  (\bibinfo{year}{2012}{\natexlab{b}}), \eprint{1207.1717}.

\bibitem[{\citenamefont{Carmi et~al.}(2012)\citenamefont{Carmi, Falkowski,
  Kuflik, Volansky, and Zupan}}]{Carmi:2012in}
\bibinfo{author}{\bibfnamefont{D.}~\bibnamefont{Carmi}},
  \bibinfo{author}{\bibfnamefont{A.}~\bibnamefont{Falkowski}},
  \bibinfo{author}{\bibfnamefont{E.}~\bibnamefont{Kuflik}},
  \bibinfo{author}{\bibfnamefont{T.}~\bibnamefont{Volansky}}, \bibnamefont{and}
  \bibinfo{author}{\bibfnamefont{J.}~\bibnamefont{Zupan}}
  (\bibinfo{year}{2012}), \eprint{1207.1718}.

\bibitem[{\citenamefont{Banerjee et~al.}(2012)\citenamefont{Banerjee,
  Mukhopadhyay, and Mukhopadhyaya}}]{Banerjee:2012xc}
\bibinfo{author}{\bibfnamefont{S.}~\bibnamefont{Banerjee}},
  \bibinfo{author}{\bibfnamefont{S.}~\bibnamefont{Mukhopadhyay}},
  \bibnamefont{and}
  \bibinfo{author}{\bibfnamefont{B.}~\bibnamefont{Mukhopadhyaya}},
  \bibinfo{journal}{JHEP} \textbf{\bibinfo{volume}{1210}}, \bibinfo{pages}{062}
  (\bibinfo{year}{2012}), \eprint{1207.3588}.

\bibitem[{\citenamefont{Plehn and Rauch}(2012)}]{Plehn:2012iz}
\bibinfo{author}{\bibfnamefont{T.}~\bibnamefont{Plehn}} \bibnamefont{and}
  \bibinfo{author}{\bibfnamefont{M.}~\bibnamefont{Rauch}}
  (\bibinfo{year}{2012}), \eprint{1207.6108}.

\bibitem[{\citenamefont{{ATLAS
  Collaboration}}(2012{\natexlab{a}})}]{CONF-2012-127}
\bibinfo{author}{\bibnamefont{{ATLAS Collaboration}}}
  (\bibinfo{year}{2012}{\natexlab{a}}), \eprint{Note CONF-2012-127}.

\bibitem[{\citenamefont{Agashe et~al.}(2003)\citenamefont{Agashe, Delgado, May,
  and Sundrum}}]{Agashe:2003zs}
\bibinfo{author}{\bibfnamefont{K.}~\bibnamefont{Agashe}},
  \bibinfo{author}{\bibfnamefont{A.}~\bibnamefont{Delgado}},
  \bibinfo{author}{\bibfnamefont{M.~J.} \bibnamefont{May}}, \bibnamefont{and}
  \bibinfo{author}{\bibfnamefont{R.}~\bibnamefont{Sundrum}},
  \bibinfo{journal}{JHEP} \textbf{\bibinfo{volume}{0308}}, \bibinfo{pages}{050}
  (\bibinfo{year}{2003}), \eprint{hep-ph/0308036}.

\bibitem[{\citenamefont{Agashe et~al.}(2006{\natexlab{b}})\citenamefont{Agashe,
  Contino, Da~Rold, and Pomarol}}]{Agashe:2006at}
\bibinfo{author}{\bibfnamefont{K.}~\bibnamefont{Agashe}},
  \bibinfo{author}{\bibfnamefont{R.}~\bibnamefont{Contino}},
  \bibinfo{author}{\bibfnamefont{L.}~\bibnamefont{Da~Rold}}, \bibnamefont{and}
  \bibinfo{author}{\bibfnamefont{A.}~\bibnamefont{Pomarol}},
  \bibinfo{journal}{Phys.Lett.} \textbf{\bibinfo{volume}{B641}},
  \bibinfo{pages}{62} (\bibinfo{year}{2006}{\natexlab{b}}),
  \eprint{hep-ph/0605341}.

\bibitem[{\citenamefont{Djouadi et~al.}(2007)\citenamefont{Djouadi, Moreau, and
  Richard}}]{Djouadi:2006rk}
\bibinfo{author}{\bibfnamefont{A.}~\bibnamefont{Djouadi}},
  \bibinfo{author}{\bibfnamefont{G.}~\bibnamefont{Moreau}}, \bibnamefont{and}
  \bibinfo{author}{\bibfnamefont{F.}~\bibnamefont{Richard}},
  \bibinfo{journal}{Nucl.Phys.} \textbf{\bibinfo{volume}{B773}},
  \bibinfo{pages}{43} (\bibinfo{year}{2007}), \eprint{hep-ph/0610173}.

\bibitem[{\citenamefont{Carena et~al.}(2007)\citenamefont{Carena, Ponton,
  Santiago, and Wagner}}]{Carena:2007ua}
\bibinfo{author}{\bibfnamefont{M.~S.} \bibnamefont{Carena}},
  \bibinfo{author}{\bibfnamefont{E.}~\bibnamefont{Ponton}},
  \bibinfo{author}{\bibfnamefont{J.}~\bibnamefont{Santiago}}, \bibnamefont{and}
  \bibinfo{author}{\bibfnamefont{C.~E.~M.} \bibnamefont{Wagner}},
  \bibinfo{journal}{Phys.Rev.} \textbf{\bibinfo{volume}{D76}},
  \bibinfo{pages}{035006} (\bibinfo{year}{2007}), \eprint{hep-ph/0701055}.

\bibitem[{\citenamefont{Djouadi et~al.}(2008)\citenamefont{Djouadi, Moreau, and
  Singh}}]{Djouadi:2007eg}
\bibinfo{author}{\bibfnamefont{A.}~\bibnamefont{Djouadi}},
  \bibinfo{author}{\bibfnamefont{G.}~\bibnamefont{Moreau}}, \bibnamefont{and}
  \bibinfo{author}{\bibfnamefont{R.~K.} \bibnamefont{Singh}},
  \bibinfo{journal}{Nucl.Phys.} \textbf{\bibinfo{volume}{B797}},
  \bibinfo{pages}{1} (\bibinfo{year}{2008}), \eprint{0706.4191}.

\bibitem[{\citenamefont{Djouadi and Moreau}(2008)}]{Djouadi:2007fm}
\bibinfo{author}{\bibfnamefont{A.}~\bibnamefont{Djouadi}} \bibnamefont{and}
  \bibinfo{author}{\bibfnamefont{G.}~\bibnamefont{Moreau}},
  \bibinfo{journal}{Phys.Lett.} \textbf{\bibinfo{volume}{B660}},
  \bibinfo{pages}{67} (\bibinfo{year}{2008}), \eprint{0707.3800}.

\bibitem[{\citenamefont{Ledroit et~al.}(2007)\citenamefont{Ledroit, Moreau, and
  Morel}}]{Ledroit:2007ik}
\bibinfo{author}{\bibfnamefont{F.}~\bibnamefont{Ledroit}},
  \bibinfo{author}{\bibfnamefont{G.}~\bibnamefont{Moreau}}, \bibnamefont{and}
  \bibinfo{author}{\bibfnamefont{J.}~\bibnamefont{Morel}},
  \bibinfo{journal}{JHEP} \textbf{\bibinfo{volume}{0709}}, \bibinfo{pages}{071}
  (\bibinfo{year}{2007}), \eprint{hep-ph/0703262}.

\bibitem[{\citenamefont{Bouchart and
  Moreau}(2009{\natexlab{a}})}]{Bouchart:2008vp}
\bibinfo{author}{\bibfnamefont{C.}~\bibnamefont{Bouchart}} \bibnamefont{and}
  \bibinfo{author}{\bibfnamefont{G.}~\bibnamefont{Moreau}},
  \bibinfo{journal}{Nucl.Phys.} \textbf{\bibinfo{volume}{B810}},
  \bibinfo{pages}{66} (\bibinfo{year}{2009}{\natexlab{a}}),
  \eprint{arXiv:0807.4461 [hep-ph]}.

\bibitem[{\citenamefont{Bouchart and
  Moreau}(2009{\natexlab{b}})}]{Bouchart:2009vq}
\bibinfo{author}{\bibfnamefont{C.}~\bibnamefont{Bouchart}} \bibnamefont{and}
  \bibinfo{author}{\bibfnamefont{G.}~\bibnamefont{Moreau}},
  \bibinfo{journal}{Phys.Rev.} \textbf{\bibinfo{volume}{D80}},
  \bibinfo{pages}{095022} (\bibinfo{year}{2009}{\natexlab{b}}),
  \eprint{arXiv:0909.4812 [hep-ph]}.

\bibitem[{\citenamefont{Djouadi et~al.}(2010)\citenamefont{Djouadi, Moreau,
  Richard, and Singh}}]{Djouadi:2009nb}
\bibinfo{author}{\bibfnamefont{A.}~\bibnamefont{Djouadi}},
  \bibinfo{author}{\bibfnamefont{G.}~\bibnamefont{Moreau}},
  \bibinfo{author}{\bibfnamefont{F.}~\bibnamefont{Richard}}, \bibnamefont{and}
  \bibinfo{author}{\bibfnamefont{R.~K.} \bibnamefont{Singh}},
  \bibinfo{journal}{Phys.Rev.} \textbf{\bibinfo{volume}{D82}},
  \bibinfo{pages}{071702} (\bibinfo{year}{2010}), \eprint{0906.0604}.

\bibitem[{\citenamefont{Casagrande et~al.}(2010)\citenamefont{Casagrande,
  Goertz, Haisch, Neubert, and Pfoh}}]{Casagrande:2010si}
\bibinfo{author}{\bibfnamefont{S.}~\bibnamefont{Casagrande}},
  \bibinfo{author}{\bibfnamefont{F.}~\bibnamefont{Goertz}},
  \bibinfo{author}{\bibfnamefont{U.}~\bibnamefont{Haisch}},
  \bibinfo{author}{\bibfnamefont{M.}~\bibnamefont{Neubert}}, \bibnamefont{and}
  \bibinfo{author}{\bibfnamefont{T.}~\bibnamefont{Pfoh}},
  \bibinfo{journal}{JHEP} \textbf{\bibinfo{volume}{1009}}, \bibinfo{pages}{014}
  (\bibinfo{year}{2010}), \eprint{1005.4315}.

\bibitem[{\citenamefont{Djouadi et~al.}(2011)\citenamefont{Djouadi, Moreau, and
  Richard}}]{Djouadi:2011aj}
\bibinfo{author}{\bibfnamefont{A.}~\bibnamefont{Djouadi}},
  \bibinfo{author}{\bibfnamefont{G.}~\bibnamefont{Moreau}}, \bibnamefont{and}
  \bibinfo{author}{\bibfnamefont{F.}~\bibnamefont{Richard}},
  \bibinfo{journal}{Phys.Lett.} \textbf{\bibinfo{volume}{B701}},
  \bibinfo{pages}{458} (\bibinfo{year}{2011}), \eprint{1105.3158}.

\bibitem[{\citenamefont{del Aguila et~al.}(2003)\citenamefont{del Aguila,
  Perez-Victoria, and Santiago}}]{delAguila:2003bh}
\bibinfo{author}{\bibfnamefont{F.}~\bibnamefont{del Aguila}},
  \bibinfo{author}{\bibfnamefont{M.}~\bibnamefont{Perez-Victoria}},
  \bibnamefont{and} \bibinfo{author}{\bibfnamefont{J.}~\bibnamefont{Santiago}},
  \bibinfo{journal}{JHEP} \textbf{\bibinfo{volume}{0302}}, \bibinfo{pages}{051}
  (\bibinfo{year}{2003}), \eprint{hep-th/0302023}.

\bibitem[{\citenamefont{Cabrer et~al.}(2012)\citenamefont{Cabrer, von
  Gersdorff, and Quiros}}]{Cabrer:2011qb}
\bibinfo{author}{\bibfnamefont{J.~A.} \bibnamefont{Cabrer}},
  \bibinfo{author}{\bibfnamefont{G.}~\bibnamefont{von Gersdorff}},
  \bibnamefont{and} \bibinfo{author}{\bibfnamefont{M.}~\bibnamefont{Quiros}},
  \bibinfo{journal}{JHEP} \textbf{\bibinfo{volume}{1201}}, \bibinfo{pages}{033}
  (\bibinfo{year}{2012}), \eprint{1110.3324}.

\bibitem[{\citenamefont{Dawson and Furlan}(2012)}]{Dawson:2012di}
\bibinfo{author}{\bibfnamefont{S.}~\bibnamefont{Dawson}} \bibnamefont{and}
  \bibinfo{author}{\bibfnamefont{E.}~\bibnamefont{Furlan}},
  \bibinfo{journal}{Phys.Rev.} \textbf{\bibinfo{volume}{D86}},
  \bibinfo{pages}{015021} (\bibinfo{year}{2012}), \eprint{1205.4733}.

\bibitem[{\citenamefont{Carena et~al.}(2012{\natexlab{a}})\citenamefont{Carena,
  Low, and Wagner}}]{Carena:2012xa}
\bibinfo{author}{\bibfnamefont{M.}~\bibnamefont{Carena}},
  \bibinfo{author}{\bibfnamefont{I.}~\bibnamefont{Low}}, \bibnamefont{and}
  \bibinfo{author}{\bibfnamefont{C.~E.} \bibnamefont{Wagner}},
  \bibinfo{journal}{JHEP} \textbf{\bibinfo{volume}{1208}}, \bibinfo{pages}{060}
  (\bibinfo{year}{2012}{\natexlab{a}}), \eprint{1206.1082}.

\bibitem[{\citenamefont{Azatov et~al.}(2012{\natexlab{c}})\citenamefont{Azatov,
  Bondu, Falkowski, Felcini, Gascon-Shotkin et~al.}}]{Azatov:2012rj}
\bibinfo{author}{\bibfnamefont{A.}~\bibnamefont{Azatov}},
  \bibinfo{author}{\bibfnamefont{O.}~\bibnamefont{Bondu}},
  \bibinfo{author}{\bibfnamefont{A.}~\bibnamefont{Falkowski}},
  \bibinfo{author}{\bibfnamefont{M.}~\bibnamefont{Felcini}},
  \bibinfo{author}{\bibfnamefont{S.}~\bibnamefont{Gascon-Shotkin}},
  \bibnamefont{et~al.}, \bibinfo{journal}{Phys.Rev.}
  \textbf{\bibinfo{volume}{D85}}, \bibinfo{pages}{115022}
  (\bibinfo{year}{2012}{\natexlab{c}}), \eprint{1204.0455}.

\bibitem[{\citenamefont{Bonne and Moreau}(2012)}]{Bonne:2012im}
\bibinfo{author}{\bibfnamefont{N.}~\bibnamefont{Bonne}} \bibnamefont{and}
  \bibinfo{author}{\bibfnamefont{G.}~\bibnamefont{Moreau}},
  \bibinfo{journal}{Phys.Lett.} \textbf{\bibinfo{volume}{B717}},
  \bibinfo{pages}{409} (\bibinfo{year}{2012}), \eprint{1206.3360}.

\bibitem[{\citenamefont{Joglekar et~al.}(2012)\citenamefont{Joglekar,
  Schwaller, and Wagner}}]{Joglekar:2012vc}
\bibinfo{author}{\bibfnamefont{A.}~\bibnamefont{Joglekar}},
  \bibinfo{author}{\bibfnamefont{P.}~\bibnamefont{Schwaller}},
  \bibnamefont{and} \bibinfo{author}{\bibfnamefont{C.~E.} \bibnamefont{Wagner}}
  (\bibinfo{year}{2012}), \eprint{1207.4235}.

\bibitem[{\citenamefont{Kearney et~al.}(2012)\citenamefont{Kearney, Pierce, and
  Weiner}}]{Kearney:2012zi}
\bibinfo{author}{\bibfnamefont{J.}~\bibnamefont{Kearney}},
  \bibinfo{author}{\bibfnamefont{A.}~\bibnamefont{Pierce}}, \bibnamefont{and}
  \bibinfo{author}{\bibfnamefont{N.}~\bibnamefont{Weiner}}
  (\bibinfo{year}{2012}), \eprint{1207.7062}.

\bibitem[{\citenamefont{Voloshin}(2012)}]{Voloshin:2012tv}
\bibinfo{author}{\bibfnamefont{M.}~\bibnamefont{Voloshin}}
  (\bibinfo{year}{2012}), \eprint{1208.4303}.

\bibitem[{\citenamefont{Batell et~al.}(2012)\citenamefont{Batell, Gori, and
  Wang}}]{Batell:2012ca}
\bibinfo{author}{\bibfnamefont{B.}~\bibnamefont{Batell}},
  \bibinfo{author}{\bibfnamefont{S.}~\bibnamefont{Gori}}, \bibnamefont{and}
  \bibinfo{author}{\bibfnamefont{L.-T.} \bibnamefont{Wang}}
  (\bibinfo{year}{2012}), \eprint{1209.6382}.

\bibitem[{\citenamefont{Arkani-Hamed et~al.}(2012)\citenamefont{Arkani-Hamed,
  Blum, D'Agnolo, and Fan}}]{ArkaniHamed:2012kq}
\bibinfo{author}{\bibfnamefont{N.}~\bibnamefont{Arkani-Hamed}},
  \bibinfo{author}{\bibfnamefont{K.}~\bibnamefont{Blum}},
  \bibinfo{author}{\bibfnamefont{R.~T.} \bibnamefont{D'Agnolo}},
  \bibnamefont{and} \bibinfo{author}{\bibfnamefont{J.}~\bibnamefont{Fan}}
  (\bibinfo{year}{2012}), \eprint{1207.4482}.

\bibitem[{\citenamefont{Almeida et~al.}(2012)\citenamefont{Almeida, Bertuzzo,
  Machado, and Funchal}}]{Almeida:2012bq}
\bibinfo{author}{\bibfnamefont{L.~G.} \bibnamefont{Almeida}},
  \bibinfo{author}{\bibfnamefont{E.}~\bibnamefont{Bertuzzo}},
  \bibinfo{author}{\bibfnamefont{P.~A.} \bibnamefont{Machado}},
  \bibnamefont{and} \bibinfo{author}{\bibfnamefont{R.~Z.}
  \bibnamefont{Funchal}} (\bibinfo{year}{2012}), \eprint{1207.5254}.

\bibitem[{\citenamefont{Djouadi}(2008)}]{Djouadi:2005gi}
\bibinfo{author}{\bibfnamefont{A.}~\bibnamefont{Djouadi}},
  \bibinfo{journal}{Phys.Rept.} \textbf{\bibinfo{volume}{457}},
  \bibinfo{pages}{1} (\bibinfo{year}{2008}), \eprint{hep-ph/0503172}.

\bibitem[{\citenamefont{Nakamura et~al.}(2010)}]{Beringer:1900zz}
\bibinfo{author}{\bibfnamefont{K.}~\bibnamefont{Nakamura}} \bibnamefont{et~al.}
  (\bibinfo{collaboration}{Particle Data Group}), \bibinfo{journal}{J.Phys.}
  \textbf{\bibinfo{volume}{G37}}, \bibinfo{pages}{075021}
  (\bibinfo{year}{2010}).

\bibitem[{\citenamefont{Cacciapaglia et~al.}(2009)\citenamefont{Cacciapaglia,
  Deandrea, and Llodra-Perez}}]{Cacciapaglia:2009ky}
\bibinfo{author}{\bibfnamefont{G.}~\bibnamefont{Cacciapaglia}},
  \bibinfo{author}{\bibfnamefont{A.}~\bibnamefont{Deandrea}}, \bibnamefont{and}
  \bibinfo{author}{\bibfnamefont{J.}~\bibnamefont{Llodra-Perez}},
  \bibinfo{journal}{JHEP} \textbf{\bibinfo{volume}{0906}}, \bibinfo{pages}{054}
  (\bibinfo{year}{2009}), \eprint{0901.0927}.

\bibitem[{\citenamefont{Cacciapaglia et~al.}(2010)\citenamefont{Cacciapaglia,
  Deandrea, Harada, and Okada}}]{Cacciapaglia:2010vn}
\bibinfo{author}{\bibfnamefont{G.}~\bibnamefont{Cacciapaglia}},
  \bibinfo{author}{\bibfnamefont{A.}~\bibnamefont{Deandrea}},
  \bibinfo{author}{\bibfnamefont{D.}~\bibnamefont{Harada}}, \bibnamefont{and}
  \bibinfo{author}{\bibfnamefont{Y.}~\bibnamefont{Okada}},
  \bibinfo{journal}{JHEP} \textbf{\bibinfo{volume}{1011}}, \bibinfo{pages}{159}
  (\bibinfo{year}{2010}), \eprint{1007.2933}.

\bibitem[{\citenamefont{Cacciapaglia et~al.}(2012)\citenamefont{Cacciapaglia,
  Deandrea, Panizzi, Gaur, Harada et~al.}}]{Cacciapaglia:2011fx}
\bibinfo{author}{\bibfnamefont{G.}~\bibnamefont{Cacciapaglia}},
  \bibinfo{author}{\bibfnamefont{A.}~\bibnamefont{Deandrea}},
  \bibinfo{author}{\bibfnamefont{L.}~\bibnamefont{Panizzi}},
  \bibinfo{author}{\bibfnamefont{N.}~\bibnamefont{Gaur}},
  \bibinfo{author}{\bibfnamefont{D.}~\bibnamefont{Harada}},
  \bibnamefont{et~al.}, \bibinfo{journal}{JHEP}
  \textbf{\bibinfo{volume}{1203}}, \bibinfo{pages}{070} (\bibinfo{year}{2012}),
  \eprint{1108.6329}.

\bibitem[{\citenamefont{{LHC Higgs Cross Section Working
  Group}}(2012)}]{XsectionWepPage}
\bibinfo{author}{\bibnamefont{{LHC Higgs Cross Section Working Group}}}
  (\bibinfo{year}{2012}), \eprint{webpage, \\ {\tt
  https://twiki.cern.ch/twiki/bin/view/LHCPhysics/CrossSections}}.

\bibitem[{\citenamefont{Group et~al.}(2012)\citenamefont{Group, Collaborations,
  the Tevatron New~Physics, and Working}}]{:2012zzl}
\bibinfo{author}{\bibfnamefont{C.}~\bibnamefont{Group}},
  \bibinfo{author}{\bibfnamefont{D.}~\bibnamefont{Collaborations}},
  \bibinfo{author}{\bibnamefont{the Tevatron New~Physics}}, \bibnamefont{and}
  \bibinfo{author}{\bibfnamefont{H.}~\bibnamefont{Working}}
  (\bibinfo{collaboration}{Tevatron New Physics Higgs Working Group, CDF
  Collaboration, D0 Collaboration}) (\bibinfo{year}{2012}), \eprint{1207.0449}.

\bibitem[{\citenamefont{{ATLAS
  Collaboration}}(2012{\natexlab{b}})}]{CONF-2012-091}
\bibinfo{author}{\bibnamefont{{ATLAS Collaboration}}}
  (\bibinfo{year}{2012}{\natexlab{b}}), \eprint{Note CONF-2012-091}.

\bibitem[{\citenamefont{{CMS
  Collaboration}}(2012{\natexlab{a}})}]{PAS-HIG-12-015}
\bibinfo{author}{\bibnamefont{{CMS Collaboration}}}
  (\bibinfo{year}{2012}{\natexlab{a}}), \eprint{Note PAS HIG-12-015}.

\bibitem[{\citenamefont{Chatrchyan et~al.}(2012{\natexlab{a}})}]{:2012gu}
\bibinfo{author}{\bibfnamefont{S.}~\bibnamefont{Chatrchyan}}
  \bibnamefont{et~al.} (\bibinfo{collaboration}{CMS Collaboration}),
  \bibinfo{journal}{Phys.Lett.} \textbf{\bibinfo{volume}{B716}},
  \bibinfo{pages}{30} (\bibinfo{year}{2012}{\natexlab{a}}), \eprint{1207.7235}.

\bibitem[{\citenamefont{Aad et~al.}(2012)}]{:2012gk}
\bibinfo{author}{\bibfnamefont{G.}~\bibnamefont{Aad}} \bibnamefont{et~al.}
  (\bibinfo{collaboration}{ATLAS Collaboration}), \bibinfo{journal}{Phys.Lett.}
  \textbf{\bibinfo{volume}{B716}}, \bibinfo{pages}{1} (\bibinfo{year}{2012}),
  \eprint{1207.7214}.

\bibitem[{\citenamefont{{ATLAS
  Collaboration}}(2012{\natexlab{c}})}]{CONF-2012-093}
\bibinfo{author}{\bibnamefont{{ATLAS Collaboration}}}
  (\bibinfo{year}{2012}{\natexlab{c}}), \eprint{Note CONF-2012-093}.

\bibitem[{\citenamefont{{CMS
  Collaboration}}(2012{\natexlab{b}})}]{PAS-HIG-12-020}
\bibinfo{author}{\bibnamefont{{CMS Collaboration}}}
  (\bibinfo{year}{2012}{\natexlab{b}}), \eprint{Note PAS HIG-12-020}.

\bibitem[{\citenamefont{{ATLAS
  Collaboration}}(2012{\natexlab{d}})}]{CONF-2012-019}
\bibinfo{author}{\bibnamefont{{ATLAS Collaboration}}}
  (\bibinfo{year}{2012}{\natexlab{d}}), \eprint{Note CONF-2012-019}.

\bibitem[{\citenamefont{{ATLAS
  Collaboration}}(2012{\natexlab{e}})}]{CONF-2012-098}
\bibinfo{author}{\bibnamefont{{ATLAS Collaboration}}}
  (\bibinfo{year}{2012}{\natexlab{e}}), \eprint{Note CONF-2012-098}.

\bibitem[{\citenamefont{Dawson and Jaiswal}(2010)}]{Dawson:2010yz}
\bibinfo{author}{\bibfnamefont{S.}~\bibnamefont{Dawson}} \bibnamefont{and}
  \bibinfo{author}{\bibfnamefont{P.}~\bibnamefont{Jaiswal}},
  \bibinfo{journal}{Phys.Rev.} \textbf{\bibinfo{volume}{D81}},
  \bibinfo{pages}{073008} (\bibinfo{year}{2010}), \eprint{1002.2672}.

\bibitem[{\citenamefont{Carena et~al.}(2012{\natexlab{b}})\citenamefont{Carena,
  Gori, Juste, Menon, Wagner et~al.}}]{Carena:2012rw}
\bibinfo{author}{\bibfnamefont{M.}~\bibnamefont{Carena}},
  \bibinfo{author}{\bibfnamefont{S.}~\bibnamefont{Gori}},
  \bibinfo{author}{\bibfnamefont{A.}~\bibnamefont{Juste}},
  \bibinfo{author}{\bibfnamefont{A.}~\bibnamefont{Menon}},
  \bibinfo{author}{\bibfnamefont{C.~E.} \bibnamefont{Wagner}},
  \bibnamefont{et~al.}, \bibinfo{journal}{JHEP}
  \textbf{\bibinfo{volume}{1207}}, \bibinfo{pages}{091}
  (\bibinfo{year}{2012}{\natexlab{b}}), \eprint{1203.1041}.

\bibitem[{\citenamefont{Chatrchyan
  et~al.}(2012{\natexlab{b}})}]{Chatrchyan:2012yea}
\bibinfo{author}{\bibfnamefont{S.}~\bibnamefont{Chatrchyan}}
  \bibnamefont{et~al.} (\bibinfo{collaboration}{CMS Collaboration}),
  \bibinfo{journal}{JHEP} \textbf{\bibinfo{volume}{1205}}, \bibinfo{pages}{123}
  (\bibinfo{year}{2012}{\natexlab{b}}), \eprint{1204.1088}.

\bibitem[{\citenamefont{{XLVIIIth ``Rencontres de Moriond''}}(2013)}]{Moriond}
\bibinfo{author}{\bibnamefont{{XLVIIIth ``Rencontres de Moriond''}}}
  (\bibinfo{year}{2013}), \eprint{Session devoted to ELECTROWEAK INTERACTIONS
  AND UNIFIED THEORIES, March 2nd - 16th 2013, La Thuile, Italy.}

\bibitem[{\citenamefont{{ATLAS
  Collaboration}}(2013{\natexlab{a}})}]{CONF-2013-012}
\bibinfo{author}{\bibnamefont{{ATLAS Collaboration}}}
  (\bibinfo{year}{2013}{\natexlab{a}}), \eprint{Note CONF-2013-012}.

\bibitem[{\citenamefont{{ATLAS
  Collaboration}}(2013{\natexlab{b}})}]{CONF-2013-013}
\bibinfo{author}{\bibnamefont{{ATLAS Collaboration}}}
  (\bibinfo{year}{2013}{\natexlab{b}}), \eprint{Note CONF-2013-013}.

\bibitem[{\citenamefont{{ATLAS
  Collaboration}}(2013{\natexlab{c}})}]{CONF-2013-030}
\bibinfo{author}{\bibnamefont{{ATLAS Collaboration}}}
  (\bibinfo{year}{2013}{\natexlab{c}}), \eprint{Note CONF-2013-030}.

\bibitem[{\citenamefont{{ATLAS
  Collaboration}}(2012{\natexlab{f}})}]{CONF-2012-170}
\bibinfo{author}{\bibnamefont{{ATLAS Collaboration}}}
  (\bibinfo{year}{2012}{\natexlab{f}}), \eprint{Note CONF-2012-170}.

\bibitem[{\citenamefont{{ATLAS
  Collaboration}}(2012{\natexlab{g}})}]{CONF-2012-160}
\bibinfo{author}{\bibnamefont{{ATLAS Collaboration}}}
  (\bibinfo{year}{2012}{\natexlab{g}}), \eprint{Note CONF-2012-160}.

\bibitem[{\citenamefont{{ATLAS
  Collaboration}}(2013{\natexlab{d}})}]{CONF-2013-014}
\bibinfo{author}{\bibnamefont{{ATLAS Collaboration}}}
  (\bibinfo{year}{2013}{\natexlab{d}}), \eprint{Note CONF-2013-014}.

\bibitem[{\citenamefont{{ATLAS
  Collaboration}}(2013{\natexlab{e}})}]{CONF-2013-034}
\bibinfo{author}{\bibnamefont{{ATLAS Collaboration}}}
  (\bibinfo{year}{2013}{\natexlab{e}}), \eprint{Note CONF-2013-034}.

\bibitem[{\citenamefont{{Ochando, Christophe}}(2013)}]{Ochando}
\bibinfo{author}{\bibnamefont{{Ochando, Christophe}}} (\bibinfo{year}{2013}),
  \eprint{Talk [on behalf of the CMS Collaboration] at the Moriond
  Conference~\cite{Moriond}.}

\bibitem[{\citenamefont{{CMS
  Collaboration}}(2013{\natexlab{a}})}]{PAS-HIG-13-002}
\bibinfo{author}{\bibnamefont{{CMS Collaboration}}}
  (\bibinfo{year}{2013}{\natexlab{a}}), \eprint{Note PAS HIG-13-002}.

\bibitem[{\citenamefont{{CMS
  Collaboration}}(2013{\natexlab{b}})}]{PAS-HIG-13-003}
\bibinfo{author}{\bibnamefont{{CMS Collaboration}}}
  (\bibinfo{year}{2013}{\natexlab{b}}), \eprint{Note PAS HIG-13-003}.

\bibitem[{\citenamefont{{CMS
  Collaboration}}(2012{\natexlab{c}})}]{PAS-HIG-12-045}
\bibinfo{author}{\bibnamefont{{CMS Collaboration}}}
  (\bibinfo{year}{2012}{\natexlab{c}}), \eprint{Note PAS HIG-12-045}.

\bibitem[{\citenamefont{{CMS
  Collaboration}}(2013{\natexlab{c}})}]{PAS-HIG-13-004}
\bibinfo{author}{\bibnamefont{{CMS Collaboration}}}
  (\bibinfo{year}{2013}{\natexlab{c}}), \eprint{Note PAS HIG-13-004}.

\end{thebibliography}

\vspace{1.5cm} 
\appendix 
\noindent \textbf{\Large Appendix} 
\vspace{0.5cm}


\renewcommand{\thesubsection}{A.\arabic{subsection}} 
\renewcommand{\theequation}{A.\arabic{equation}} 
\setcounter{subsection}{0} 
\setcounter{equation}{0} 

\section{The updated results}
\label{Updated}

\begin{figure}[h!]
\begin{center}
\vspace{2.cm}
\begin{tabular}{cc}
\includegraphics[width=0.47\textwidth,height=8.cm]{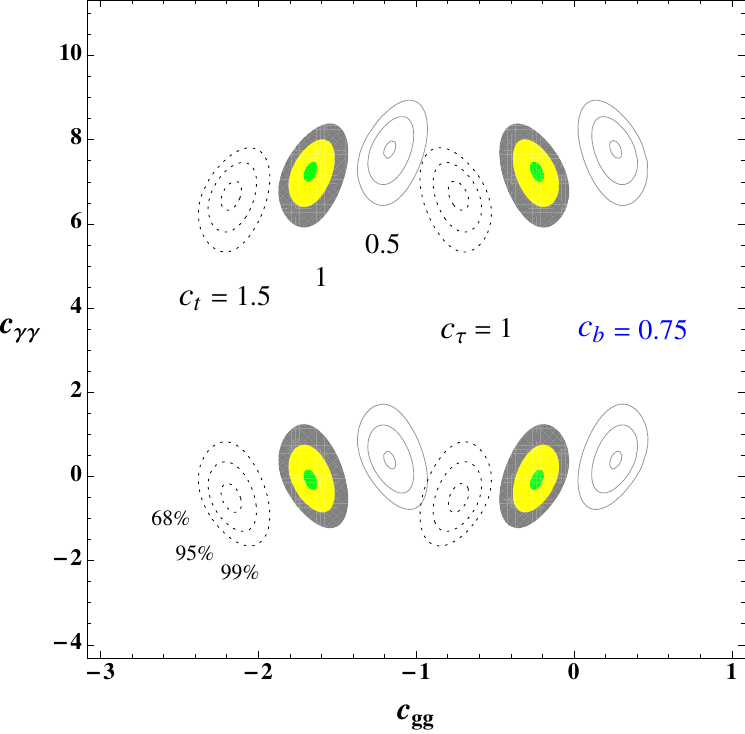}
\hspace{0.5cm}
\includegraphics[width=0.47\textwidth,height=8.cm]{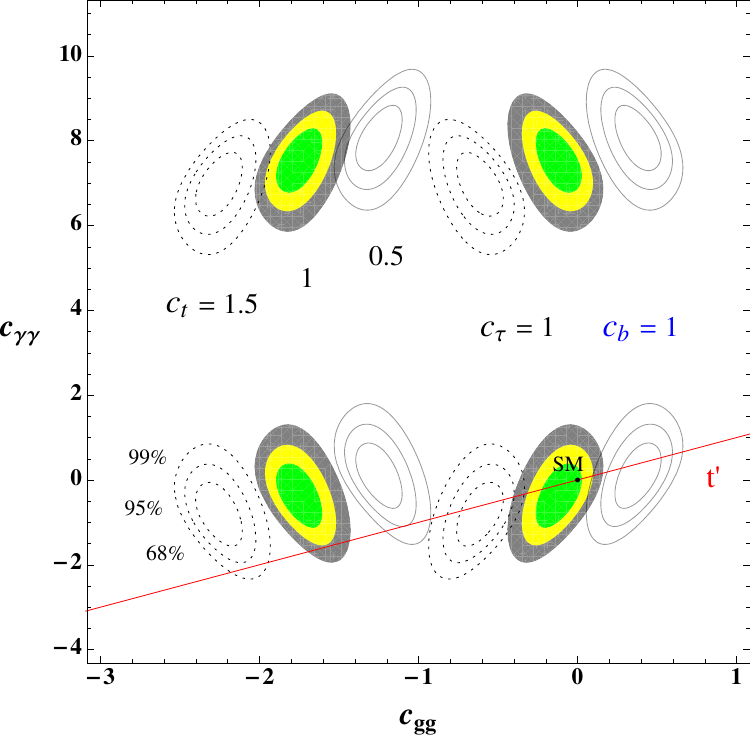}
\vspace{0.1cm}
\\
\includegraphics[width=0.47\textwidth,height=8cm]{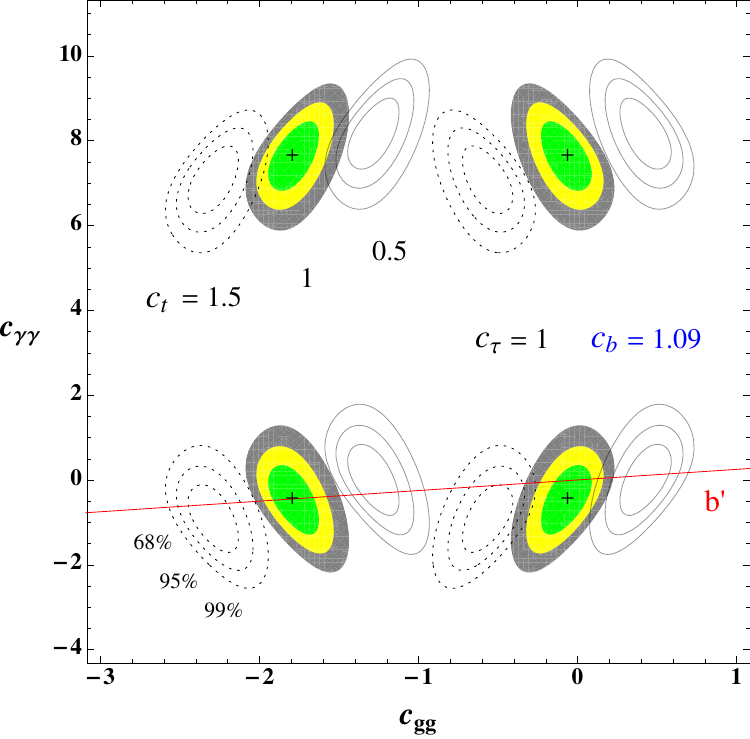}
\hspace{0.5cm}
\includegraphics[width=0.47\textwidth,height=8cm]{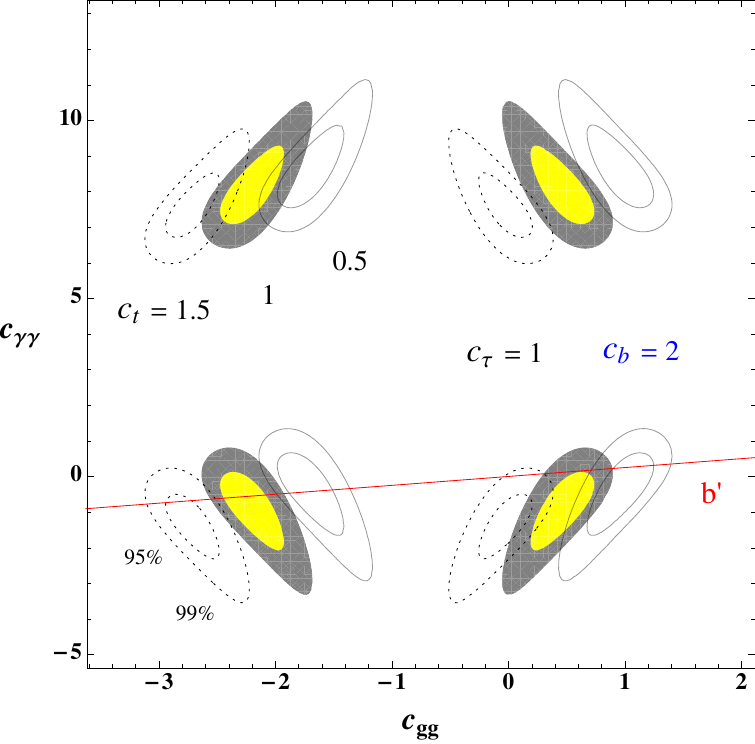}
\\
\end{tabular}
\caption{Update of Fig.(\ref{Fig:cbVAR}).}
\label{Fig:cbVARnew}
\end{center}
\end{figure}

\begin{figure}[h!]
\begin{center}
\vspace{2.cm}
\begin{tabular}{cc}
\includegraphics[width=0.33\textwidth,height=6.cm]{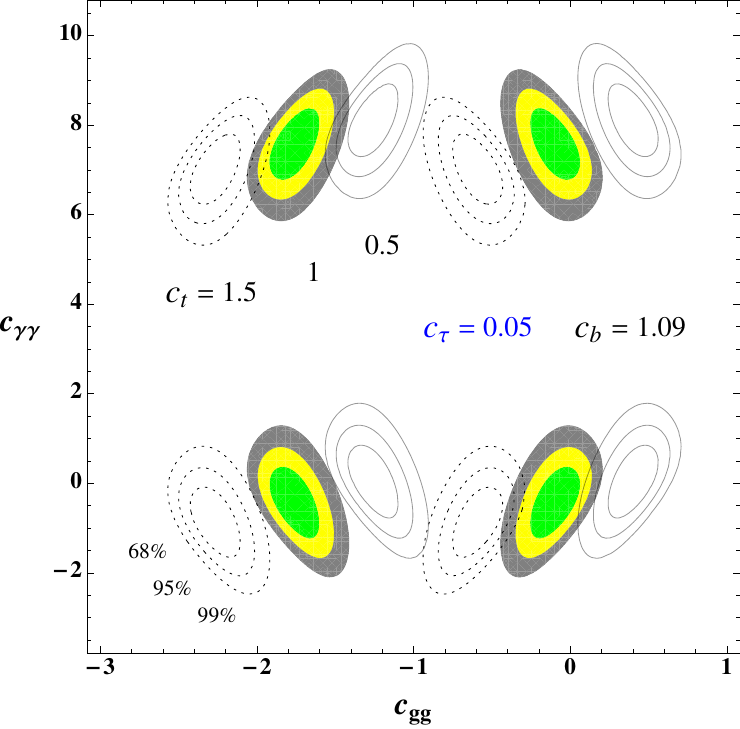}
\includegraphics[width=0.33\textwidth,height=6.cm]{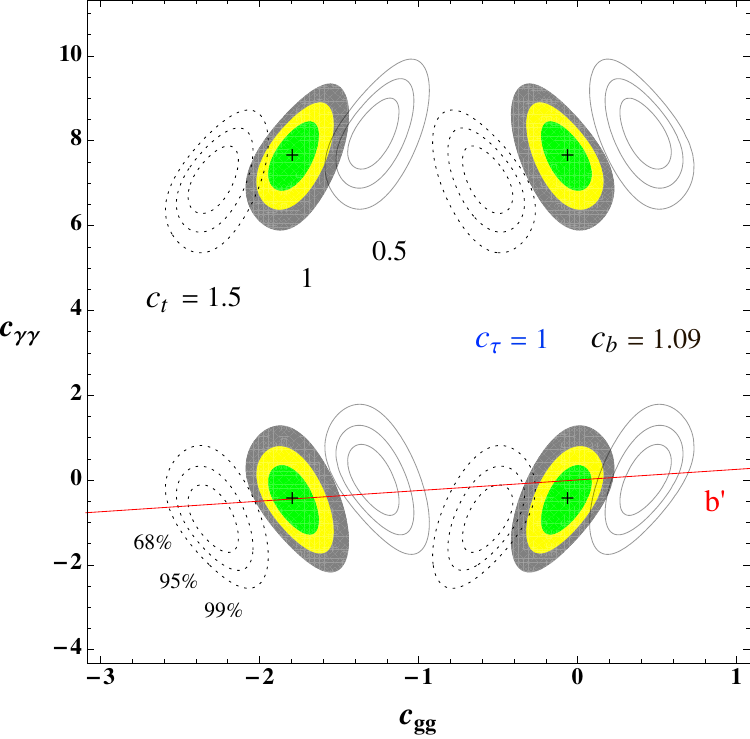}
\includegraphics[width=0.33\textwidth,height=6.cm]{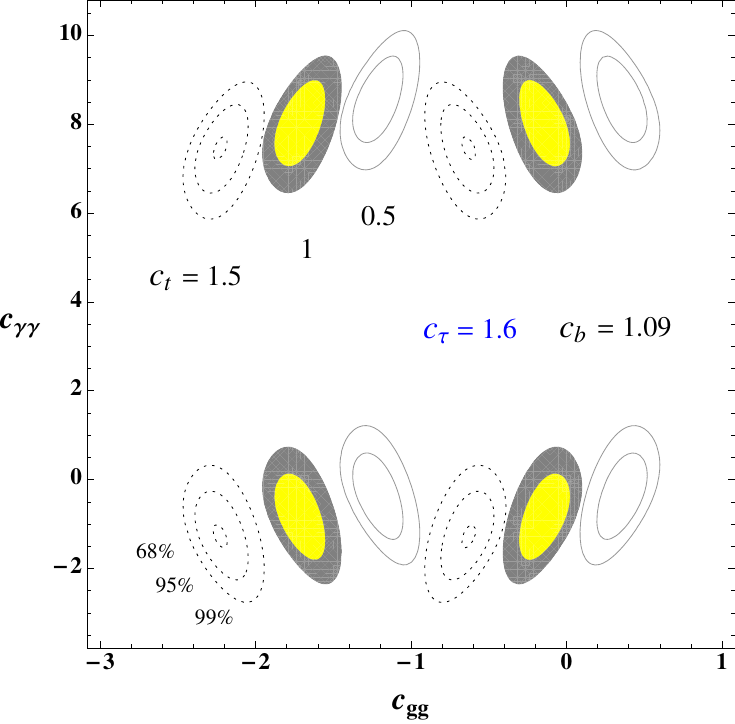}
\end{tabular}
\caption{Update of Fig.(\ref{Fig:ctauVAR}).}
\label{Fig:ctauVARnew}
\end{center}
\end{figure}

\begin{figure}[h!]
\begin{center}
\includegraphics[width=0.44\textwidth,height=7.7cm]{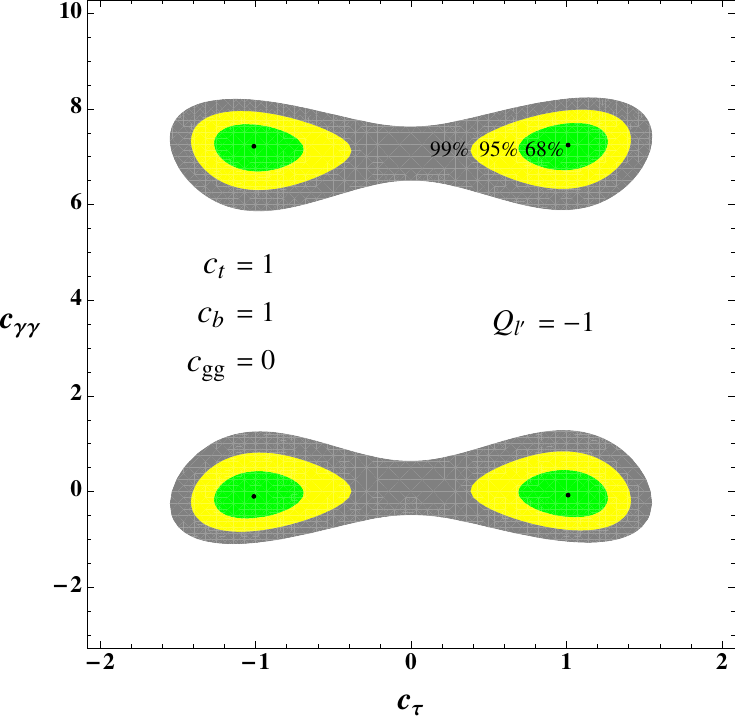}
\caption{Update of Fig.(\ref{Fig:EFcgg-lepton}).}
\label{Fig:EFcgg-leptonnew}
\end{center}
\end{figure}

\begin{figure}[h!]
\begin{center}
\vspace{3.5cm}
\includegraphics[width=0.47\textwidth,height=8cm]{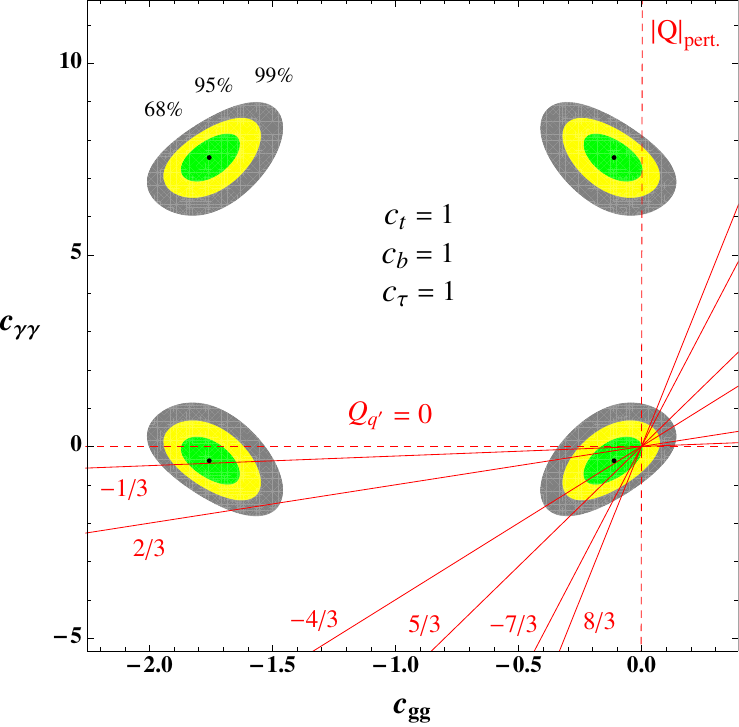}
\caption{Update of Fig.(\ref{Fig:EFcgg}).}
\label{Fig:EFcggnew}
\end{center}
\end{figure}

\begin{figure}[h!]
\begin{center}
\vspace{2.5cm}
\begin{tabular}{cc}
\includegraphics[width=0.42\textwidth,height=6.7cm]{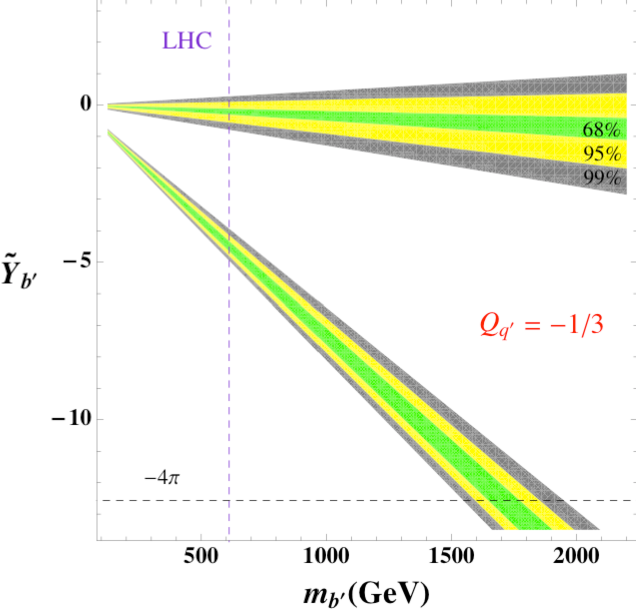}
\hspace{0.6cm}
\includegraphics[width=0.42\textwidth,height=6.7cm]{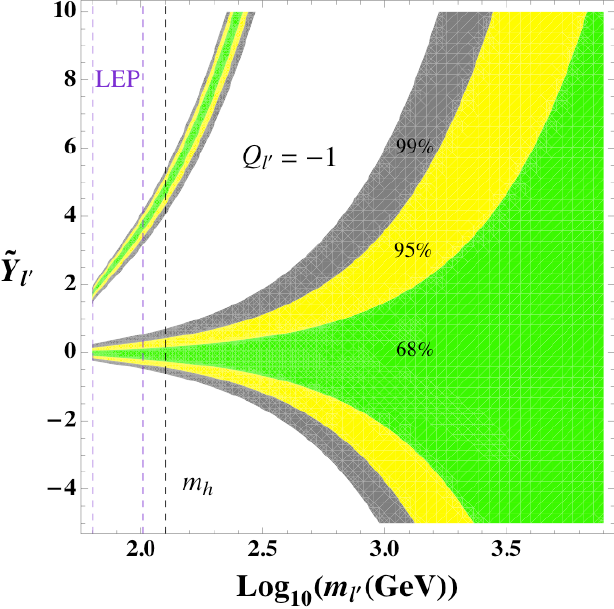}
\vspace{0.5cm}
\\
\includegraphics[width=0.42\textwidth,height=6.7cm]{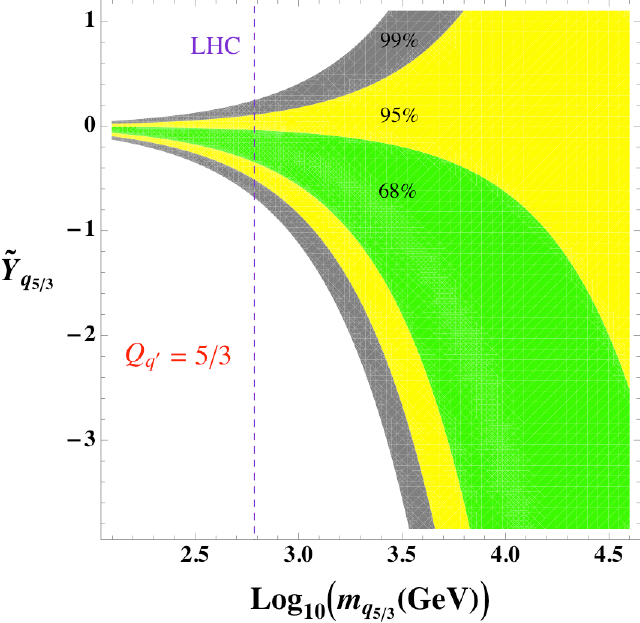}
\hspace{0.6cm}
\includegraphics[width=0.45\textwidth,height=6.7cm]{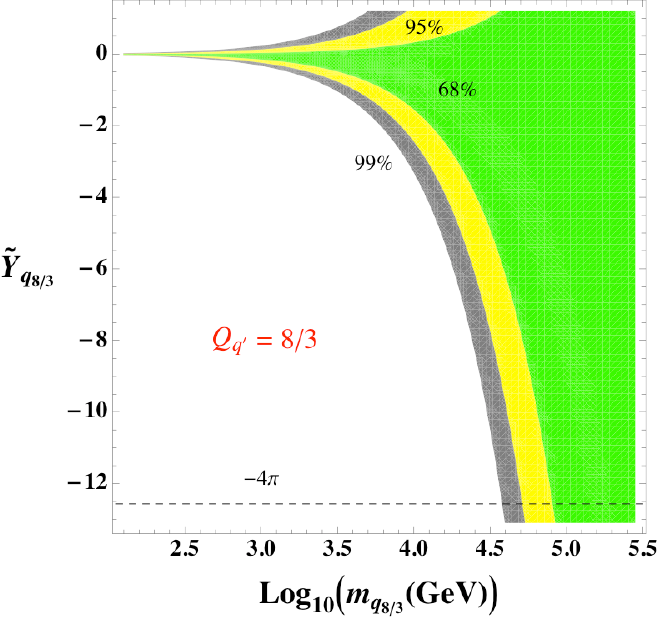}
\\
\end{tabular}
\caption{Update of Fig.(\ref{Fig:Ym}).}
\label{Fig:Ymnew}
\end{center}
\end{figure}

In Fig.(\ref{Fig:cbVARnew})-(\ref{Fig:ctauVARnew})-(\ref{Fig:EFcgg-leptonnew})-(\ref{Fig:EFcggnew})-(\ref{Fig:Ymnew}), 
we present the new numerical results of this paper based on the latest experimental data  
about the various Higgs boson rates provided by the LHC Collaborations -- including the ones shown at the
Moriond 2013 winter conference~\cite{Moriond}. It turns out that most of the Higgs channels have 
been updated.

Let us summarize here the latest references per channel and experiment;
Regarding the ATLAS data, 
the diphoton final state results are taken from Ref.~\cite{CONF-2013-012}, 
the $ZZ$ channel is from Ref.~\cite{CONF-2013-013}, 
the $WW$ channel from Ref.~\cite{CONF-2013-030}, 
the $b \bar b$ from Ref.~\cite{CONF-2012-170} and 
the $\tau\bar \tau$ from Ref.~\cite{CONF-2012-160}
(see also the combined channels in Ref.~\cite{CONF-2013-014,CONF-2013-034}).
\\ 
As for the CMS results,
the diphoton final state has just been presented in Ref.~\cite{Ochando}, 
the $ZZ$ channel measurements are provided in Ref.~\cite{PAS-HIG-13-002}, 
the $WW$ channel ones in Ref.~\cite{PAS-HIG-13-003},
the $b \bar b$ in Ref.~\cite{PAS-HIG-12-020,PAS-HIG-12-045} and 
the $\tau\bar \tau$ in Ref.~\cite{PAS-HIG-13-004}
(combined analyses in Ref.~\cite{PAS-HIG-12-045}).
\\
Finally, the latest results from the Tevatron (D0 and CDF Collaborations) can be found
in Ref.~\cite{:2012zzl}.

\end{document}